%% file: nova_xsec_tuning.tex
\documentclass[nofootinbib,superscriptaddress,10pt,preprintnumbers
]{revtex4-1}
\usepackage[final]{graphicx}
\usepackage{float}
\usepackage{natbib}
\usepackage{cprotect}
\usepackage{hyperref}
\usepackage{placeins}
\usepackage{refcount}
\usepackage[caption=false]{subfig}
\usepackage[margin=1in]{geometry}
\usepackage[utf8]{inputenc}
\usepackage{dcolumn}
\usepackage{bm}
\usepackage[normalem]{ulem}
\usepackage{units}
\usepackage{multirow}
\hypersetup{final}
\usepackage{fixme}
\fxsetup{theme=color,inline,nomargin}
\FXRegisterAuthor{jw}{ajw}{JW}
\FXRegisterAuthor{kb}{akb}{KB}

\def\valencia/{Val\`{e}ncia}

\newcommand{\numu}{\ensuremath{\nu_{\mu}}}
\newcommand{\ehad}{\ensuremath{E_{\text{had}}}}
\newcommand{\ehadvis}{\ensuremath{E_{\text{had}}^{\text{vis}}}}
\newcommand{\qzero}{\ensuremath{q_{0}}}
\newcommand{\qmag}{\ensuremath{|\vec{q}|}}
\newcommand{\qmagreco}{\ensuremath{|\vec{q}|_{\text{reco}}}}
\newcommand{\qZqThree}{\ensuremath{(\qzero, \qmag)}}
\newcommand{\recospace}{\ensuremath{(\ehadvis, \qmagreco)}}

\begin{document}

\preprint{FERMILAB-PUB-20-243-ND}

\graphicspath{{figs/}}

\title{Adjusting Neutrino Interaction Models and Evaluating Uncertainties using NOvA Near Detector Data}

\input{authorlist.inc}

\date{\today}

\begin{abstract}
The two-detector design of the NOvA neutrino oscillation experiment, in which two functionally identical detectors are exposed to an intense neutrino beam, aids in canceling leading order effects of cross-section uncertainties.
However, limited knowledge of neutrino interaction cross sections still
gives rise to some of the largest systematic uncertainties in current oscillation
measurements.
We show contemporary models of neutrino interactions to be discrepant with data from NOvA, consistent with discrepancies seen in other experiments.
Adjustments to neutrino interaction models in GENIE are presented, creating an effective model that improves agreement with our data.
We also describe systematic uncertainties on these models, including uncertainties on multi-nucleon interactions from a newly developed procedure using NOvA near detector data.
\end{abstract}

\maketitle

\section{Introduction}
\label{sec:intro}

The non-zero value of the reactor mixing angle $\theta_{13}$~\cite{t2k-th13, doublechooz, dayabay, reno} has enabled searches for leptonic CP violation and measurements of the neutrino mass ordering using long-baseline neutrino oscillation experiments with pion-decay-in-flight beams~\cite{novaosc2019, t2k2018, minos}.
Such experiments can also constrain or measure other standard neutrino oscillation model parameters, such as $\Delta m^2_{32}$ and $\theta_{23}$.

Long-baseline experiments generally utilize a two-detector design.
A smaller near detector (ND) close to the neutrino production target constrains neutrino flux and interaction cross sections.
A larger far detector (FD) is positioned to observe the neutrinos after oscillations.
Measurements are based on reconstructed neutrino energy spectra observed in the FD, which are compared to simulated predictions for various oscillation parameter values with systematic uncertainties taken into account.  ND data are used to adjust FD predictions and constrain systematic uncertainties, via either a simultaneous fit of ND and FD simulation to the respective data samples~\cite{t2kprd}, or by using differences between ND data and simulation to adjust FD simulation~\cite{t2kprd, wolcott-nufact}.  In either case, this process relies on simulation to account for oscillations and the differing beam flux and geometric acceptances between the detectors, making the ND constraint on the FD model-dependent.  Interactions of neutrinos with nuclei at neutrino energies around \unit[1]{GeV}, and the resulting final states, are challenging both to describe theoretically and to measure experimentally.  As a consequence, systematic uncertainties in neutrino interaction cross sections are typically among the largest uncertainties affecting long-baseline neutrino oscillation measurements, even with the two-detector approach~\cite{novaosc2019,t2k2018}.

NOvA is a long-baseline neutrino oscillation experiment, utilizing a \unit[14]{kton} FD located \unit[810]{km} downstream of the beam source and a functionally identical \unit[0.3]{kton} ND located approximately 1 km from beam target.  The detectors are made from PVC cells of cross section $\unit[3.9 \times 6.6]{cm^{2}}$ and of length $\unit[3.9]{m}$ (ND) or $\unit[15.5]{m}$ (FD), which are filled with organic liquid scintillator.
This results in detectors with 63\% active material by mass and a radiation length of \unit[38]{cm}.
Cells are extruded together in units and joined edgewise along the long dimension to produce square planes, which are then stacked perpendicular to the beam direction in alternating vertical and horizontal cell orientations to permit three-dimensional event reconstruction.
The near detector additionally has at its downstream end a ``muon catcher'' composed of a stack of ten sets of planes in which a pair of one vertically oriented and one horizontally oriented scintillator plane is interleaved with one \unit[10]{cm}-thick plane of steel.
Including the muon catcher, the ND can stop muons up to about \unit[3]{GeV}.
The FD is approximately four times longer, wider, and taller than the ND.

High-purity neutrino or antineutrino beams are produced by the NuMI facility at Fermilab~\cite{numibeam} according to the current polarity of two magnetic horns that focus and charge-select the parent hadrons.
The detectors are located \unit[14.6]{mrad} off-axis which results in an incoming neutrino energy spectrum narrowly peaked at \unit[2]{GeV}.
This neutrino energy is chosen to optimize sensitivity to oscillations, since $\nu_{e}$ appearance and $\nu_{\mu}$ disappearance probabilities both experience local maxima at an $L/E$ of around \unit[500]{km/GeV}.
The full NOvA experimental setup, including estimates for the neutrino flux from NuMI, is described in Refs.~\cite{novanumufa,novanuefa,novanumusa,novanuesa,novata,novaosc2019}.

This paper details adjustments made to the neutrino interaction models used in NOvA's simulation and the construction of associated systematic uncertainties.
NOvA's 2019 measurements of oscillation parameters~\cite{novaosc2019} use this work\footnote{The code used to produce these modifications for GENIE 2.12.2 is available at \url{https://github.com/novaexperiment/NOvARwgt-public}. There are minor differences between the full set of changes used in the oscillation measurements (\texttt{CVTune2018} in the code release) and what is shown in this paper (\texttt{CVTune2018$\_$RPAfix}).    The effect on the oscillation results is negligible.  See footnotes in  Sec.~\ref{subsec:ext}.}.

The data samples and observables used in the analysis are described in Sec.~\ref{sec:data}.
Details of the models in the simulation are given in Sec.~\ref{sec:base sim} and the adjustment procedure is developed in Sec.~\ref{sec:tune CV}.
We discuss systematic uncertainties associated with the adjustments and how we treat them in Sec.~\ref{sec:systs}.
Finally, we compare our findings to those of other experiments in Sec.~\ref{sec:other obs}.

\section{Data sample and reconstruction}
\label{sec:data}

The NOvA data presented here are from a near detector exposure of $8.03 \times 10^{20}$ protons on target with the neutrino beam and $3.10 \times 10^{20}$ protons on target with the antineutrino beam, totaling $1.48 \times 10^{6}$ selected neutrino interactions and $3.33 \times 10^{5}$ selected antineutrino interactions.  The events used here are the same events selected in the 2019 NOvA $\nu_{\mu}$ disappearance oscillation results~\cite{novaosc2019}, where the selection criteria, efficiencies, and purities are detailed.
After selection in the ND, we expect the neutrino beam candidate sample to be composed of 97.1\% muon neutrinos and 2.9\% muon antineutrinos, with negligible contributions from neutral-current (NC) or other charged-current (CC) neutrino flavors.
For the antineutrino beam we expect 90.2\% muon antineutrinos and 9.8\% muon neutrinos\footnote{The given fractions are with the adjustments described in the subsequent sections.  Without the adjustments, the selected sample in neutrino beam is predicted to be {$97.2\% $} {$\nu_{\mu}$} and {2.8\% $\bar{\nu}_{\mu}$}; the antineutrino beam sample is {90.0\%} {$\bar{\nu}_{\mu}$} and 10.0\% {$\nu_{\mu}$}.}.

Throughout this paper we compare various observables in our data and quantities we derive from them to the predictions we obtain from simulation.  For simulated observables, we distinguish the ``true,'' or generated, value from the ``reco'' value reconstructed in the detector.  The energy of muons that stop in the detectors ($E_{\mu}$) is measured with a resolution of about 3\% using track length, while the energy of all other particles, which collectively make up the hadronic recoil system, is measured using calorimetry.  For muon neutrino charged-current interactions in NOvA, the visible hadronic energy (\ehadvis{} or visible \ehad{}) is the sum of the calibrated observed hadronic energy deposits in scintillator.  This is distinct from the fully reconstructed hadronic energy, \ehad{}, which also accounts for unseen energy, such as that lost to dead material in the detector or to escaping invisible neutral particles. We measure \ehad{} with an energy resolution of about 30\%.  The reconstructed neutrino energy $E_{\nu}$ is the sum of $E_{\mu}$ and \ehad{}.

The variables $E_{\nu}$, $E_{\mu}$, $p_{\mu}$ (the muon momentum), and $\cos \theta_{\mu}$ (the opening angle between the muon and the neutrino beam directions) are estimated as in the $\nu_\mu$ disappearance analysis~\cite{novata}, as is the method for calculating visible \ehad{}. 
We use these, along with the muon mass $m_{\mu}$, to estimate the square of the four-momentum transferred from the initial neutrino to the nuclear system as
\[ Q_\mathrm{reco}^{2} = 2 E_{\nu} \left( E_{\mu} - p_{\mu} \cos \theta_{\mu} \right) - m_{\mu}^2. \]
In conjunction with the energy transfer $q_0$, which we measure as \ehad{}, $Q_\mathrm{reco}^2$ can then be used to approximate the three-momentum transfer as
\[ |\vec{q} |_\mathrm{reco} = \sqrt{Q_\mathrm{reco}^{2} + E_{\text{had}}^2}. \]
Finally, we combine \ehad{}, $Q_\mathrm{reco}^{2}$, and the proton mass $m_{p}$ to estimate the invariant mass of the hadronic system as
\[ W_\mathrm{reco} = \sqrt{m_{p}^{2} + 2m_{p} \ehad - Q_\mathrm{reco}^{2}}. \]

\section{Simulation}
\label{sec:base sim}

The NuMI beamline, including the 120 GeV protons and the hadrons produced by their interactions, is simulated using Geant4~\cite{geant4}, as is the flux of resultant neutrinos.  This neutrino flux is corrected using tools developed by the MINERvA collaboration for the NuMI beam, adding constraints on the hadron spectrum~\cite{ppfx}.  GENIE 2.12.2~\cite{genie-nim, geniemanual} (hereinafter referred to as GENIE) is used to predict the interactions of the neutrinos with the detector.  Adjustments to GENIE as used by NOvA are the focus of this paper.  GENIE prior to our adjustments will be referred to as the ``default'' simulation.

The simulation of neutrino interactions is separated into distinct parts within GENIE: the initial nuclear state, the hard scatter, and reinteractions of the resultant particles within the nuclear medium.
The initial state in the default GENIE configuration is a global relativistic Fermi gas (RFG) model based on the work of Smith and Moniz~\cite{smith-moniz} and modified by adding a high-momentum tail~\cite{bodek-ritchie} to account for potential short-range nuclear correlations~\cite{src-electron}. 
GENIE classifies the hard scatter into four primary interaction types.  At neutrino energies around \unit[2]{GeV}, the three most common are: quasi-elastic (QE) interactions, predicted according to the formalism attributed to Llewellyn Smith~\cite{llewellynsmith}, which result in a single baryon; resonance (RES) processes, predicted according to the model by Rein and Sehgal~\cite{rein-sehgal}, which result in baryons and mesons via an intermediate hadronic excited state; and what GENIE calls deep inelastic scattering (DIS), predicted using the Bodek-Yang scaling formalism~\cite{bodek-yang} together with a custom hadronization model~\cite{agky} and PYTHIA6~\cite{pythia}, which results in a broad spectrum of hadrons from inelastic scattering over a large range of hadronic invariant masses.  The fourth primary process is the rare instance where a neutrino scatters from the entire nucleus as a coherent whole (COH).

\begin{figure}[ht]
	\centering
    \subfloat[]{
		\centering
		\includegraphics[width=0.48\textwidth]{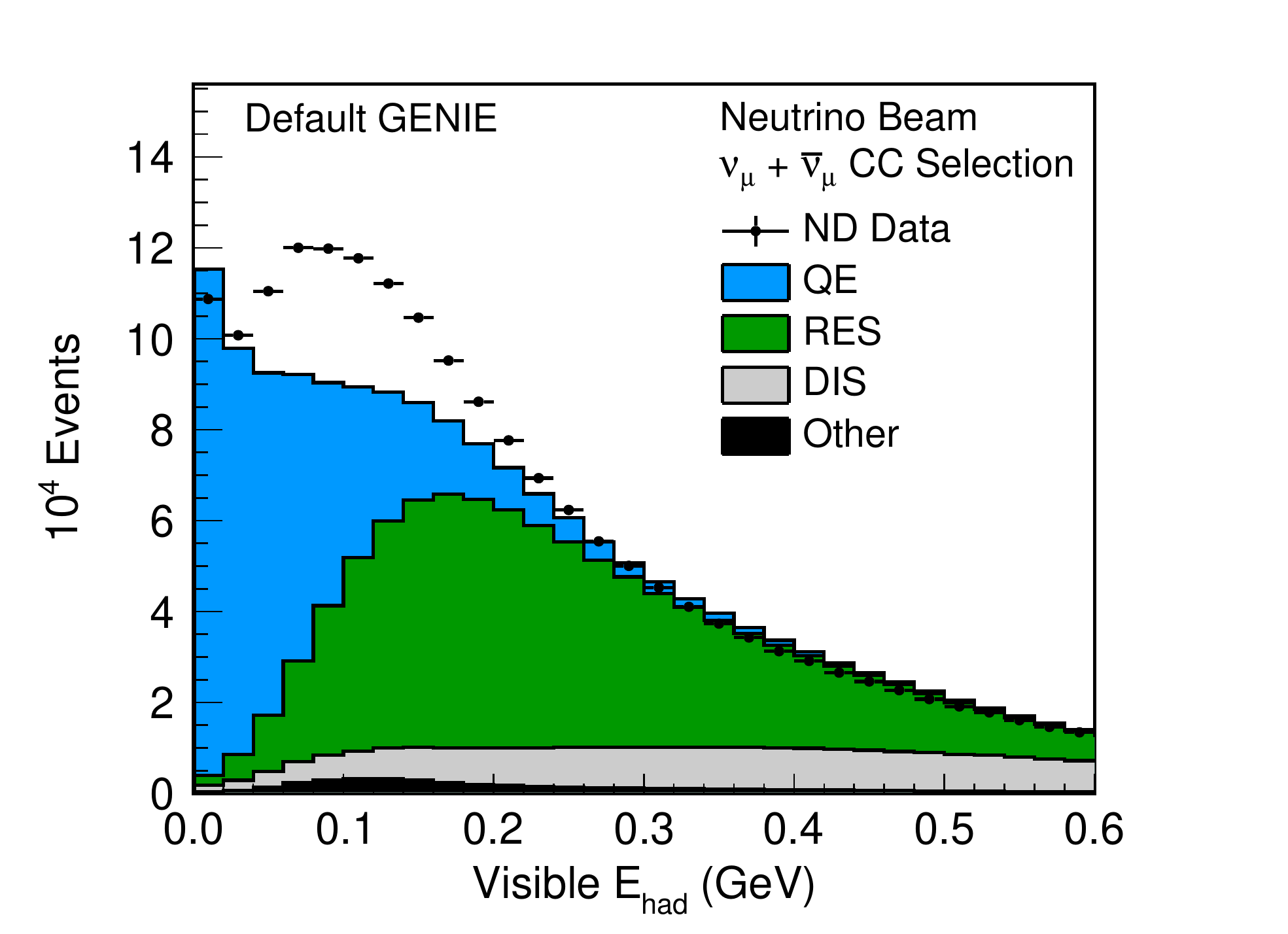}
		\label{fig:fhcdefaultgenie} }
    \subfloat[]{
		\centering
		\includegraphics[width=0.48\textwidth]{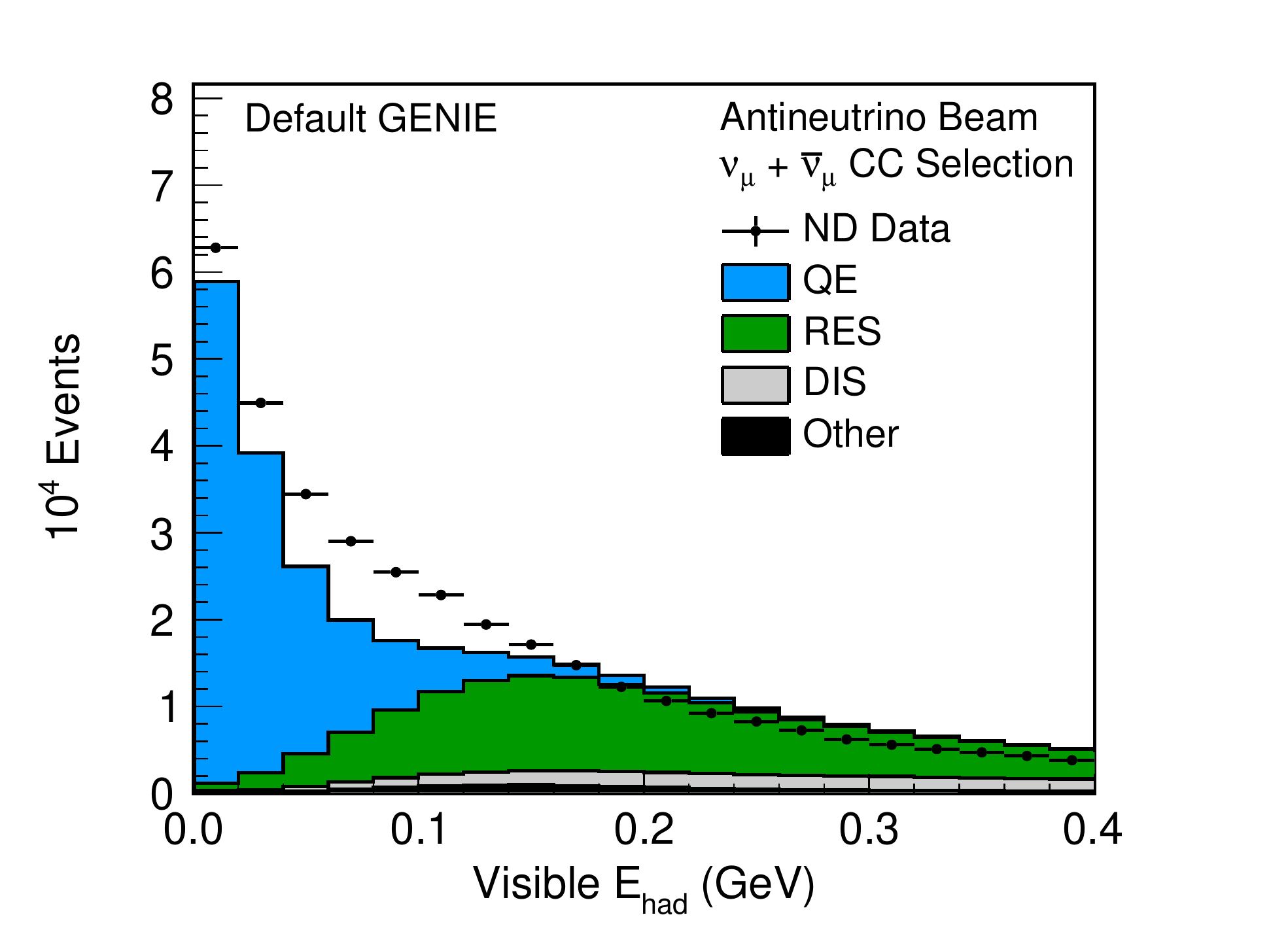}
		\label{fig:rhcdefaultgenie} }
	\caption{Reconstructed visible hadronic energy distributions for neutrino beam (left) and antineutrino beam (right), comparing NOvA near detector data and default GENIE 2.12.2 simulation.  Data are indicated by black points with statistical error bars; the stacked histogram is the sum of the GENIE predictions for the the various interaction types.}
	\label{fig:defaultgenie}
\end{figure}

As Fig.~\ref{fig:defaultgenie} shows, the default GENIE configuration does not reproduce the visible hadronic energy distribution in the ND neutrino or antineutrino data, undershooting by as much as 25\% in the range from 50 to \unit[250]{MeV}.
GENIE, however, does have optional support for the simulation of meson exchange currents (MEC), a process modeled as a neutrino interacting on a nucleon coupled to another nucleon via a meson.
Such a process knocks out multiple nucleons from the nuclear ground state in an otherwise QE-like interaction.
Two MEC models were available in GENIE that we considered for use, including ``Empirical MEC''\cite{teppeimecmodels}, and the model by the \valencia/ group (Nieves \textit{et al.})~\cite{nieves-mec}.
Other models exist but are not implemented in GENIE 2.12.2~\cite{martini-mec, susa-mec, gibuu}.
None of these models explicitly predict the kinematics of the resulting hadrons.
Instead, a separate model is necessary to specify how the momentum transfer should be assigned to individual nucleons.
The model GENIE uses for all MEC simulation is a so-called ``nucleon cluster'' model, in which an intermediate nucleon pair whose initial momenta are drawn from the Fermi sea is assigned the total momentum transfer and then allowed to decay isotropically \cite{teppeimecmodels}.

GENIE also considers final state interactions (FSI) that can occur as the resultant particles traverse the nuclear medium.
These are modeled with the hA-INTRANUKE effective cascade model~\cite{genie-hA-FSI, genie-hA-FSI-2}.
More discussion and further references regarding neutrino-nucleus scattering theory, experiments, and implementation of neutrino interaction software can be found in  Ref.~\cite{nustecwhitepaper}.

\section{Cross-section model adjustment methodology}
\label{sec:tune CV}

As each of the interaction types produced by GENIE has independent degrees of freedom and separate uncertainties, it is essential to consider carefully how each model might be adjusted in order to improve data-MC agreement.  We first modify the GENIE predictions by incorporating new advances motivated by theory or external data and corroborate them with NOvA ND data in regions where the various modes are expected to be well separated (see Secs. \ref{subsec:ext} and \ref{subsec:res}).
After these adjustments, the prediction still disagrees with the ND data, which we attribute to 
the considerable uncertainty on the spectral shape of MEC events.
We reshape and rescale the MEC component so that its sum with the otherwise adjusted simulation matches NOvA ND data, as described in Sec.~\ref{subsec:2p2h}.

While this procedure explicitly accounts for two-body knockout via MEC, interactions on nuclear pairs formed by short-range correlations between nucleons in the nuclear ground state can also result in a similar final state. The default simulation does not model this explicitly, but our work reshapes the MEC kinematics to match data, effectively adding such missing processes. We thus use the more inclusive term ``2p2h'' (two-particle two-hole, describing the ejected particles and the final-state nucleus) to refer to that channel after our model adjustment.

The neutrino and antineutrino beams are simulated separately but the same model adjustments are made to both unless otherwise noted.  No adjustments are made to the COH interaction model or to FSI.  

\subsection{Incorporating constraints on quasi-elastic and deep inelastic scattering interactions}
\label{subsec:ext}

Three modifications to the GENIE default configuration are based on work external to NOvA:

\begin{enumerate}
\item \emph{Adjustment to CCQE $M_{A}$} 

GENIE uses the dipole approximation for the axial form factor, with the only free parameter, $M_{A}$, equal to $\unit[0.99]{GeV/}c^2$.
Recent reanalysis of the original deuterium data suggests $M_{A}$ should be larger.
We use the error-weighted mean of the ANL and BNL results cited in that work: $M_{A} = \unit[1.04]{GeV/}c^2$~\cite{zexpma}.

\item \emph{Nucleon momentum distribution and long-range nuclear mean field effects in CCQE}
\label{subsubsec:LFG+RPA}

The more sophisticated local Fermi gas model of the nuclear ground state employed by Nieves \textit{et al.}~\cite{valencia-rpa} predicts a different initial nucleon momentum distribution than the RFG model.  This difference, when combined with the effect of Pauli suppression, changes the available kinematic space in QE reactions.
Long-range internucleon interactions analogous to charge screening in electromagnetism also modify the kinematics of QE reactions.
A popular approach to account for the latter dynamic in calculations uses the random phase approximation (RPA)~\cite{valencia-rpa,rpa-ghent}. 
The combination of these effects significantly suppresses QE reactions at low invariant four-momentum transferred to the nucleus ($Q^{2}$), and mildly enhances them at higher $Q^2$, relative to the RFG prediction.
To approximate the result of these two phenomena, we employ the weighting functions based on the \valencia/ model constructed by MINERvA~\cite{rpa-valencia-unc}, hereinafter referred to as ``QE nuclear model weights.''
These weights are parameterized in a two-dimensional space of energy and momentum transfer to the nucleus \qZqThree{}, and are calculated separately for neutrinos and antineutrinos\footnote{\label{note1}In Ref.~\cite{novaosc2019}, the QE nuclear model weights were incorrectly applied to interactions on hydrogen targets. Studies showed that the oscillation results were negligibly affected.}.

\item \emph{Soft non-resonant single pion production}
\label{subsubsec:nonres1pi}

We also reweight GENIE single pion DIS events with invariant hadronic mass $W < \unit[1.7]{GeV/}c^2$ to reduce their rate by 57\%\footnote{GENIE's definition of `DIS' can differ from that of others, who typically require larger $W$.  We will hereafter refer to these events as ``soft non-resonant single pion production,'' since they are at low $W$, do not occur through a resonant channel, and are only $1\pi$ final states.} according to the results of recent reanalysis of bubble chamber data~\cite{nonressinglepib}.  This is compatible with MINERvA's recent findings \cite{minerva-pi-tuning}.

Since that analysis applies only to neutrinos and the analogous GENIE prediction for antineutrinos is very different, we do not apply this correction to antineutrino soft non-resonant single pion production\footnote{\label{note2}A 10\% normalization increase was also applied to DIS events with $W > \unit[1.7]{GeV/}c^2$ in the simulation used in the oscillation analysis (neutrino beam only). This normalization increase has negligible effect on the final oscillation results, and it is not applied here.}.
Similarly, no correction is made to NC channels, as the bubble chamber analysis was for CC channels only.
\end{enumerate}

\FloatBarrier

\subsection{\texorpdfstring{Low-$Q^2$ resonance suppression}{}}
\label{subsec:res}

\begin{figure}[ht]
    \centering
 \subfloat[]{
        \centering
        \includegraphics[width=.48\textwidth]{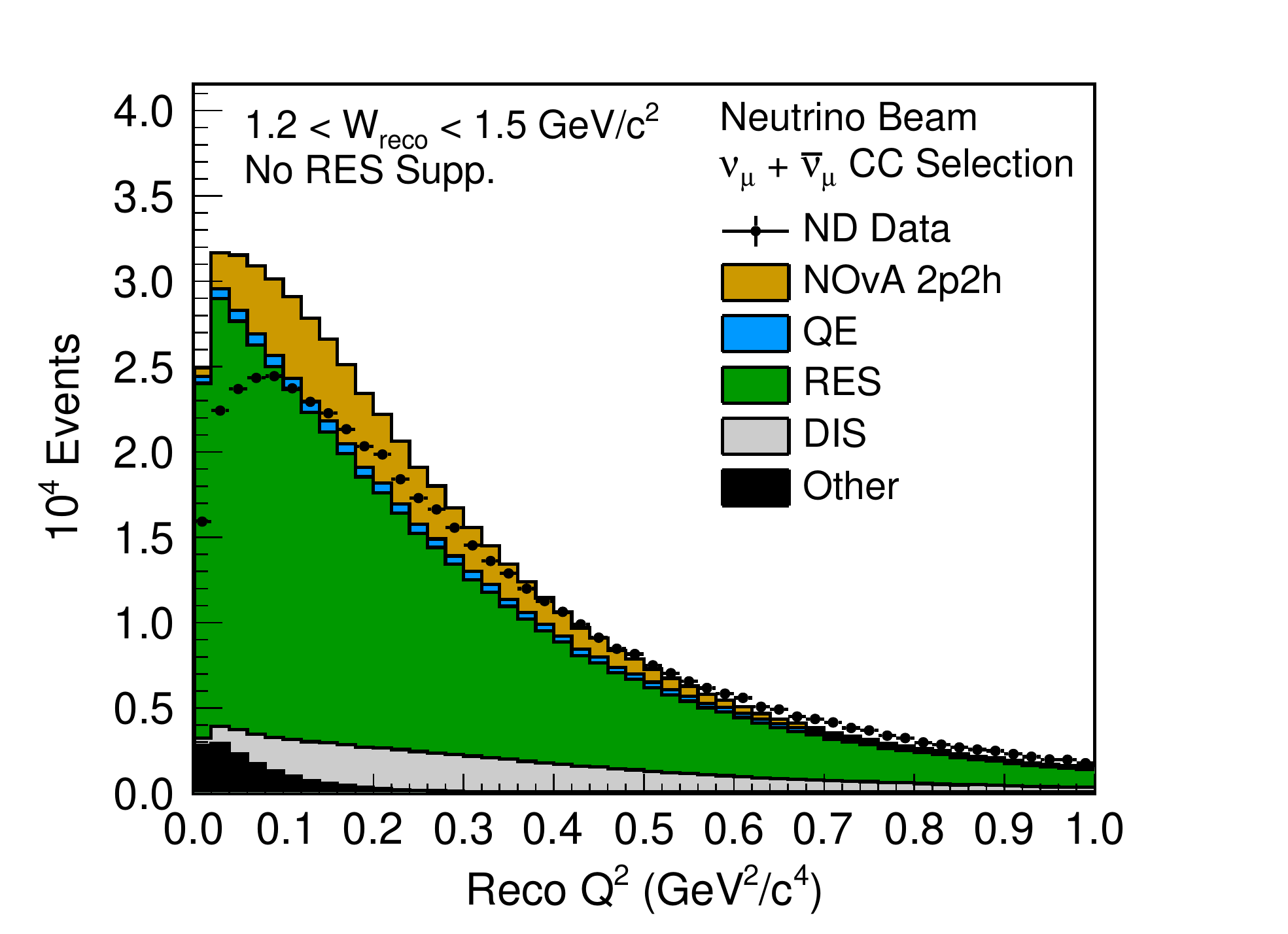}
        \label{fig:resrpaprooffhca} }
 \subfloat[]{
        \centering
        \includegraphics[width=.48\textwidth]{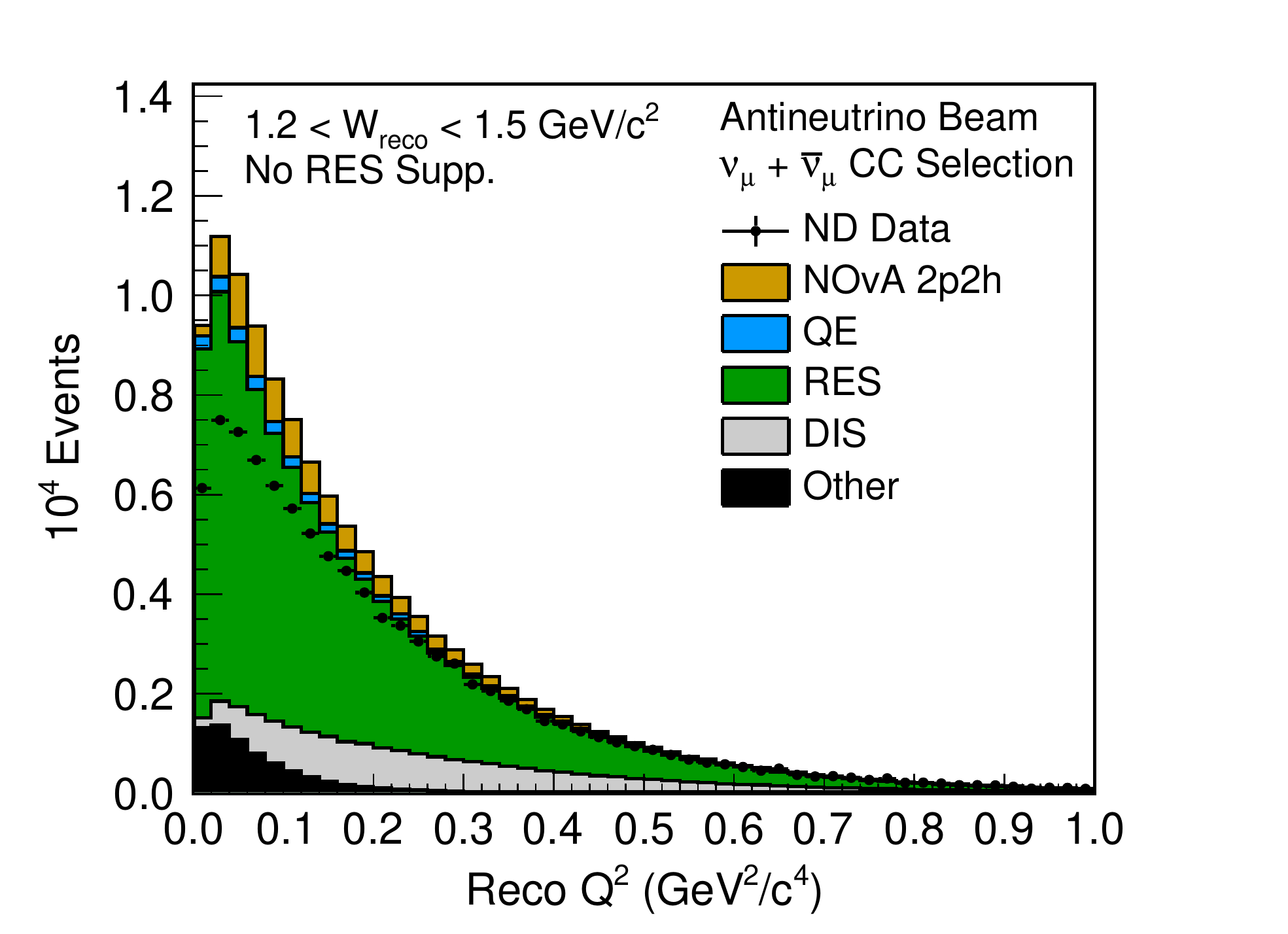}
        \label{fig:resrpaproofrhca} }
\quad
 \subfloat[]{
    \centering
    \includegraphics[width=.48\textwidth]{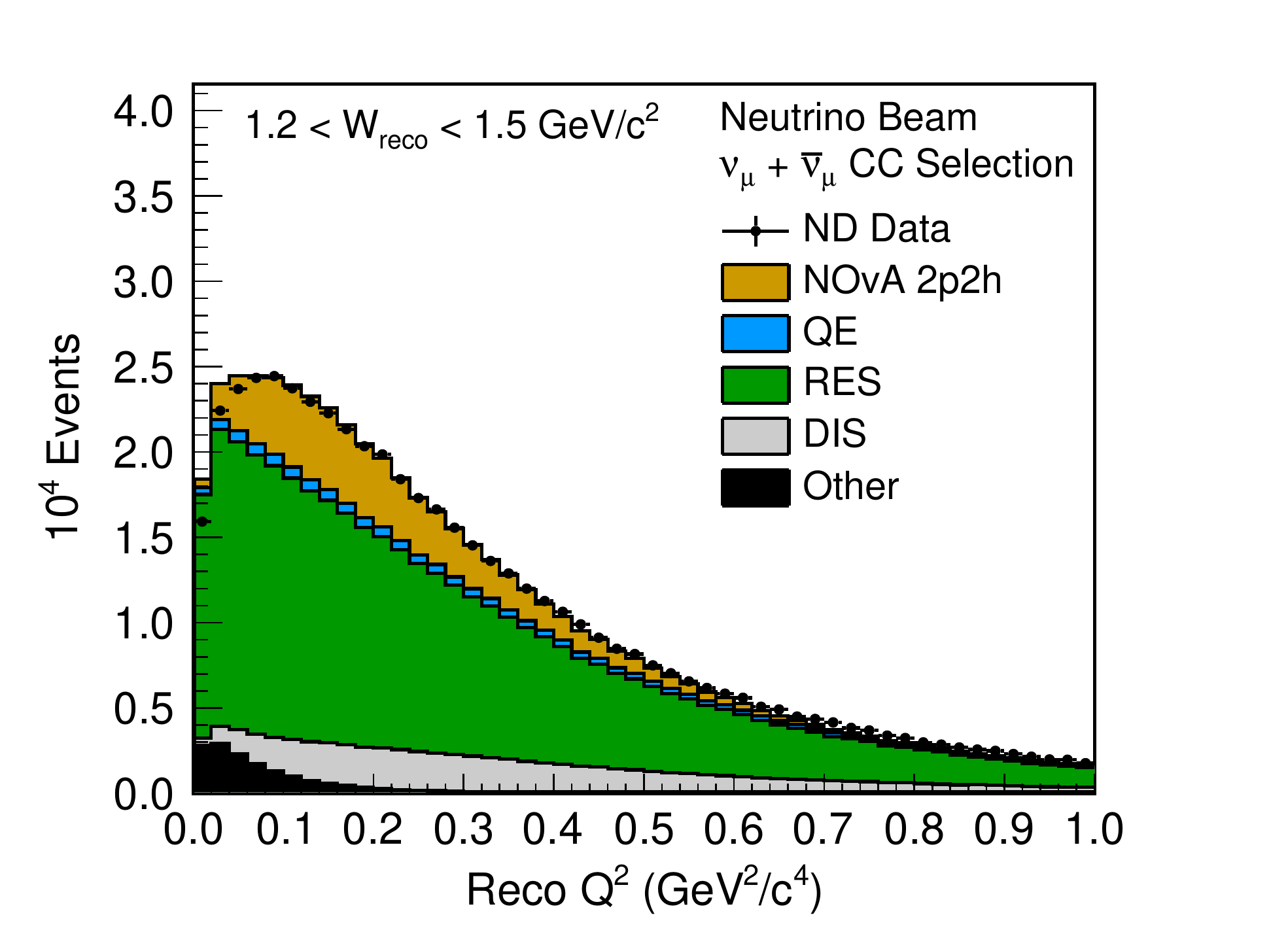}
    \label{fig:resrpaprooffhcb} }
 \subfloat[]{
    \centering
    \includegraphics[width=.48\textwidth]{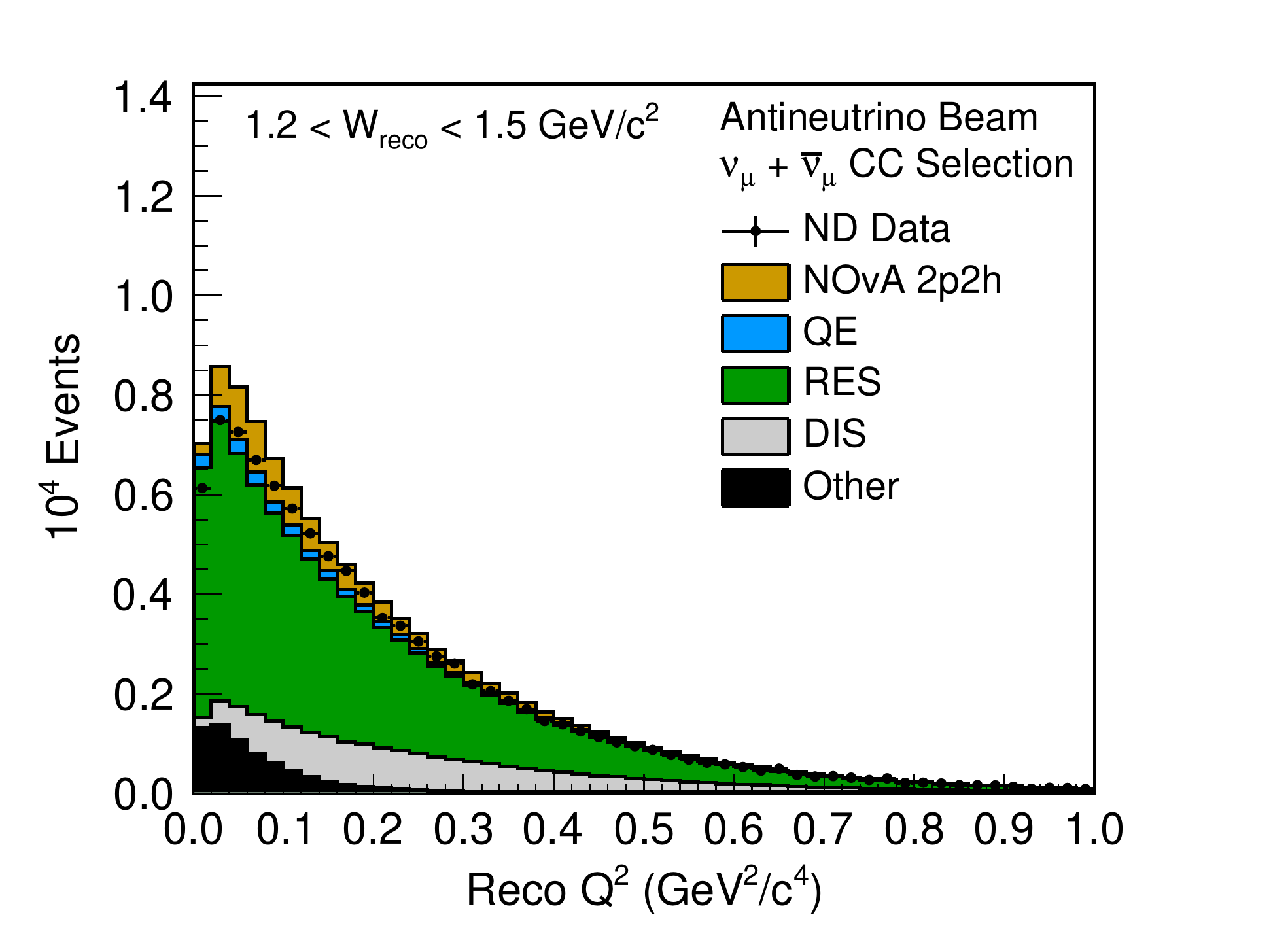}
    \label{fig:resrpaproofrhcb} }
    \caption{Reconstructed $Q^2$ distributions in the reconstructed $W$ range of 1.2~to~$\unit[1.5]{GeV/}c^2$, where RES events dominate.  Data are shown with statistical error bars, while simulation is shown as histograms stacked by interaction type.  All cross section adjustments described in this paper are applied, including the addition of the fitted 2p2h described in Sec.~\ref{subsec:2p2h}, except that the RPA-like low-$Q^{2}$ suppression is not applied to RES interactions in the top plots.  Neutrino beam is shown at left, antineutrino beam at right.}
    \label{fig:resrpaproof}
\end{figure}

Measurements of neutrino-induced $\Delta(1232)$ resonance production~\cite{miniboone-res, minos-qe, minerva-pi-1, minerva-pi-2, t2k1pi} suggest a suppression at low $Q^2$ relative to the Rein-Sehgal free-nucleon prediction.
Our own ND data reproduces this phenomenon, as seen in the top of Fig.~\ref{fig:resrpaproof}.
To our knowledge, there is no phenomenology predicting such an effect, though it superficially resembles the effect that the QE nuclear model weights have on QE interactions.
We find that applying an alternate parameterization\footnote{This parameterization, which is in $Q^2$ instead of \qZqThree, is also available from Ref.~\cite{rpa-valencia-unc}.} of the QE nuclear model weights discussed in Sec.~\ref{subsec:ext} to RES interactions significantly reduces the tension we observe with our data, as shown in the bottom of Fig.~\ref{fig:resrpaproof}.

We therefore reweight all RES events according to this prescription.
Formally, the RPA phenomenology may not apply directly to baryon resonance excitation, which requires significant three-momentum transfer to the nucleus even at $Q^2 = 0$, and thus places RES interactions out of the regime where current RPA calculations apply.
However, we employ this procedure as a placeholder for whatever the true effect may be, and invite further input from the theoretical community as to what ingredients may be missing from the model.

\FloatBarrier

\subsection{Multi-nucleon knockout (2p2h)}
\label{subsec:2p2h}
\begin{figure}[ht]
	\centering
 \subfloat[]{
		\includegraphics[width=.48\textwidth]{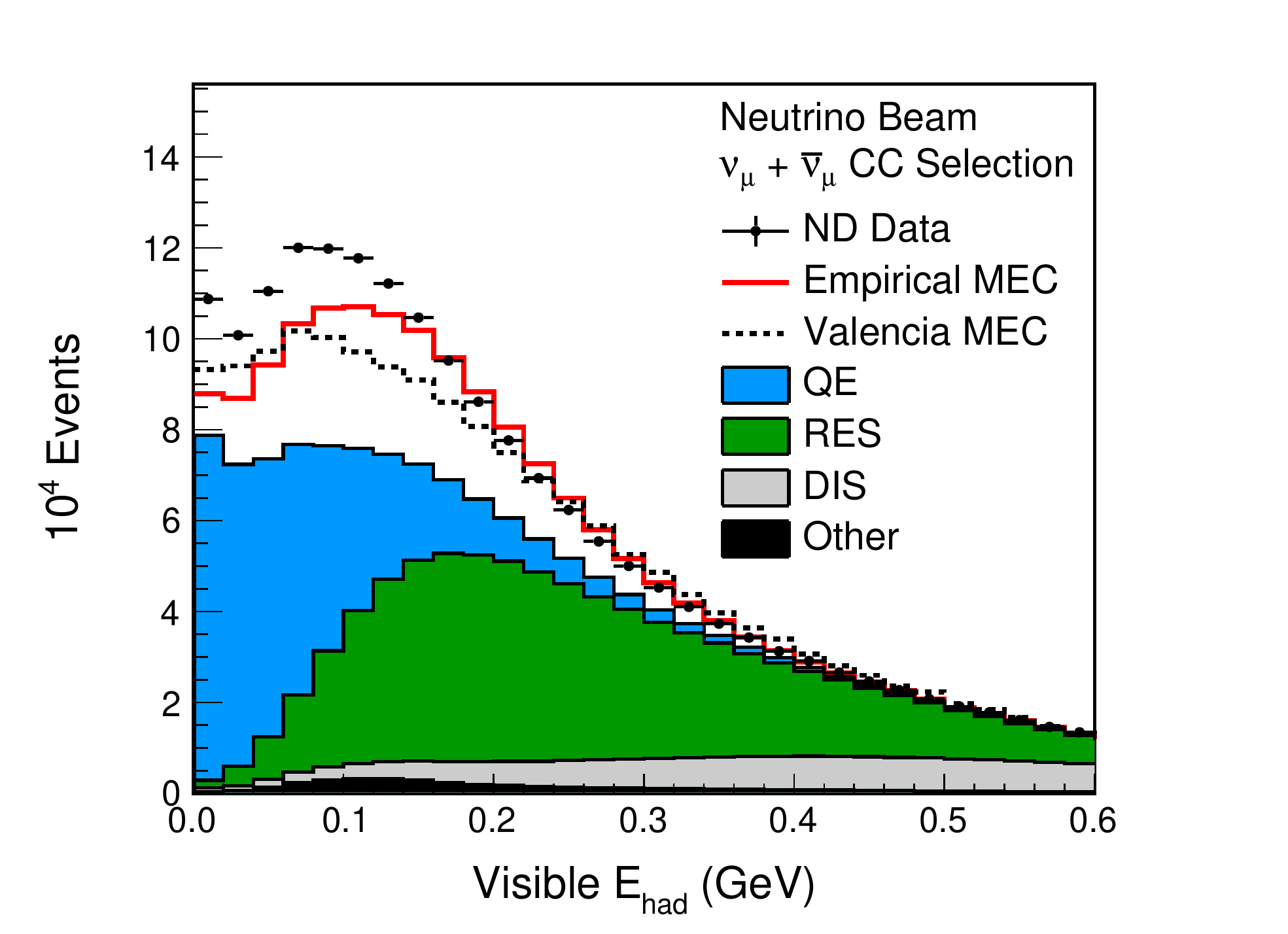} }
 \subfloat[]{
		\includegraphics[width=.48\textwidth]{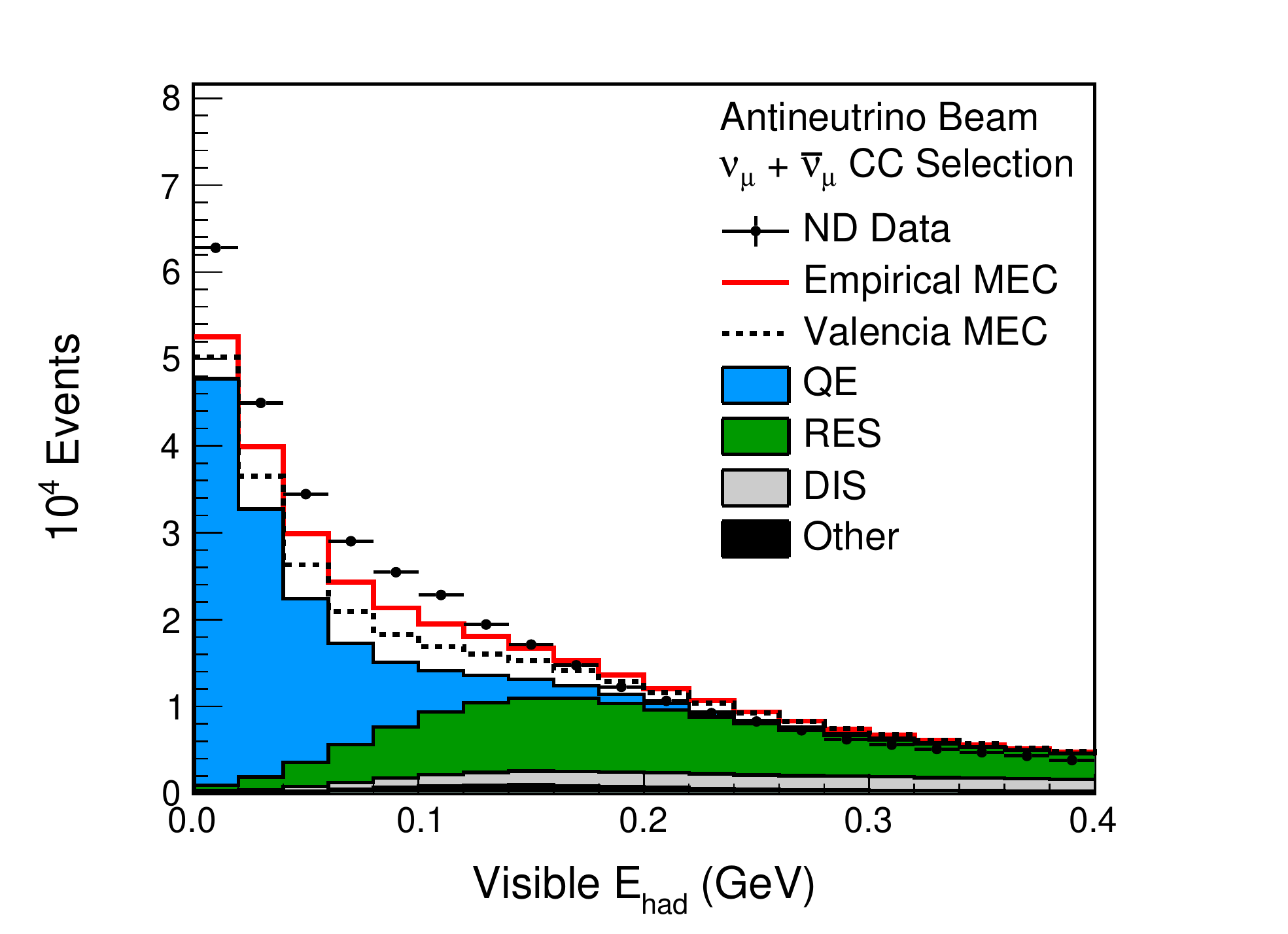} }
	\caption{Comparison of ND data to simulation in reconstructed visible hadronic energy using the default GENIE empirical MEC model (solid red curve) or the \valencia/ MEC model (dotted black curve), in neutrino beam (left) and antineutrino beam (right).  The filled, stacked histograms indicate the non-MEC components of the prediction, to which all the modifications described in Sec.~\ref{sec:tune CV} have been applied.}
	\label{fig:2p2hdefaults}
\end{figure}

Significant disagreement with the ND data remains even after combining any of the MEC models available in GENIE with the prediction after the modifications described above, as can be seen in Fig.~\ref{fig:2p2hdefaults}. Both the Empirical and Valencia MEC models produce too low of an overall rate, especially at low values of hadronic energy. The visible hadronic energy shapes of Empirical and Valencia MEC are quite different for neutrinos but similar for antineutrinos.
It is clear that any MEC model GENIE offers will require significant tuning to reproduce our data.
We choose to use the Empirical MEC model as a starting point, as it is the only model available in GENIE that includes a neutral-current component.

The Empirical MEC model is reshaped to create an ad-hoc model that matches data by modifying it in a two-dimensional space of \qZqThree{}.
Simulated GENIE Empirical MEC interactions are divided into 16 bins of energy transfer (from 0 to \unit[0.8]{GeV}) and 20 momentum transfer bins (from 0 to \unit[1]{GeV/}$c$).  Of these, 120 bins are kinematically disallowed. 
Scale factors for each of the remaining 200 bins in $(q_0, \qmag)$ are incorporated as Gaussian penalty terms into a $\chi^{2}$ fit, each with 100$\%$ uncertainty. For this fit, the non-2p2h portion of the simulation is adjusted as described in this paper, and the 2p2h component is reweighted as dictated by the penalty terms. A migration matrix is used to convert the $(q_0, \qmag)$ prediction into a binned 20x20 space of visible hadronic energy \ehadvis{} (from 0 to \unit[0.4]{GeV}) and reconstructed three-momentum transfer \qmagreco{} (from 0 to \unit[1]{GeV/}$c$). This prediction in reconstructed variables is then compared to the ND data in the fit. 
The small (2$\%$) antineutrino MC component is left in its default state when fitting the neutrino beam simulation to data.
The process is repeated for the antineutrino beam data and MC, except in this case the 2p2h fit for neutrinos is applied first to the larger (about 10$\%$) neutrino component in the antineutrino beam MC.  

\begin{figure}[ht]
	\centering
 \subfloat[]{
		\centering
		\includegraphics[width=.47\textwidth]{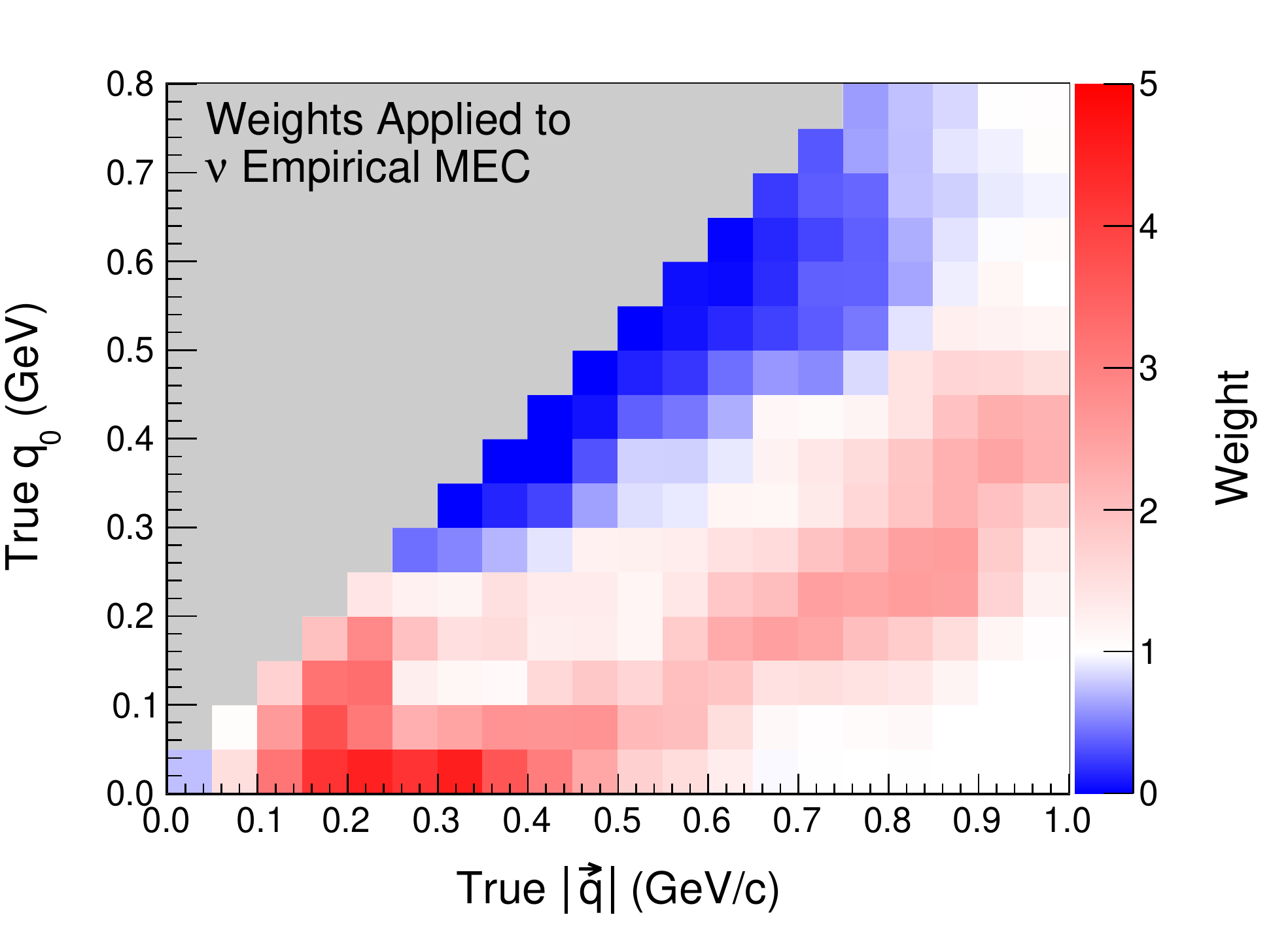} 
		\label{fig:2p2hweightsnu} }
 \subfloat[]{
		\centering
		\includegraphics[width=.47\textwidth]{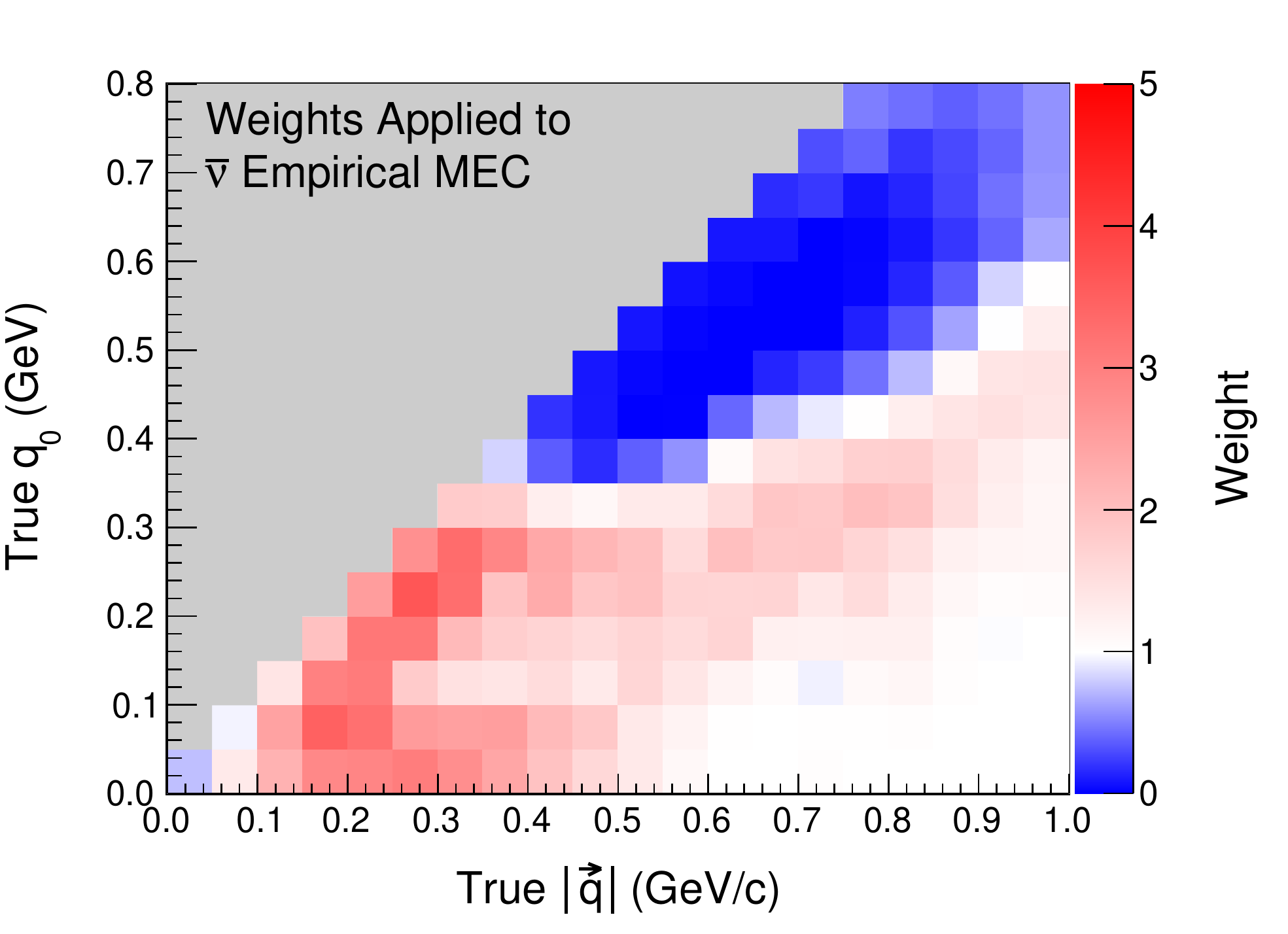}
		\label{fig:2p2hweightsnubar} }
	\caption{The weights, in three-momentum and energy transfer, applied to simulated Empirical MEC interactions to produce the fitted NOvA 2p2h predictions described in the text, for neutrinos (left) and antineutrinos (right).  Gray indicates kinematically disallowed regions, where no weights are applied.}
	\label{fig:2p2hweights}
\end{figure}

\begin{figure}[ht]
	\centering
 \subfloat[]{
		\centering
		\includegraphics[width=.47\textwidth]{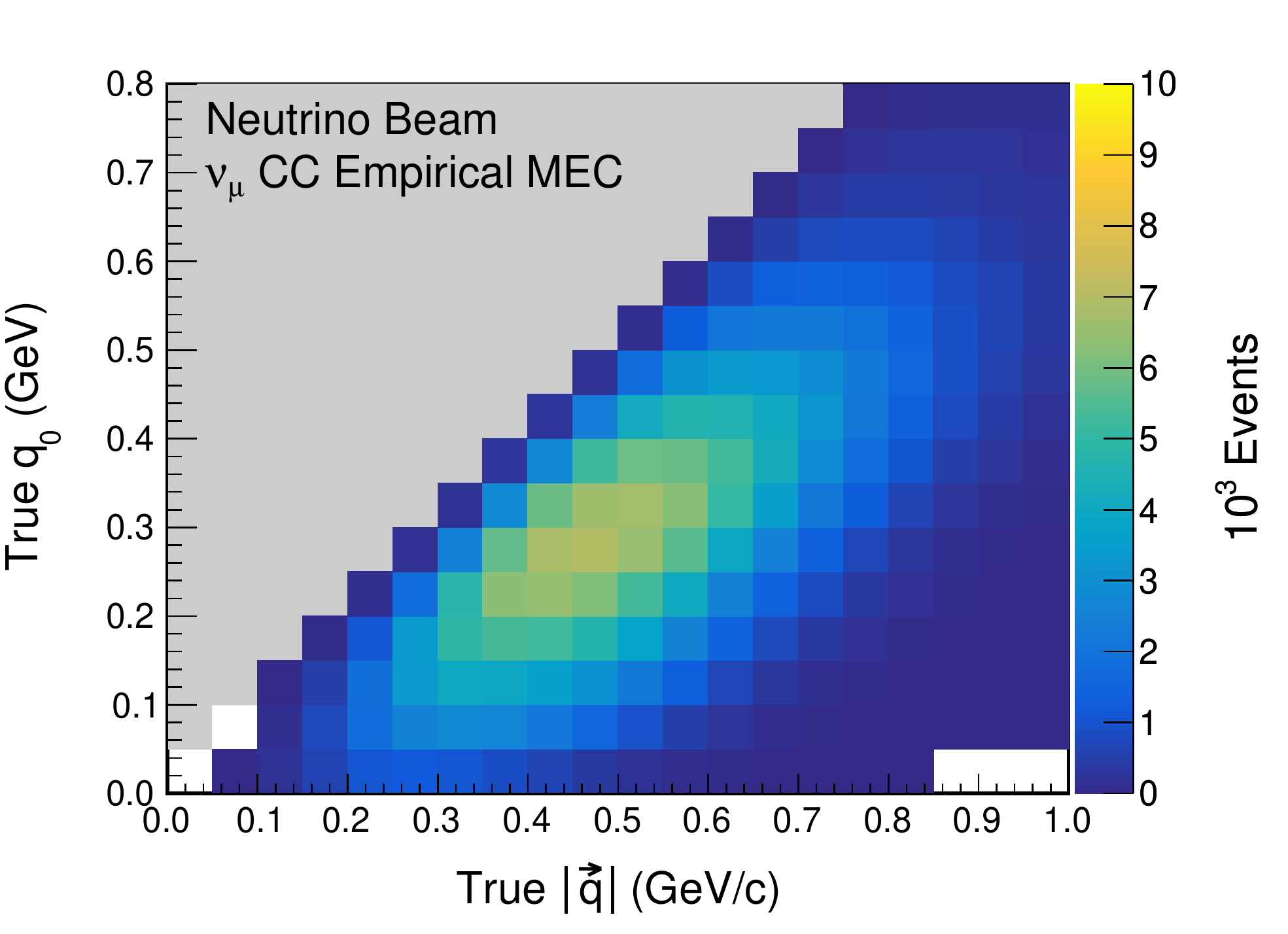}
		\label{fig:2p2hdefaultfhc} }
 \subfloat[]{
		\centering
		\includegraphics[width=.47\textwidth]{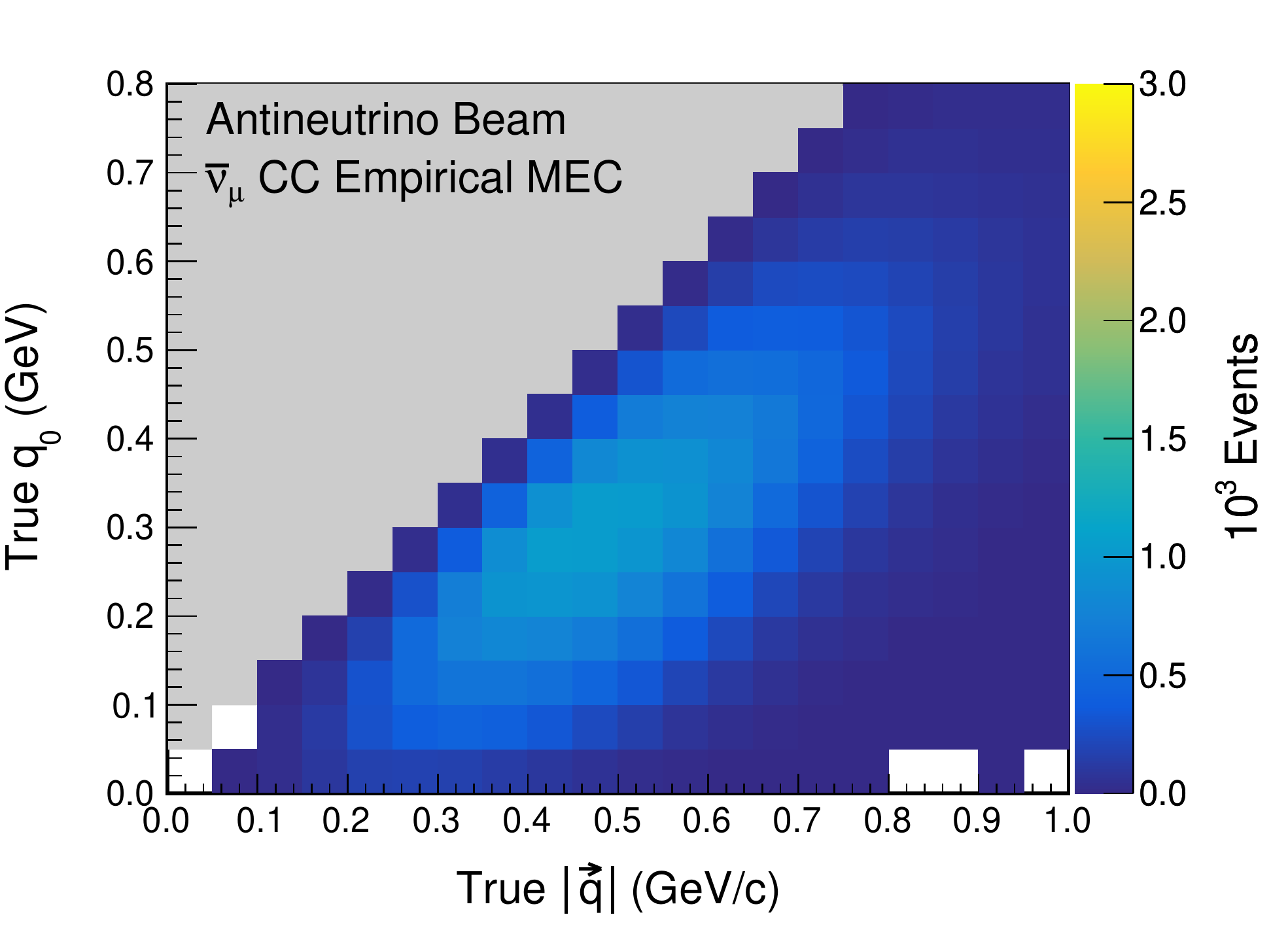}
		\label{fig:2p2hdefaultrhc} }
\quad
 \subfloat[]{
		\centering
		\includegraphics[width=.47\textwidth]{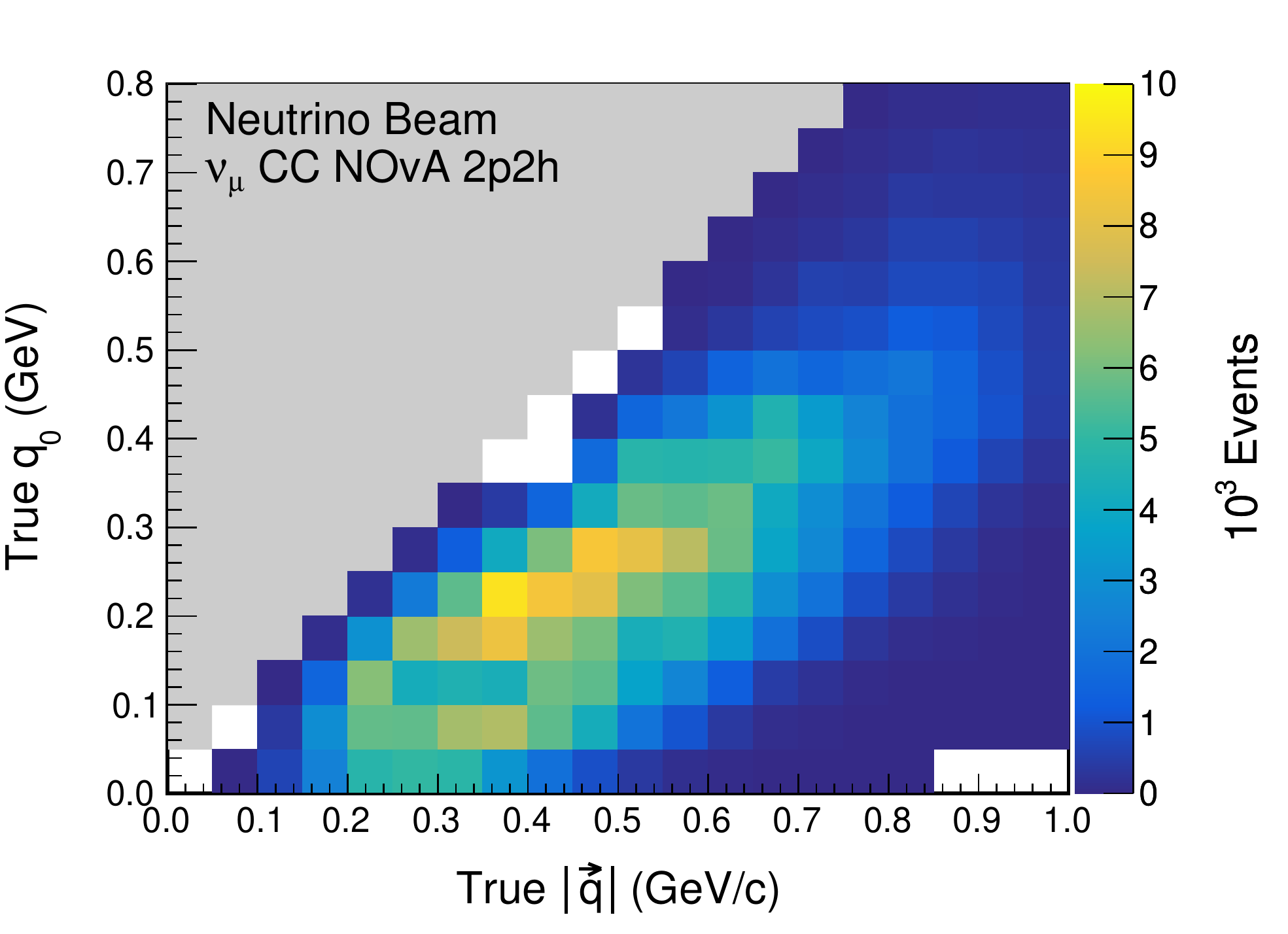}
		\label{fig:2p2htunefhc} }
 \subfloat[]{
		\centering
		\includegraphics[width=.47\textwidth]{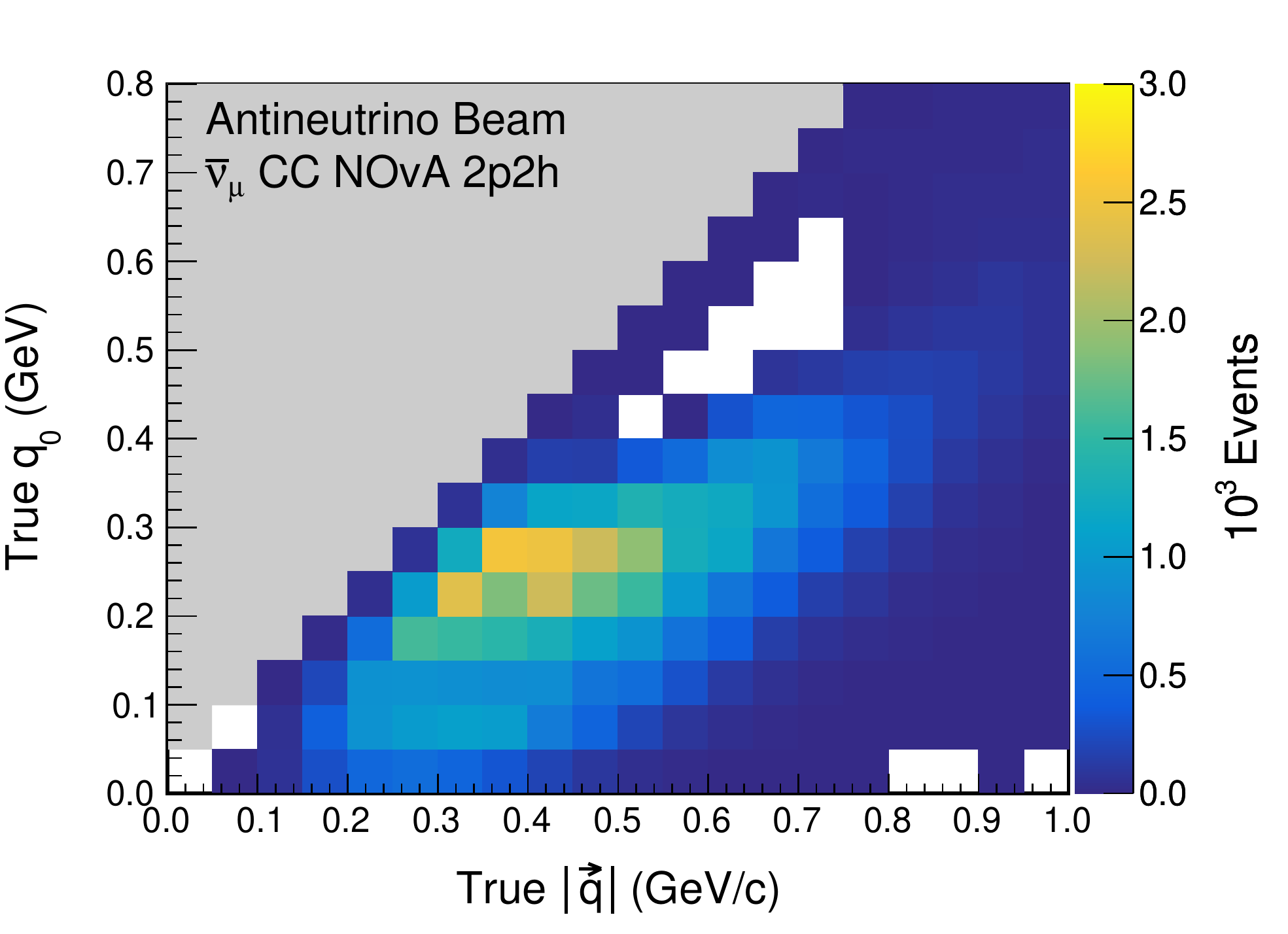}
		\label{fig:2p2htunedrhc} }
	\caption{Predicted momentum and energy transfer distributions for unmodified Empirical MEC (top row) and the result of applying the weights shown in Fig.~\ref{fig:2p2hweights} to Empirical MEC to obtain NOvA 2p2h (bottom row), for neutrino beam (left) and antineutrino beam (right).  Gray indicates the kinematically disallowed region, where no weights are applied.  White indicates weights of precisely zero where either no Empirical MEC events were generated ($\qzero < \unit[0.1]{GeV/c}$) or the fit would otherwise force the weights negative ($\qzero > \unit[0.35]{GeV/c}$).}
	\label{fig:2p2h2ddist}
\end{figure}

The resulting weights are shown in Fig.~\ref{fig:2p2hweights}.
Since true $q_{0}$ and \ehadvis{} are strongly correlated variables, the enhancement of events at low values of $q_{0}$ compensates for the deficit of simulated events at low visible hadronic energy seen in Fig.~\ref{fig:2p2hdefaults}.
In the antineutrino beam sample there is less discrepancy at low \ehadvis{} than in the neutrino beam sample, and thus the antineutrino weights show a smaller enhancement at low $q_{0}$.  Additionally, events in the higher \qzero{} tail are suppressed for antineutrinos. 
These features are evident in Fig.~\ref{fig:2p2h2ddist}, which compares the unaltered Empirical MEC distributions in energy transfer and momentum transfer to the reweighted distributions.

\FloatBarrier

\subsection{Summary of adjustments to central value prediction}

In summary, the NOvA prescription for adjusting GENIE cross-section models to incorporate external data constraints and to improve agreement with NOvA ND data is to start with GENIE, using the Empirical MEC model, and:
\begin{enumerate}
    \item Change CCQE $M_{A}$ from 0.99 to $\unit[1.04]{GeV/}c^2$;
    \item Apply \valencia/ nuclear model weights from MINERvA, using the $(q_0, |\vec{q}|)$ parameterization for QE and the $Q^{2}$ parameterization for RES;
    \item Apply a 57$\%$ reduction to soft non-resonant single pion production events from neutrinos;
    \item Apply separate $\nu$ and $\bar{\nu}$ weights in \qZqThree{} derived from NOvA ND data to Empirical MEC interactions.
\end{enumerate}

\begin{figure}[ht]
    \centering
 \subfloat[]{
        \centering
        \includegraphics[width=.48\textwidth]{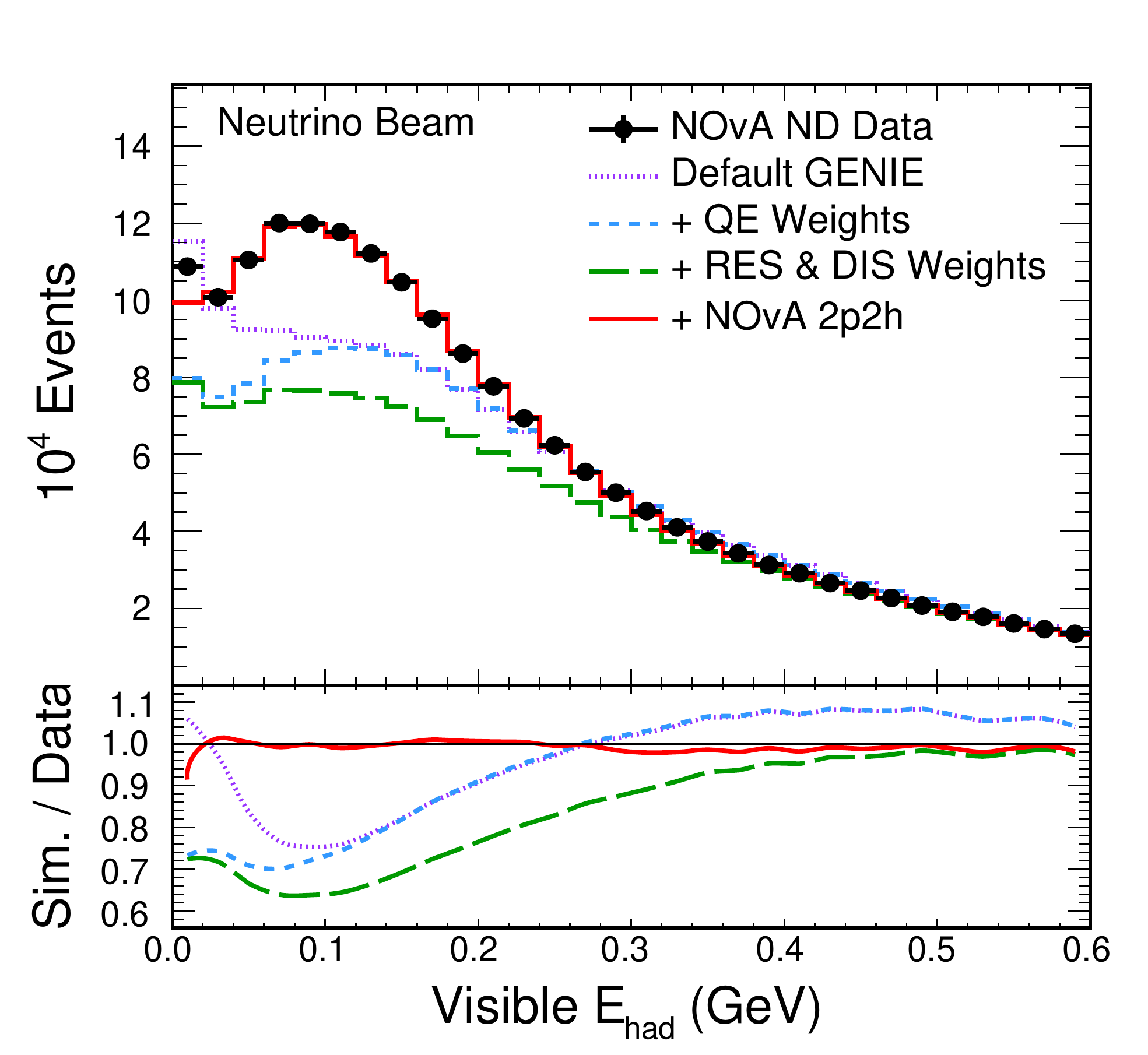}
        \label{fig:mcstagesfhc} }
 \subfloat[]{
        \centering
        \includegraphics[width=.48\textwidth]{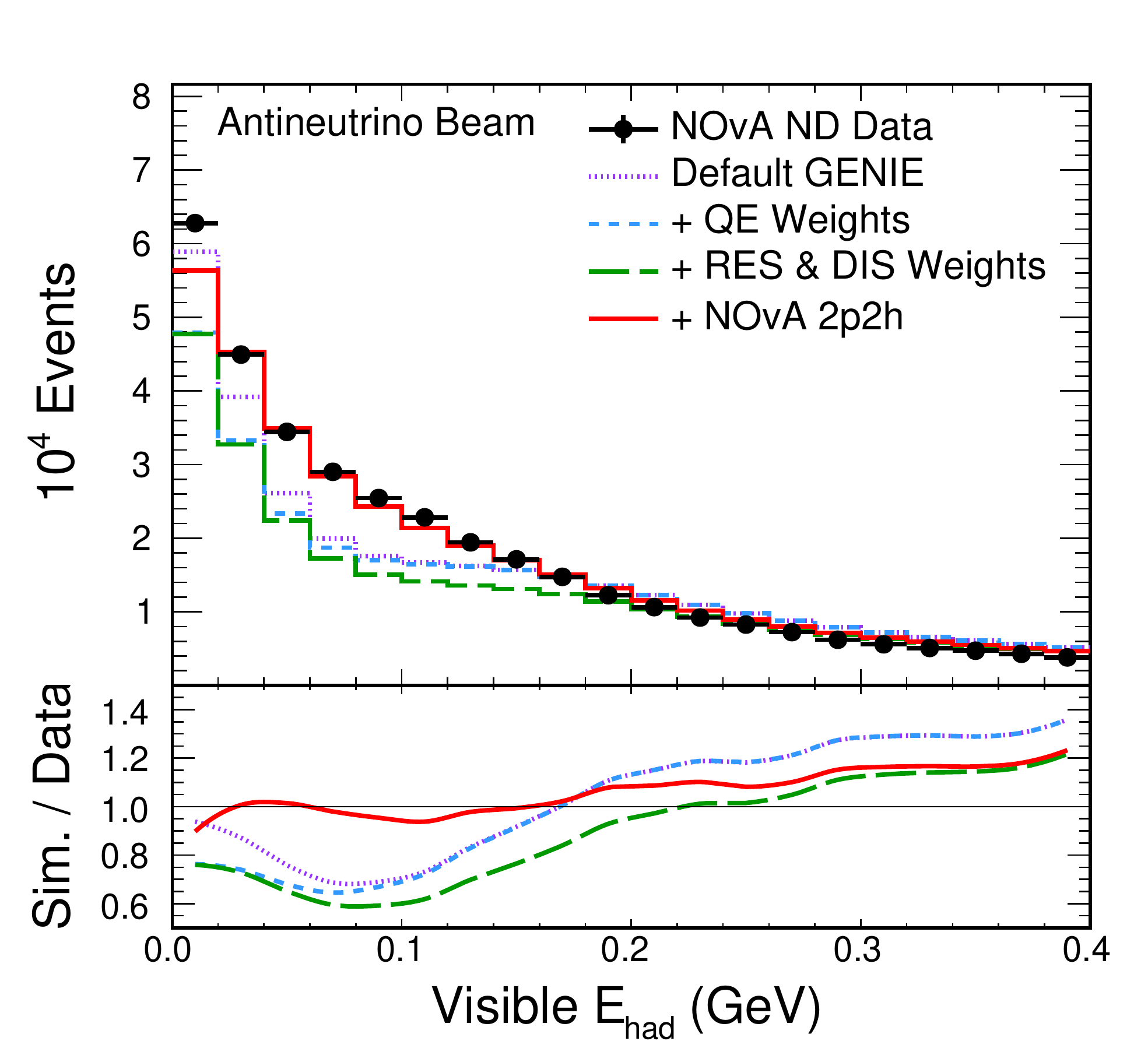}
        \label{fig:mcstagesrhc} }
    \caption{Visible hadronic energy distributions showing each step of our simulation adjustment process.  The purple dotted histogram indicates the default GENIE simulation without any 2p2h.  The blue dashed line shows the effect of adding modifications to QE (adjusting $M_{A}$ and the nuclear model).  The RES and soft non-resonant single pion production (DIS) adjustments are then also included, as shown by the green broken line.  The red solid histogram shows the final result, which further includes the fitted 2p2h contribution.  Neutrino beam is shown at left and antineutrino beam at right.}
    \label{fig:mcstages}
\end{figure}

The effect of each step is shown in Fig.~\ref{fig:mcstages}.
The default GENIE simulation has a large deficit of events in the MEC region in both beams when compared to data, though the neutrino beam prediction has a 5\% excess in the lowest hadronic energy bin.
The QE modifications particularly affect the low \ehadvis{} region due to the suppression from the nuclear model.
The adjustment to RES and DIS widens the deficit, then by design the 2p2h fit modifies the shape of this component to improve agreement.
The predicted composition of the sample before and after the tuning procedure is given in Table \ref{tab:sample fracs}.

\renewcommand{\arraystretch}{1.2}
\begin{table}
	\begin{tabular}{c||ccc|c||ccc|c}
		\multirow{2}{*}{GENIE process} & \multicolumn{4}{c}{Neutrino beam} & \multicolumn{4}{c}{Antineutrino beam} \\
		                         & Default & +MEC & Final & Before selection & Default & +MEC  & Final & Before selection \\
		\hline
		MEC/2p2h                 & ---     & 0.16 & 0.21  & 0.14          & ---     & 0.14  & 0.20  & 0.17 \\
		QE                       & 0.31    & 0.26 & 0.25  & 0.25          & 0.42    & 0.36  & 0.34  & 0.32 \\
		RES                      & 0.49    & 0.41 & 0.39  & 0.39          & 0.42    & 0.36  & 0.31  & 0.32 \\
		DIS                      & 0.17    & 0.15 & 0.13  & 0.21          & 0.13    & 0.11  & 0.12  & 0.18 \\
		Other                    & 0.02    & 0.02 & 0.02  & 0.01          & 0.04    & 0.03  & 0.03  & 0.02
	\end{tabular}
	\caption{Fraction of the predicted \numu{} CC candidate sample corresponding to each GENIE major process in the default GENIE 2.12.2 configuration (``Default''), the default configuration with the addition of unadjusted Empirical MEC (``+MEC''), and after all the adjustments of Sec. \ref{sec:tune CV} (``Final'').  The ``Before selection'' column indicates the fully adjusted fractions before selection, illustrating the important role acceptance and reconstruction efficiencies play in the ND.  Fractions may not add to precisely 1.00 due to rounding.}
	\label{tab:sample fracs}
\end{table}

\begin{figure}[ht]
    \centering
    \makebox[\linewidth][c]{%
 \subfloat[]{
	        \centering
	        \includegraphics[width=.34\textwidth]{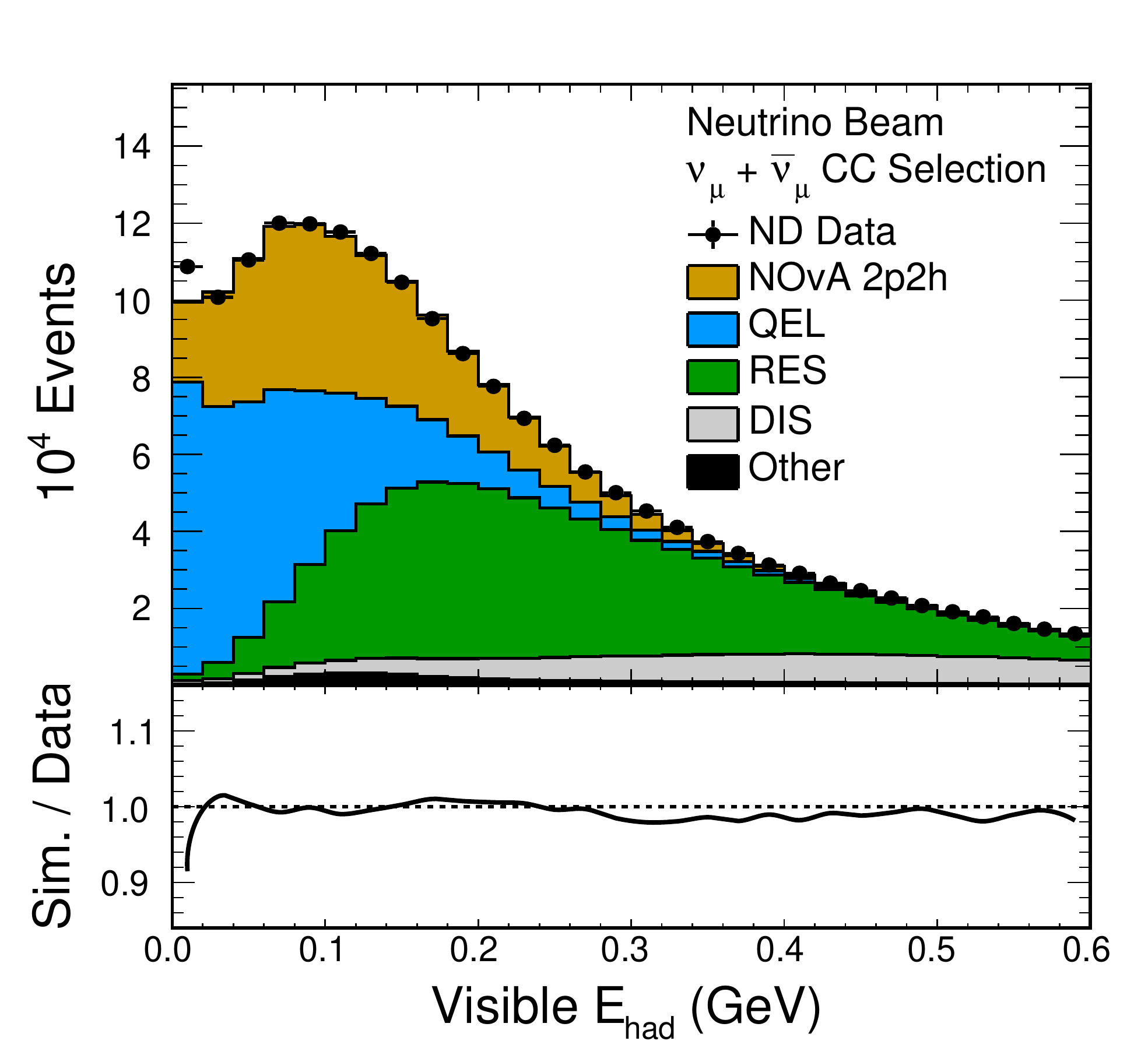}
	        \label{fig:finaltuneehadvisfhc} }
 \subfloat[]{
	        \centering
	        \includegraphics[width=.34\textwidth]{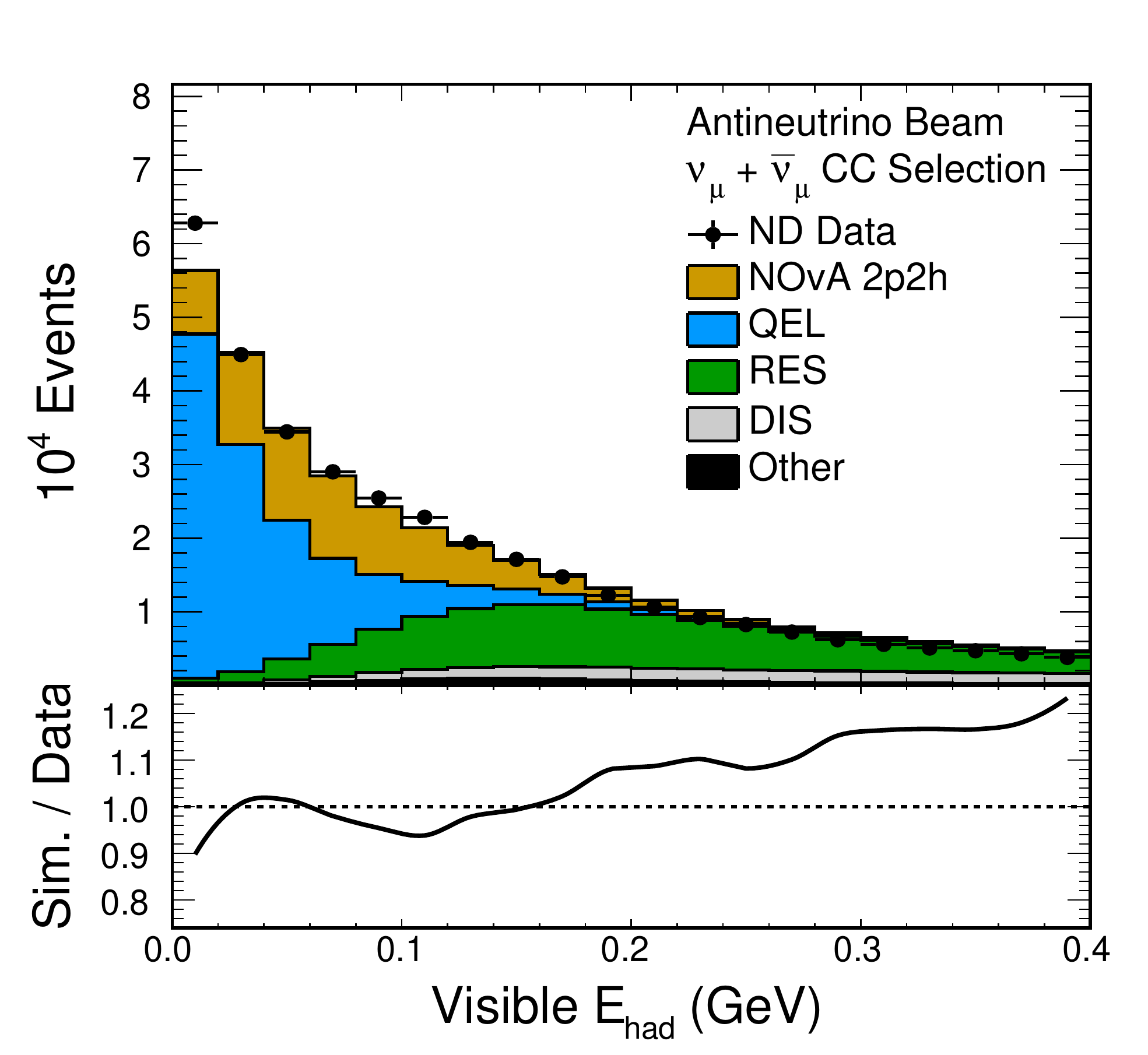}
	        \label{fig:finaltuneehadvisrhc} }
    }
    
    \makebox[\linewidth][c]{%
 \subfloat[]{
	        \centering
	        \includegraphics[width=.34\textwidth]{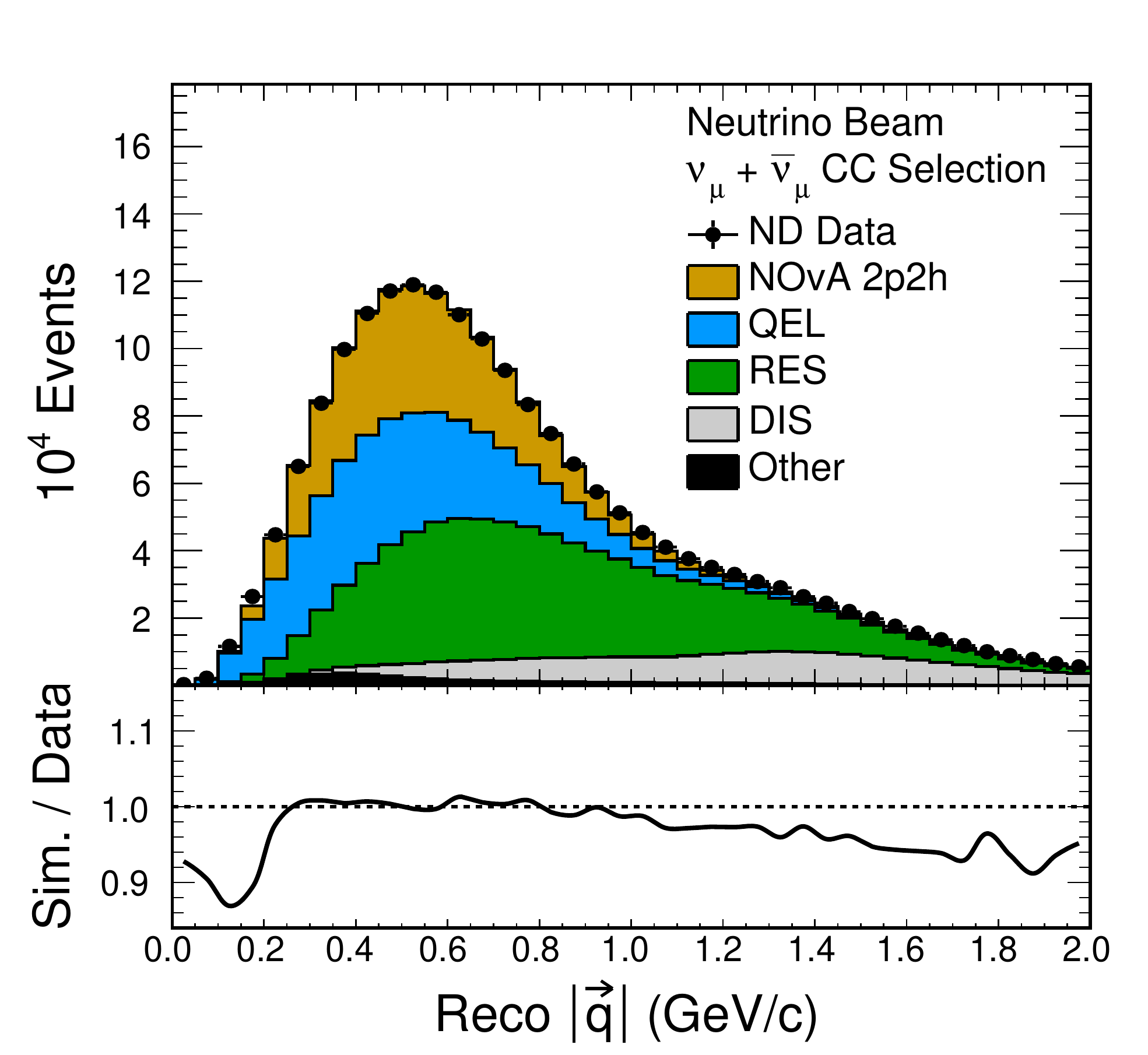}
	        \label{fig:finaltuneq3fhc} }
 \subfloat[]{
	        \centering
	        \includegraphics[width=.34\textwidth]{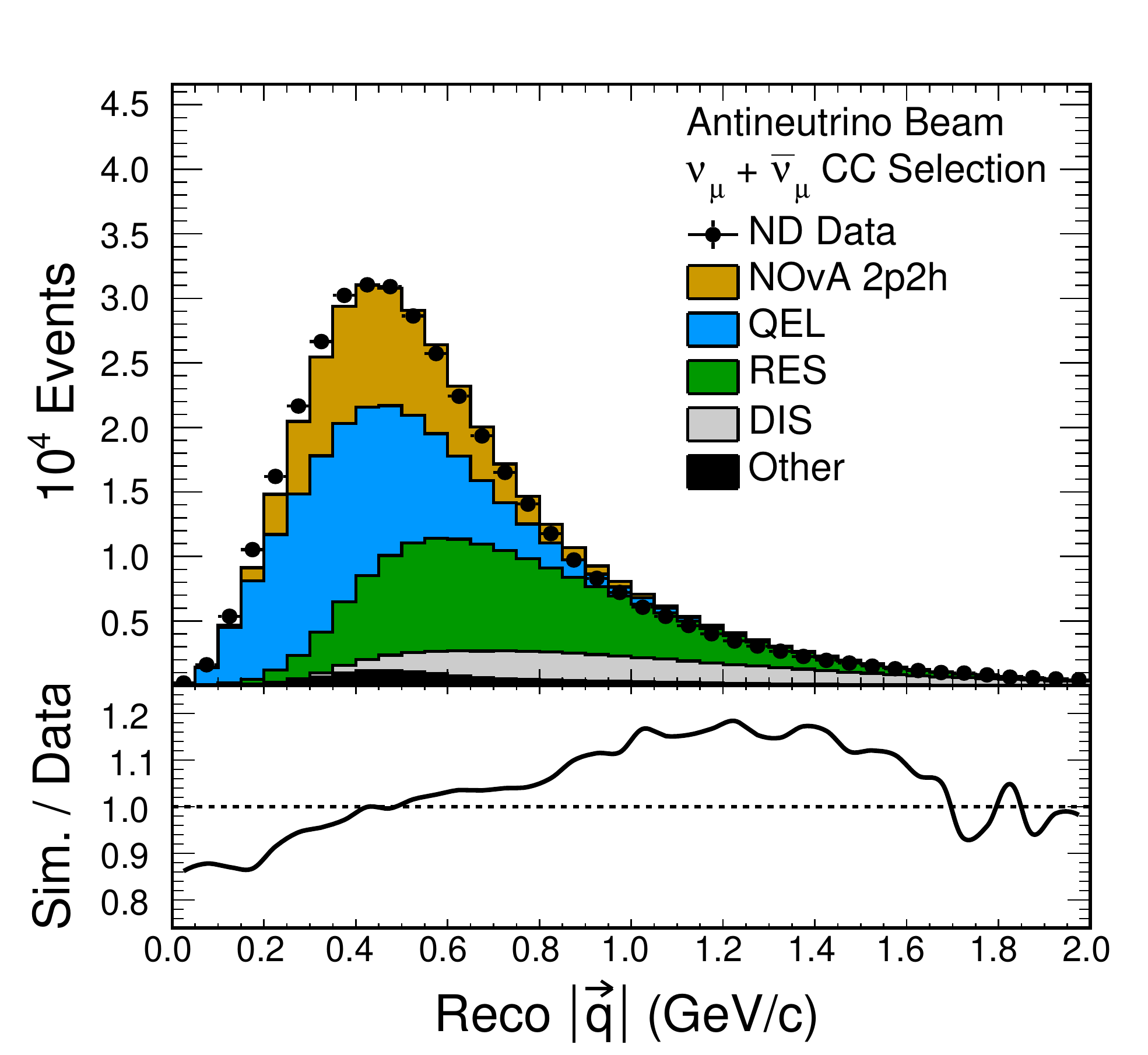}
	        \label{fig:finaltuneq3rhc} }
    }
    
    \makebox[\linewidth][c]{%
 \subfloat[]{
	        \centering
	        \includegraphics[width=.34\textwidth]{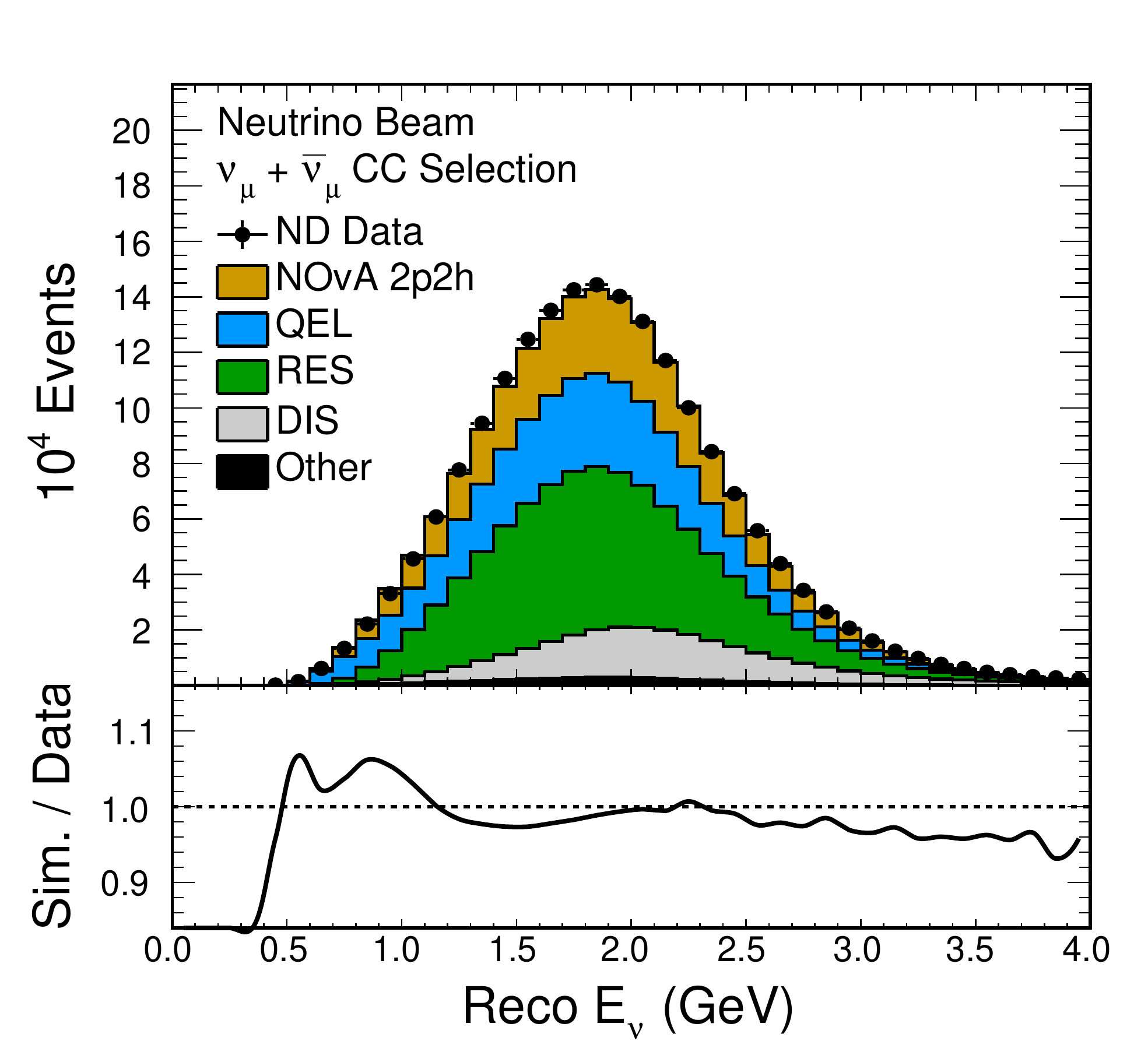}
	        \label{fig:finaltuneenufhc} }
 \subfloat[]{
	        \centering
	        \includegraphics[width=.34\textwidth]{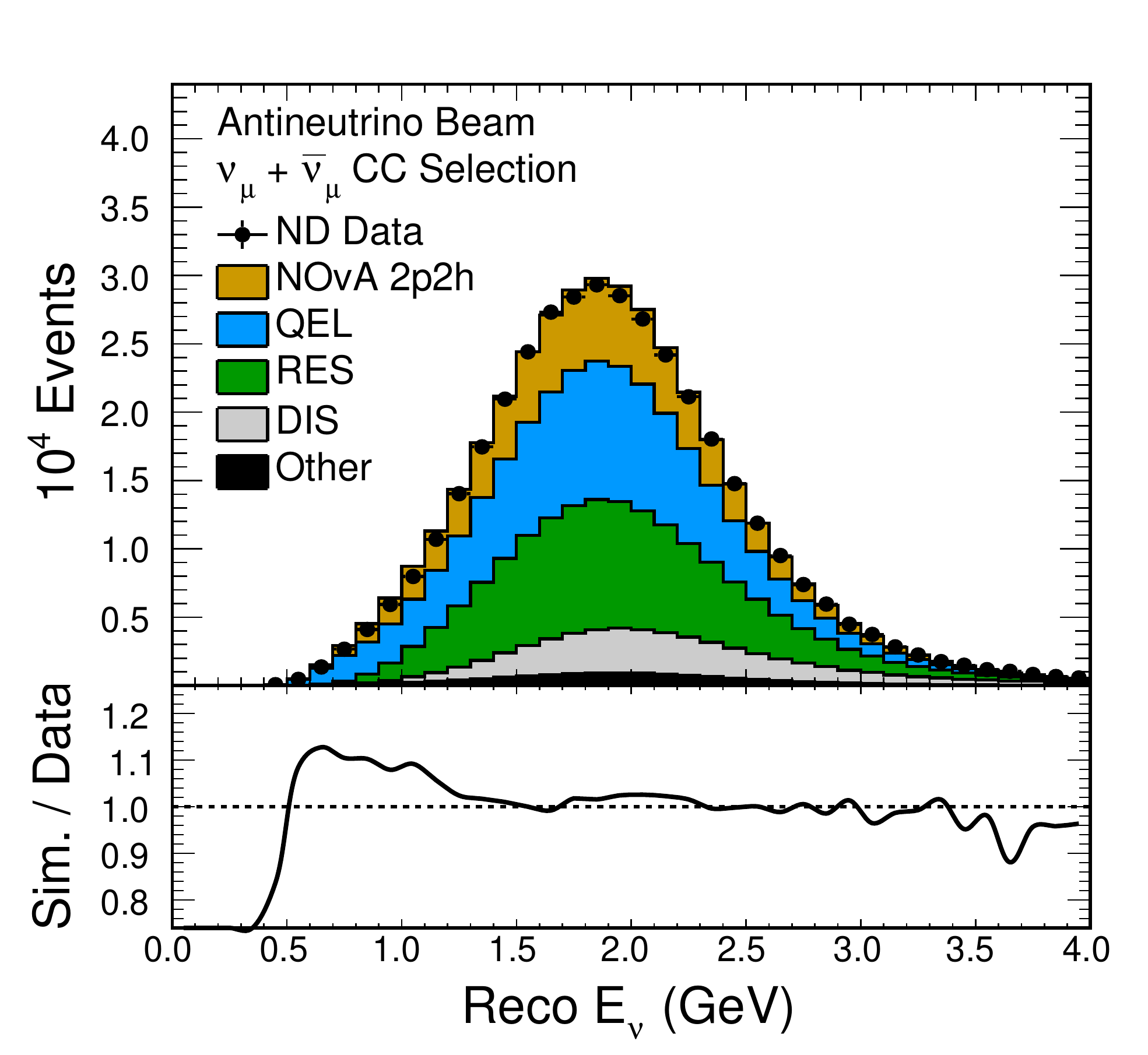}
	        \label{fig:finaltuneenurhc} }
	}
    \caption{Comparison of adjusted simulation to data in the 2p2h tuning variables $E_{\text{had}}^{\text{vis}}$ (top row) and reconstructed $|\vec{q}|$ (middle row), as well as reconstructed $E_{\nu}$ (bottom row), for neutrino beam (left) and antineutrino beam (right).  The simulation is broken up by interaction type, shown as stacked histograms.}
	\label{fig:finaltune}
\end{figure}

\begin{figure}[ht]
    \makebox[\linewidth][c]{%
 \subfloat[]{
	        \centering%
	        \includegraphics[width=.49\linewidth]{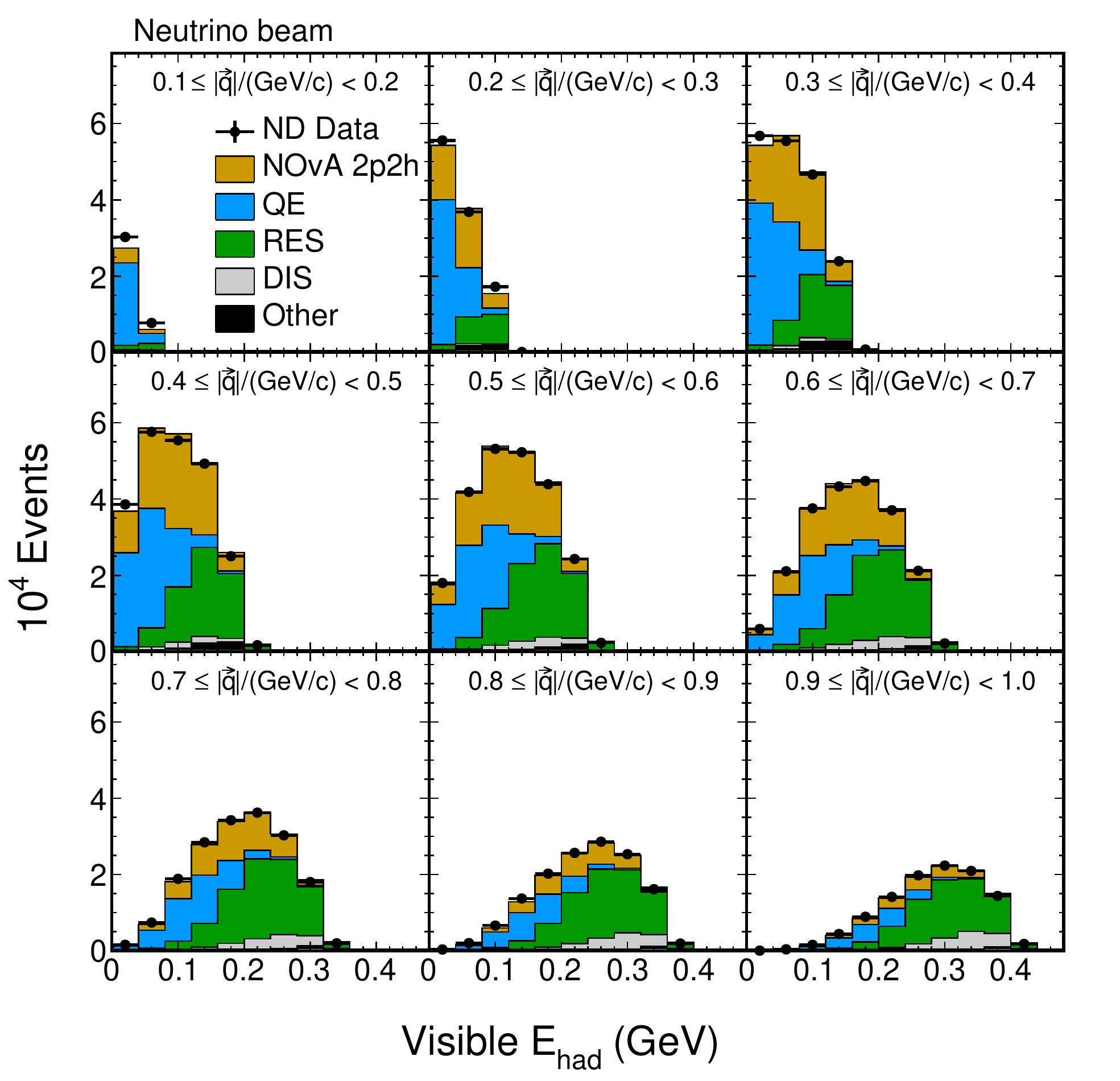}%
	        \label{fig:q0q3panela} }
 \subfloat[]{
	        \includegraphics[width=.49\linewidth]{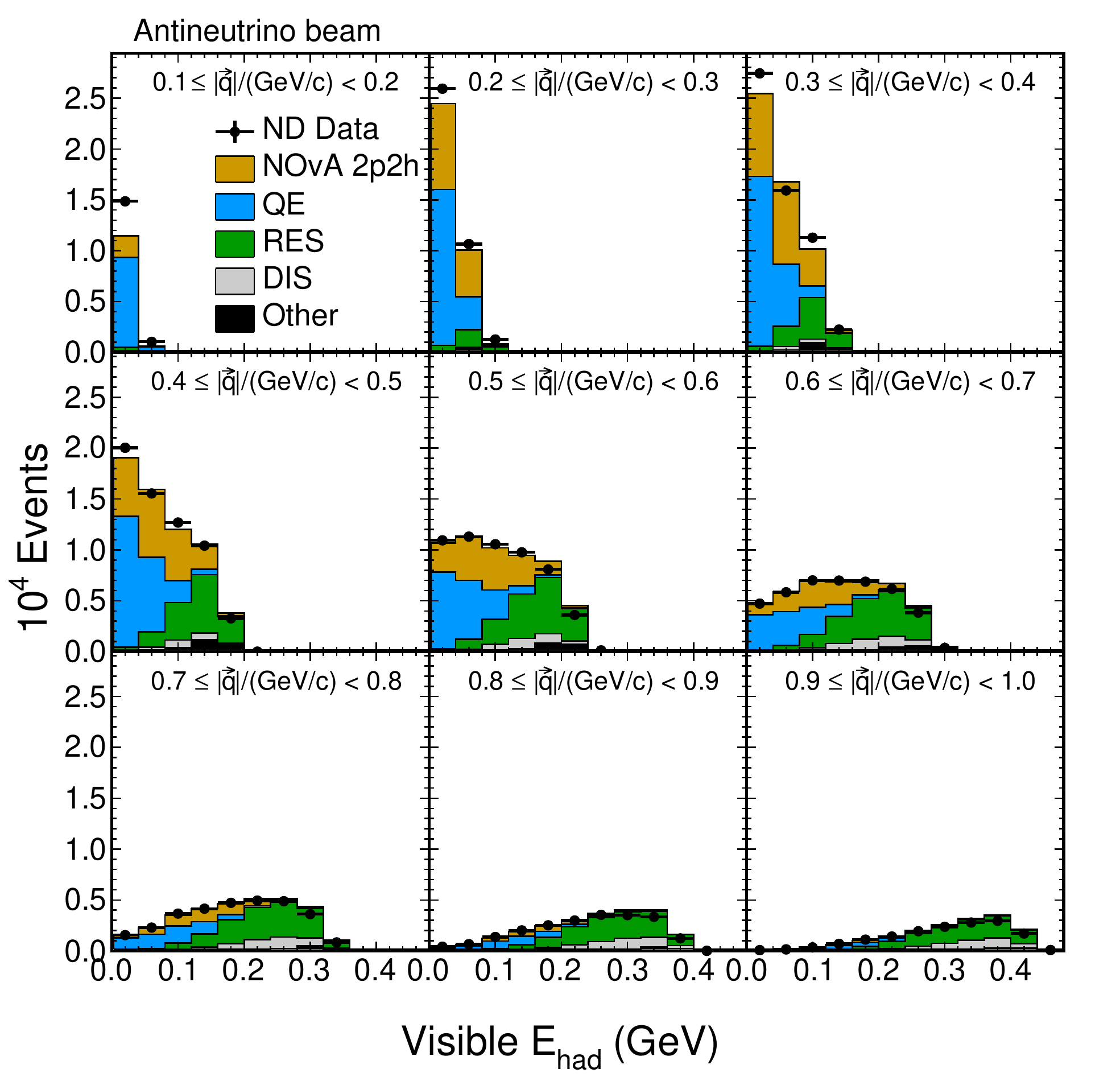}%
	        \label{fig:q0q3panelb} }
   	}
    \caption{Comparison of fully adjusted simulation to ND data in reconstructed visible hadronic energy for neutrinos (left) and antineutrinos (right).  The panels show $\unit[0.1]{GeV/}c$ wide bins of reconstructed momentum transfer from 0.1--$\unit[0.2]{GeV/}c$ (upper left) to 0.9--$\unit[1.0]{GeV/}c$ (lower right).}
	\label{fig:q0q3panels}
\end{figure}

The final distributions of \ehadvis{} and \qmagreco{} after all adjustments are shown in Fig.~\ref{fig:finaltune}.
The modified simulation largely matches data (by construction) in regions where 2p2h is significant.
The lowest visible hadronic energy bin in both beams still shows disagreement, mostly due to smearing from the quantities being modified $(q_0, \qmag)$ to the reconstructed quantities (\ehadvis{}, \qmagreco{}) used in the fit.
There are residual discrepancies in the regions dominated by pion production, which suggests further model adjustments may be warranted.  Figure~\ref{fig:finaltune} also shows the final neutrino energy distribution, which is the key variable in neutrino oscillation measurements.
The shape of this distribution, and the resolution with which NOvA measures it, is largely unchanged by the adjustment procedure, since the NOvA detectors are calorimeters and the changes do not significantly change the amount of invisible energy.
According to the simulation, the mean bias $\left\langle (E^{reco}_{\nu} - E^{true}_{\nu})/E^{true}_{\nu} \right\rangle$ is -3.6\% (-2.5\%) for neutrinos (antineutrinos) with GENIE's default prediction and -2.3\% (-2.1\%) after all the adjustments; the RMS of this variable shifts from 10.6\% (9.3\%) to 10.5\% (9.3\%).\footnote{The energy estimator is designed to replicate the peak of the neutrino energy distribution near 2 GeV, not the overall mean, which leads to a small bias in the mean of the reconstructed energy.}
Figure~\ref{fig:q0q3panels} shows the visible hadronic energy in bins of momentum transfer, illustrating that the adjusted 2p2h component resides at intermediate values of \qzero{} and \qmag{}, as expected from observations in electron scattering~\cite{susa-mec-electron} and in MINERvA~\cite{minerva-lowrecoil,minervanewantinu}.  This is a key indicator that the discrepancy between the default simulation and ND data is likely due largely to 2p2h interactions.
Other kinematic distributions comparing data and simulation can be found in Appendix A.

\section{Cross-section systematic uncertainties}
\label{sec:systs}

GENIE includes an evaluation of many cross-section uncertainties and enables corresponding adjustments to model parameters.
We employ this uncertainty model, the details of which can be found in the GENIE manual~\cite{geniemanual}, largely unchanged.
However, we substitute our own treatment in several instances where different uncertainties are warranted, as described in the following sections.

\subsection{Quasi-elastic interactions}

The default GENIE systematic uncertainty for CCQE $M_{A}$ is +25$\%$/-15$\%$.  This uncertainty was constructed to address the historical tension between bubble chamber and NOMAD measurements~\cite{nomadma}, and MiniBooNE~\cite{miniboonema}, tension which is now largely attributed to be due to multi-nucleon effects~\cite{nievesminiboone}.  As we explicitly add these multi-nucleon effects and their associated uncertainties separately, we reduce the size of the CCQE $M_{A}$ uncertainty to 5$\%$, which is a rough estimate of the free nucleon scattering uncertainty derived from bubble chamber measurements~\cite{anl1,anl2,anl3,bnl,fnal-15ft}.

In addition to the central value weights discussed in Sec.~\ref{subsec:ext}.\ref{subsubsec:LFG+RPA}, the \valencia/ CCQE nuclear model weights supplied by MINERvA include separate sets of weights that (when applied to the GENIE RFG distributions) produce alternate predictions for the \valencia/ model under enhancement and suppression uncertainties~\cite{rpa-valencia-unc}.
Separate weights are generated for neutrinos and antineutrinos.
We include these variations in the uncertainties we consider.

\subsection{Resonance interactions}

As discussed in Sec.~\ref{subsec:res}, the $Q^2$ parameterization of the QE nuclear model effect applied to RES is a placeholder for an unknown effect.  Therefore, we take a conservative 100$\%$ one-sided uncertainty on this correction.  This permits the effect to be turned off, but not increased, and it cannot change sign.  This is the largest systematic uncertainty in NOvA's measurement of $\theta_{23}$~\cite{novaosc2019}, and is correlated between neutrinos and antineutrinos.

\subsection{Deep inelastic scattering}

GENIE's uncertainty model includes uncorrelated 50$\%$ normalization uncertainties for DIS events with one- or two-pion final states (any combination of charged or neutral) and $W < \unit[1.7]{GeV/}c^2$.\footnote{The one-pion subset of these states are adjusted in sec. \ref{subsubsec:nonres1pi} based on a fit to bubble chamber data, which concludes the normalization uncertainty is approximately 10\%.  However, those authors admit that their resulting fit is poor, which suggests it may be missing important degrees of freedom.  Therefore, we use their correction to the central value since it is compatible with MINERvA's findings in their data \cite{minerva-pi-tuning} as well as our own, but we believe the uncertainty is artificially overconstrained.  We retain GENIE’s original 50\% uncertainty on the tuned value until a better model is available.}
However, there is no corresponding normalization uncertainty for DIS with $W > \unit[1.7]{GeV/}c^2$, or for any events with pion multiplicity larger than two.
Moreover, the sharp discontinuity going from 50$\%$ to 0$\%$ when crossing the $W = \unit[1.7]{GeV/}c^2$ boundary leads to unphysical variations when used to produce alternate predictions.
We therefore replace the low-$W$ GENIE DIS normalization uncertainties with 32 [4 (0$\pi$, 1$\pi$, 2$\pi$, $>2\pi$) $\times$ 2 (CC, NC) $\times$ 2 (neutrino, antineutrino) $\times$ 2 (interaction on neutron, proton)] independent, uncorrelated 50$\%$ normalization uncertainties for all DIS events up to $\unit[3]{GeV/}c^2$ in $W$.
These uncertainties drop to 10$\%$ for the $W > \unit[3]{GeV/}c^2$ region, which is more consistent with previous measurements of DIS at higher energy\footnote{While the high $W$ region does not significantly affect the NOvA CC oscillation results, which contain DIS events up to approximately $\unit[2.5]{GeV/}c^2$ in $W$, that region is important for other NOvA analyses which utilize higher energy neutrinos, such as NC disappearance measurements.}.
A comprehensive summary of the available data and corresponding theory is given in Ref.~\cite{dis-measurements}.

\subsection{2p2h}

We include three types of 2p2h uncertainty, all of which we take as uncorrelated between neutrinos and antineutrinos, for a total of 6 independent uncertainties.
Throughout, we neglect the influence of short-range correlations on the uncertainties we consider since the 2p2h contribution to the neutrino interactions considered in this work is expected to be dominated by MEC~\cite{src-neutrino}.

\begin{enumerate}
\item \emph{Target nucleon pair identities}

A CC MEC interaction always involves a target nucleon whose identity (proton or neutron) is dictated by charge conservation.
The identity of the second nucleon, coupled to the first in the interaction, is determined by the model.
We examine various theoretical models to determine the relative proportions of neutrons versus protons in the struck (initial state) nucleon pairs and use these predictions to construct an uncertainty. 
For neutrinos, we are interested in the fraction of target pairs that are neutron-proton, $R_{\nu}=np/(np+nn)$, which for the \valencia/ model included in GENIE averages 0.67 over the kinematic range of interest.
A detailed study during the development of the SuSA MEC model concluded that, over a range of kinematics, their fraction is 0.8--0.9~\cite{susa-mec}.
The Empirical MEC model in GENIE defaults to a value of 0.8.
Though the \valencia/ model predicts $R$ as a function of the momentum transfer, Empirical MEC does not, and we do not have a full simulation of the SuSA model to study the impact in our phase space.
For this analysis we therefore retain $R_{\nu}=0.8$ as a fixed central value and take the range 0.7--0.9 as a $1\sigma$ uncertainty.
In future work we plan to study the effect of the differing models' predictions as a function of kinematics in more detail.
For antineutrinos, we use the same central value and uncertainty range for the $R_{\bar{\nu}} = np/(np+pp)$ ratio.
This uncertainty has a small effect on predictions of observables; the expected visible hadronic energy shapes of $R=1$ vs $R=0$ events are shown in Fig.~\ref{fig:2p2hnpc}.

\item \emph{Energy dependence of total cross section} 

The second uncertainty addresses the difference between MEC models in cross section as a function of neutrino energy.
Four MEC models are examined: Empirical~\cite{teppeimecmodels}, \valencia/~\cite{nieves-mec}, that of the Lyon group (Martini and Ericson)~\cite{martini-mec}, and SuSA~\cite{susa-mec}.
As our tuning procedure enforces a normalization inferred from our data, we are concerned mostly with shape differences; therefore, we rescale the predictions.  In principle, we prefer to normalize at higher energies where the predicted spectra flatten, but several models do not extend this far.
Thus, we take the following approach: the \valencia/ prediction from GENIE is scaled to match Empirical MEC at \unit[10]{GeV}; the SuSA prediction is scaled to match Empirical MEC at \unit[4]{GeV} (the highest-energy prediction in \cite{susa-mec}); and the Lyon prediction is scaled so that its peak is the same as that of Empirical MEC.
Our rescaled predictions for $\sigma(E)$ from the models are shown in Fig.~\ref{fig:2p2henuunca}.
We compute the ratios of the renormalized model predictions to Empirical MEC and construct a function which approximately envelopes the variations, as shown in Fig.~\ref{fig:2p2henuuncb}.
This function becomes an energy-dependent 2p2h normalization uncertainty.

This procedure is based on neutrino MEC models.
Since fewer models that consider antineutrinos are available, the same envelope is used (uncorrelated) for antineutrinos.

\begin{figure}[ht]
    \centering
 \subfloat[]{
        \centering
        \includegraphics[width=.48\textwidth]{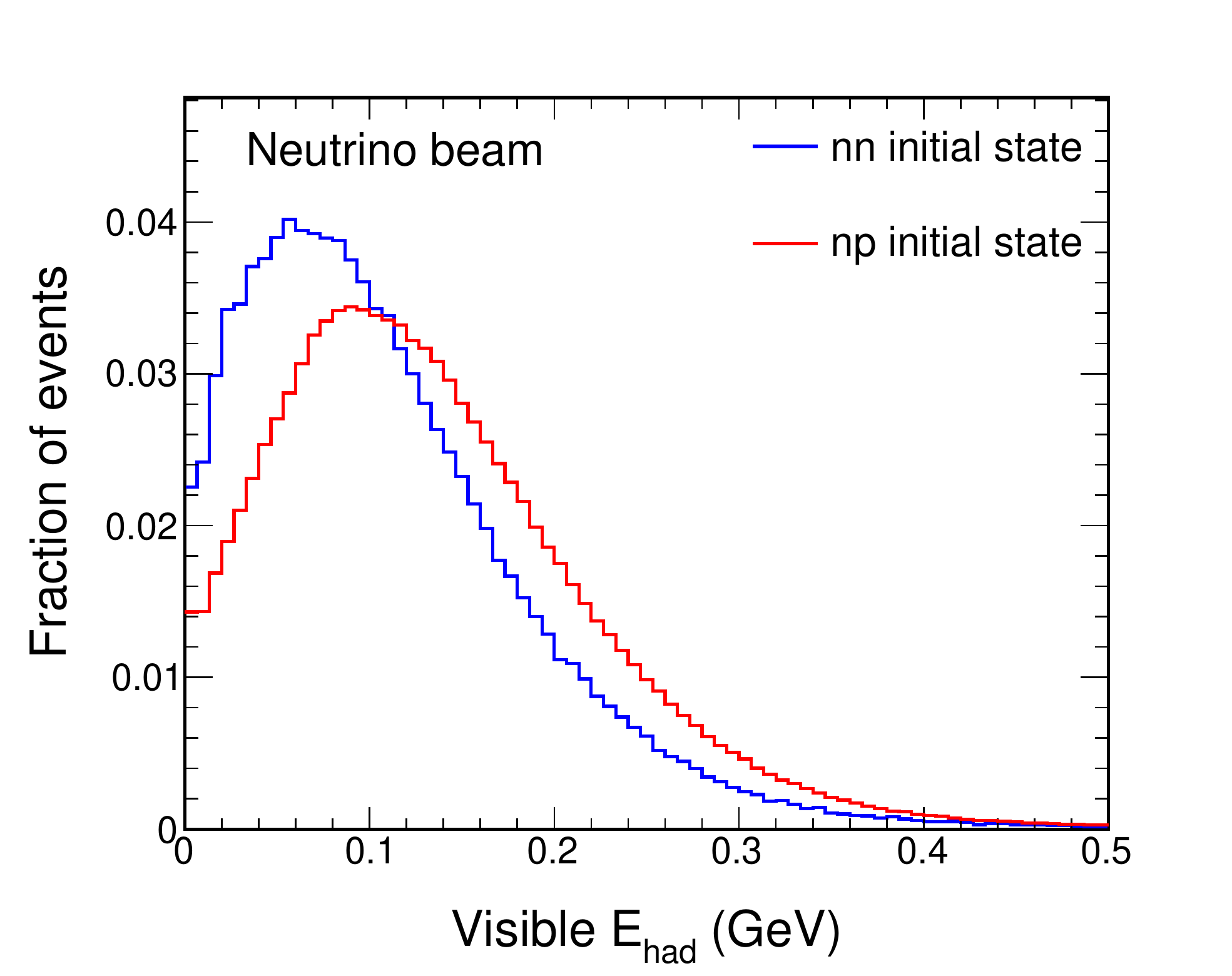}
        \label{fig:2p2hnpa} }
 \subfloat[]{
        \centering
        \includegraphics[width=.48\textwidth]{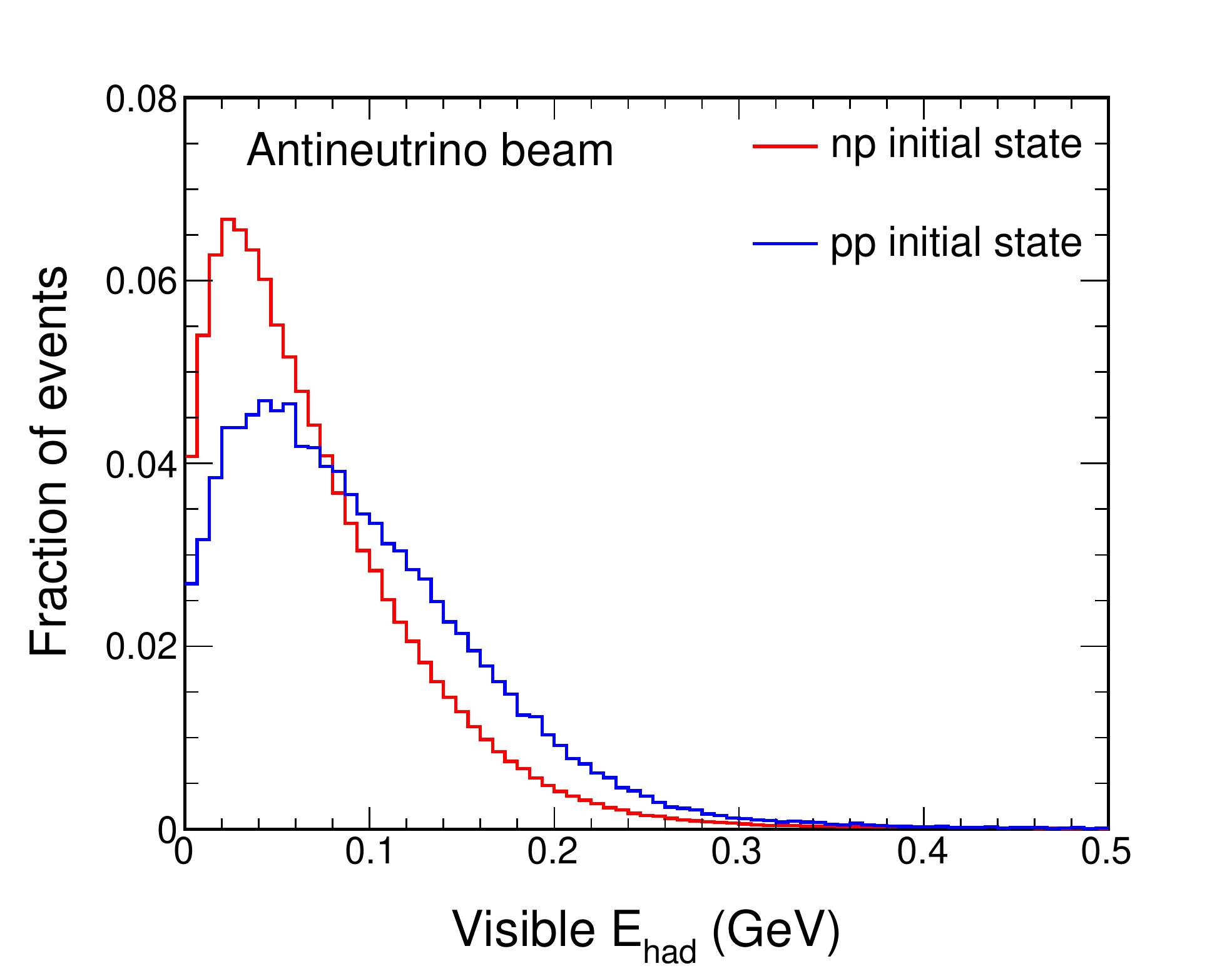}
        \label{fig:2p2hnpb} }
    \caption{Visible hadronic energy distribution for simulated Empirical MEC interactions composed of $np$ initial state pairs vs.\ $nn$ pairs in the neutrino beam (left) and $np$ vs.\ $pp$ pairs for the antineutrino beam (right).}
    \label{fig:2p2hnpc}
\end{figure}

\begin{figure}[ht]
\captionsetup[subfigure]{labelformat=parens}
    \centering
 \subfloat[]{
        \centering
        \includegraphics[width=.48\textwidth]{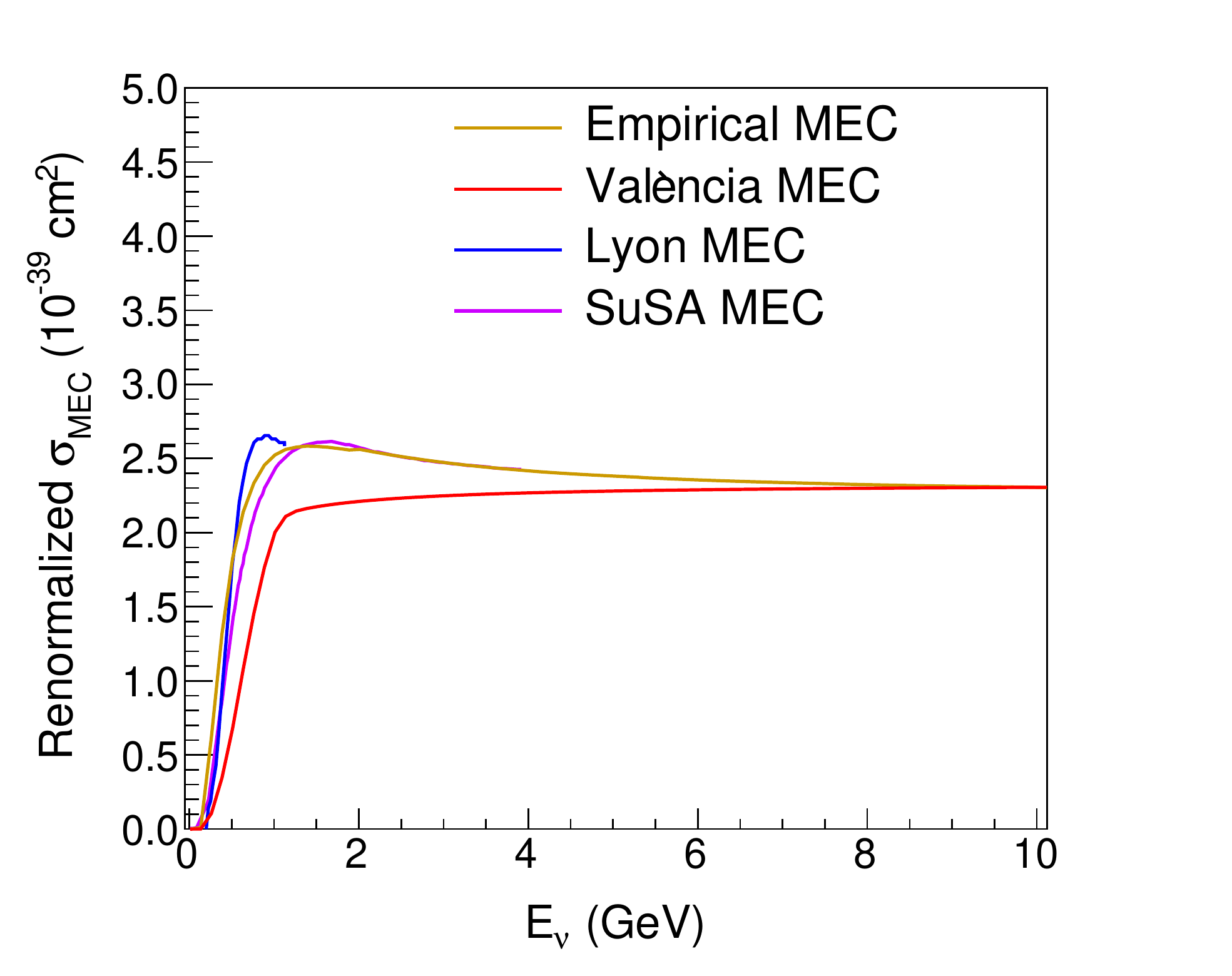}
        \label{fig:2p2henuunca} }
 \subfloat[]{
        \centering
        \includegraphics[width=.48\textwidth]{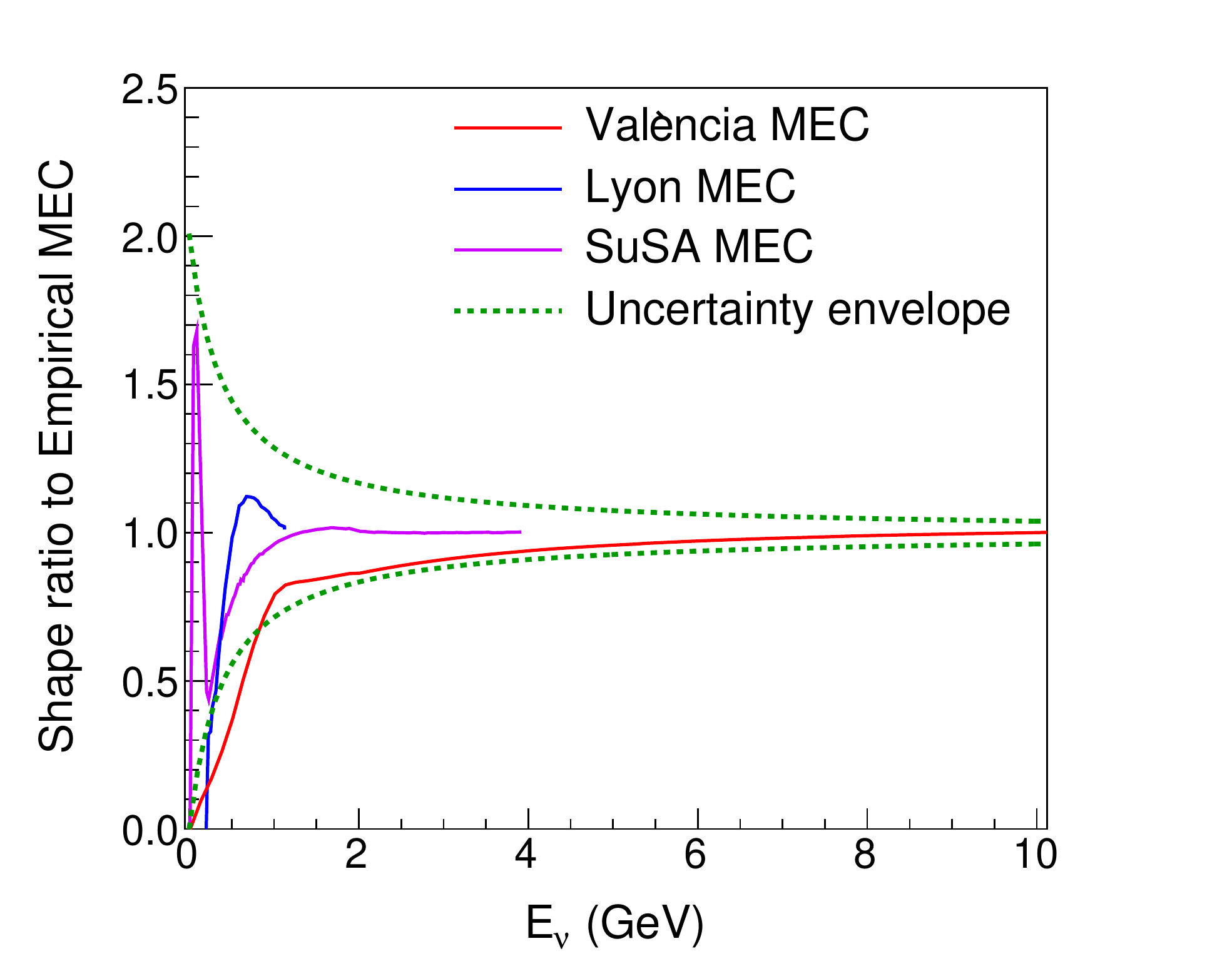}
        \label{fig:2p2henuuncb} }
    \caption{Neutrino energy distributions for various MEC neutrino models, rescaled as described in text (left), and then taken as a ratio to GENIE Empirical MEC, with systematic uncertainty envelope (dashed curve, right).}
    \label{fig:2p2henuunc}
\end{figure}

\item \emph{2p2h dependence on non-2p2h prediction} 

The 2p2h fit reshapes the Empirical MEC interactions such that the total simulation will match ND data.  Any imperfections in other parts of the simulation will consequently be absorbed into the resulting 2p2h sample. 
We can quantify this uncertainty by examining the dependence of the 2p2h fit on other systematic uncertainties.
These reactions are known to occupy a region of energy transfer in between QE interactions (at low $q_{0}$) and RES interactions (at higher $q_{0}$); this holds true in \ehadvis{} as well.
In general, uncertainties that affect the \ehadvis{} distribution of the non-2p2h prediction shift the mean to be higher or lower in \qzero{}, and thus more like a purely RES or QE spectrum.
As a result, the fitted 2p2h spectrum moves in the opposite direction in \qzero{}.
A similar effect holds in \qmag{}.
Using the largest non-2p2h cross-section systematic uncertainties, we apply correlated $1\sigma$ shifts to create the largest \qzero{}-shifting distortions allowed by our uncertainty treatment, which conservatively bound this effect.

The shifts listed in Table \ref{tab:mecsyst} are combined to distort the non-2p2h simulation to be more more ``RES-like" or ``QE-like", resulting in a fitted 2p2h prediction that is more ``QE-like” or ``RES-like” respectively.
The uncertainties in the table are either standard GENIE systematic uncertainties or are described herein.

\begin{table}[ht!]
  \begin{center}
    \caption{Correlated systematic uncertainty shifts used to make the non-2p2h 
simulation more “RES-like” or “QE-like” before fitting the 2p2h component.}
    \label{tab:mecsyst}
    \begin{tabular}{c|r|r}
      \textbf{Uncertainty} & \textbf{QE-like} & \textbf{RES-like}\\
      \hline
      QE $M_{A}$ & $+1\sigma$ & $-1\sigma$\\
      QE Nuclear Model Suppression & $+1\sigma$ & $-1\sigma$\\
      QE Nuclear Model Enhancement & $+1\sigma$ & $-1\sigma$\\
      QE Pauli Suppression & $-1\sigma$ & $+1\sigma$\\
      RES $M_{A}$ & $-1\sigma$ & $+1\sigma$\\
      RES $M_{V}$ & $-1\sigma$ & $+1\sigma$\\
      RES low-$Q^2$ suppression & on & off\\
      \hline
    \end{tabular}
  \end{center}
\end{table}
The 2p2h fitting procedure is carried out in each of these two scenarios, for both neutrinos and antineutrinos separately, to create $\pm 1\sigma$ shape uncertainties.  The differences in the fitted $q_{0}$ predictions are illustrated in Fig.~\ref{fig:2p2hshapeuncert}.  We anticipate that 2p2h predictions made using these alternative underlying model assumptions should bracket the unknown true 2p2h response. 
\end{enumerate}
\begin{figure}[ht!]
    \centering
 \subfloat[]{
        \centering
        \includegraphics[width=.48\textwidth]{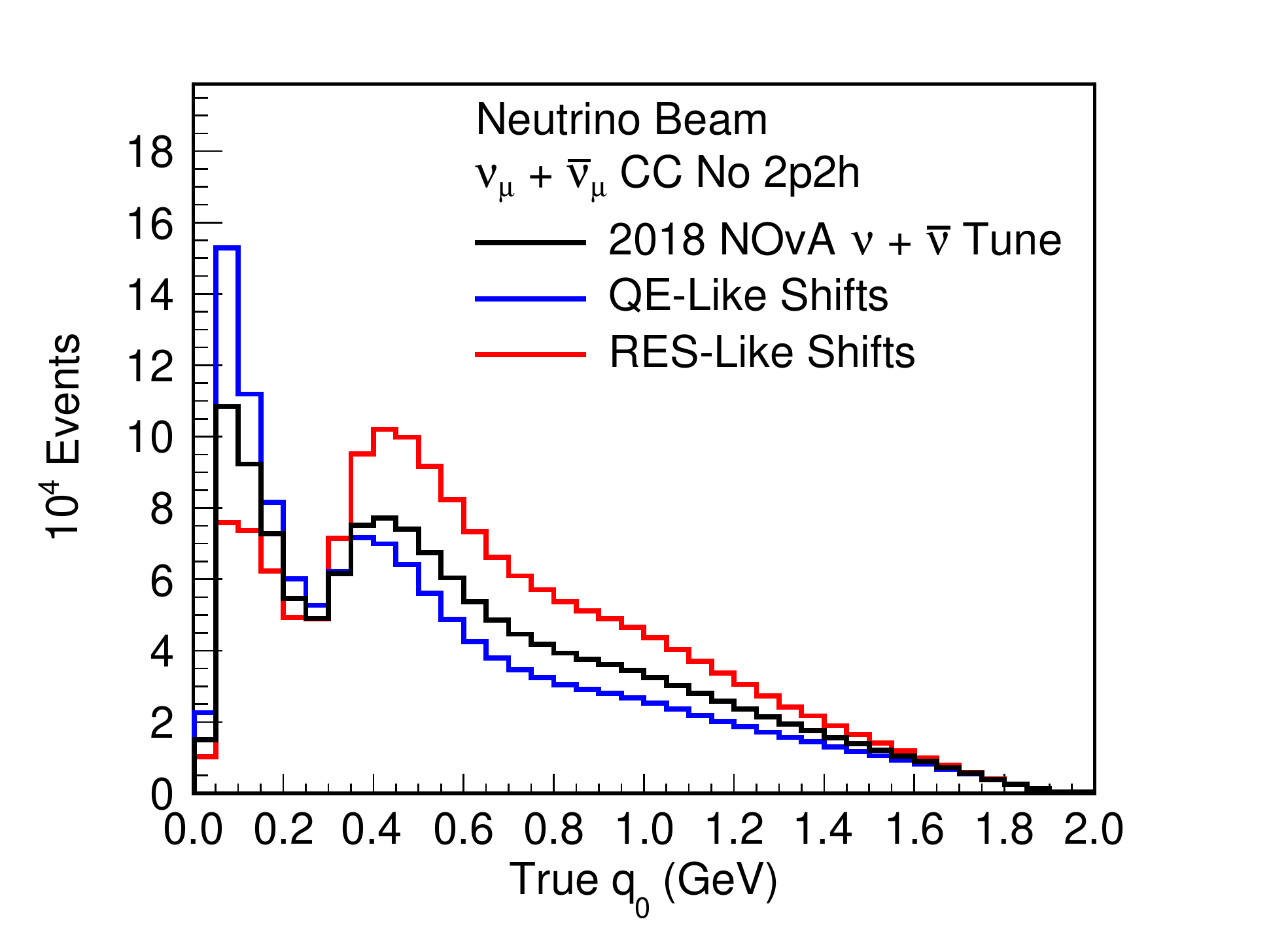}
        \label{fig:2p2huncertbasemc} }
 \subfloat[]{
        \centering
        \includegraphics[width=.48\textwidth]{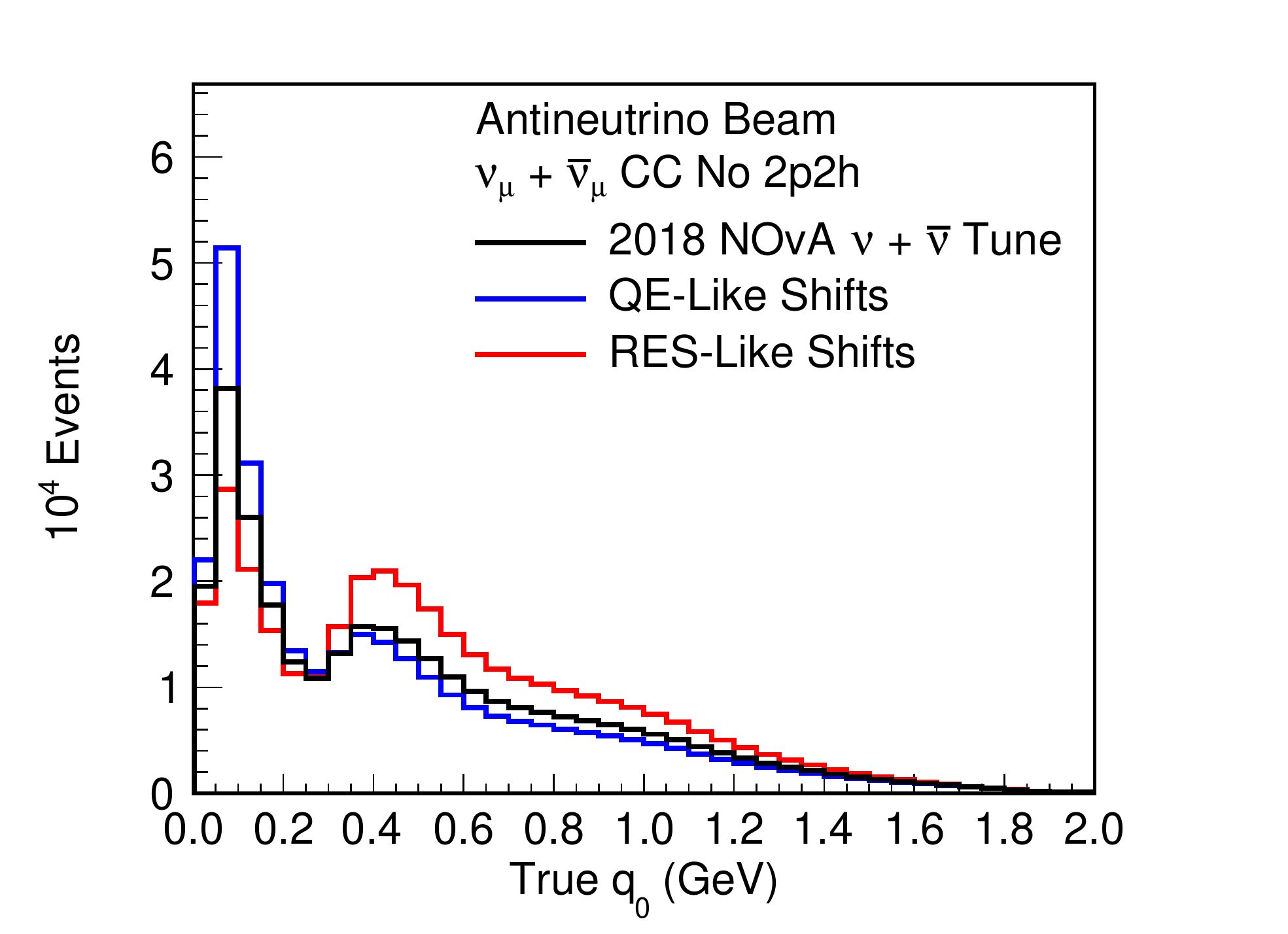}
        \label{fig:2p2huncertshifts} }
\quad
 \subfloat[]{
        \centering
        \includegraphics[width=.48\textwidth]{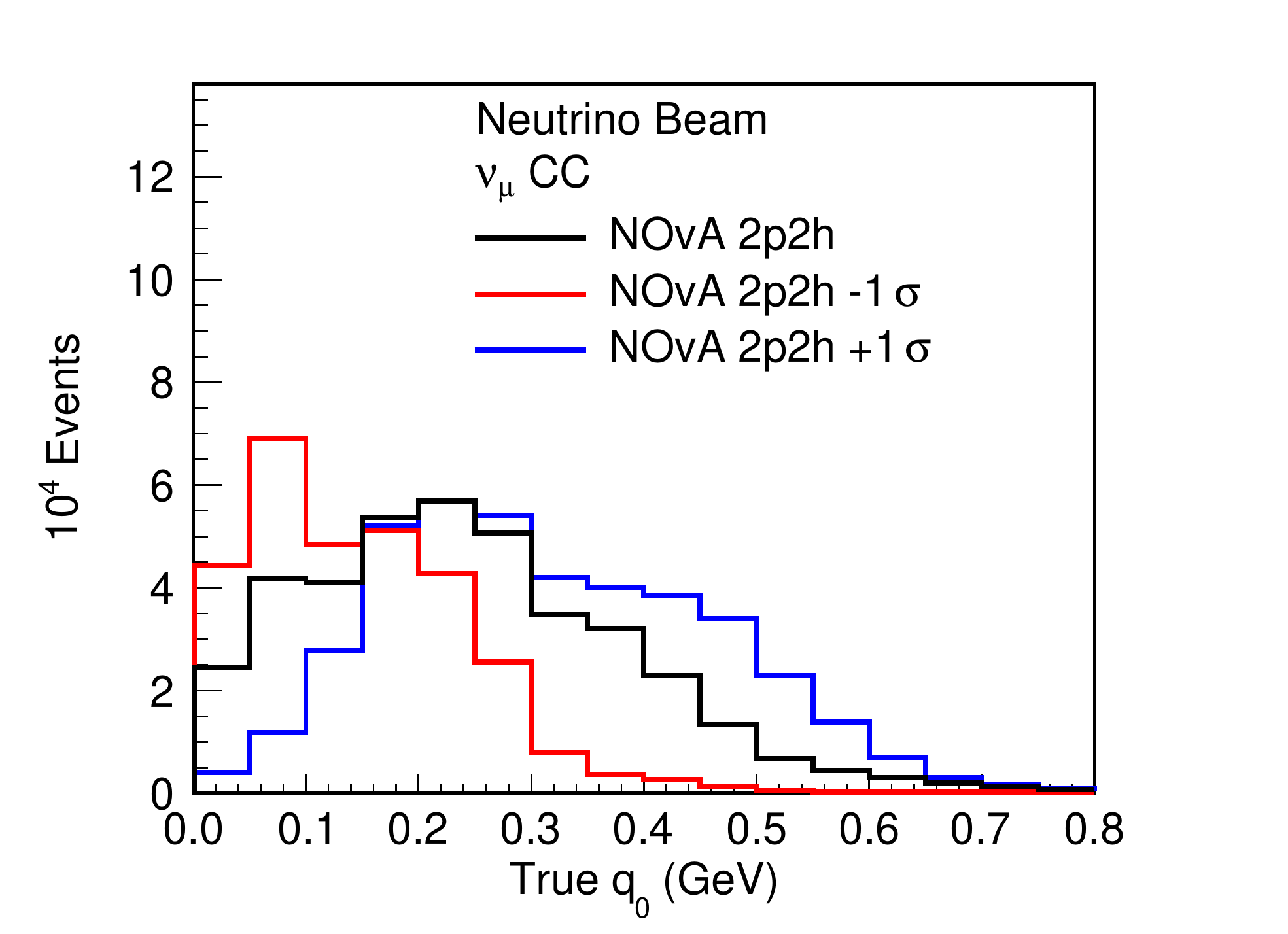}
        \label{fig:2p2huncertbasemcrhc} }
 \subfloat[]{
        \centering
        \includegraphics[width=.48\textwidth]{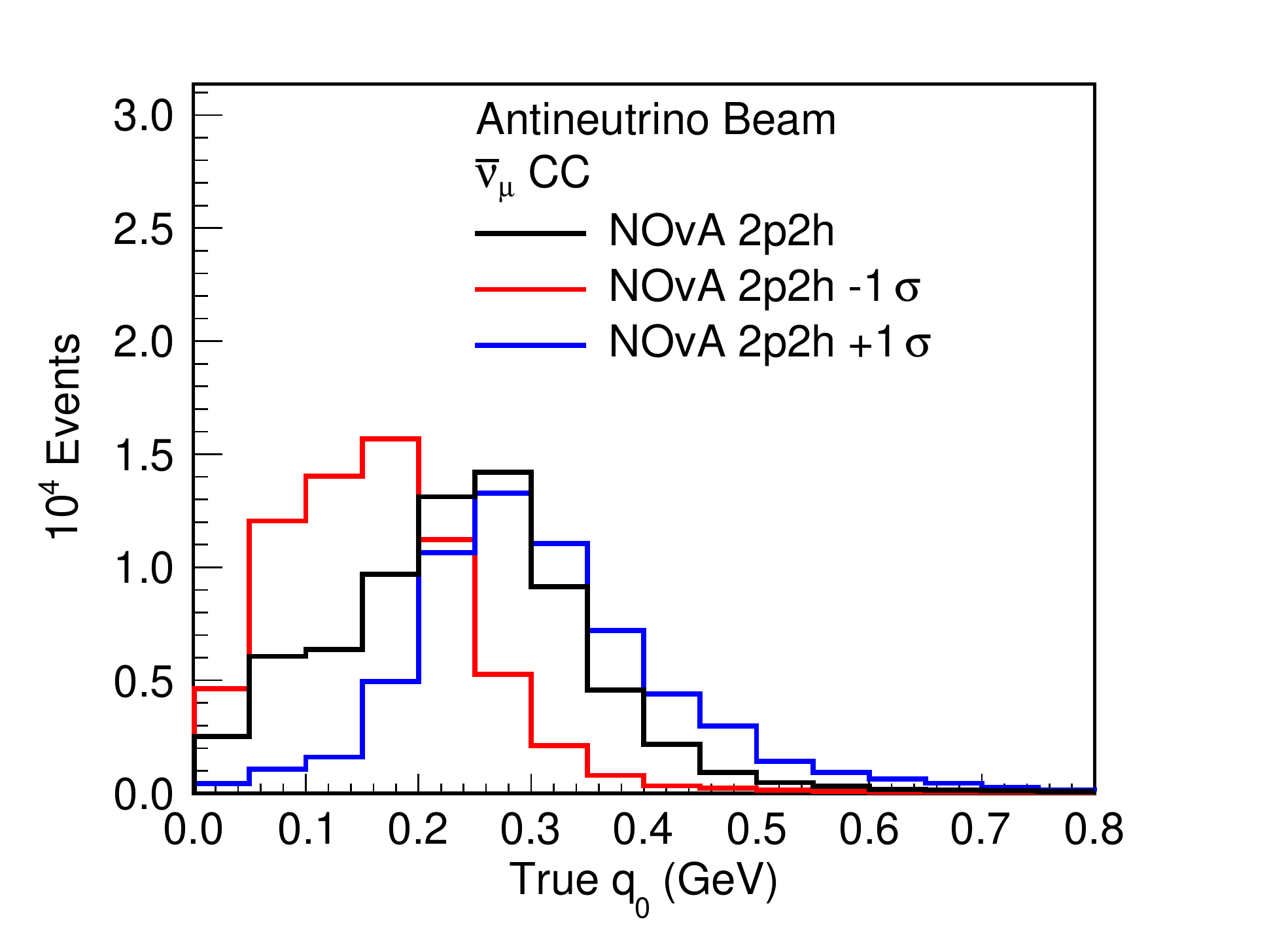}
        \label{fig:2p2huncertshiftsrhc} }
    \caption{True energy transfer distributions showing the result of shifting the fully-adjusted non-2p2h prediction to make it more QE-like or RES-like (top row; neutrino mode at left, antineutrino mode at right) and the resulting 2p2h fitted distributions we take as $1\sigma$ shape uncertainties (bottom row; neutrinos at left and antineutrinos at right).}
    \label{fig:2p2hshapeuncert}
\end{figure}

In the future we anticipate considering other potential sources of 2p2h uncertainty that we have assumed to be subdominant here, including the assignment of final-state energies to the nucleons in the nucleon cluster model in GENIE.

\subsection{Summary of cross-section model uncertainties}

Our modifications and additions to the default GENIE model uncertainties are summarized below.  In this section, ``uncorrelated'' means that parameters in the uncertainty are allowed to vary separately for neutrinos and antineutrinos; ``correlated'' indicates that neutrinos and antineutrinos use the same values.

We alter the following systematic uncertainties:
\begin{enumerate}
	\item For $M_{A}$ in the CCQE model, reduce uncertainty from +25/-15$\%$ to $\pm5\%$ (correlated);
    \item For multi-$\pi$ low-$W$ DIS, replace GENIE's default with 32 custom 50$\%$ uncertainties with expanded $W$ range (uncorrelated).
\end{enumerate}

We also introduce three additional uncertainties:

\begin{enumerate}
    \item QE nuclear model uncertainties (different for neutrino and antineutrino; uncorrelated);
    \item A 100$\%$ uncertainty on the RES low-$Q^2$ suppression, which can never go above 100$\%$ or negative (correlated);
    \item Three 2p2h uncertainties: one covering uncertainty in target nucleons, one addressing uncertainties in the energy dependence of the cross section, and one treating uncertainties in the $\qZqThree$ response (all uncorrelated).
\end{enumerate}

\begin{figure}[ht!]
    \centering
 \subfloat[]{
        \centering
        \includegraphics[width=.48\textwidth]{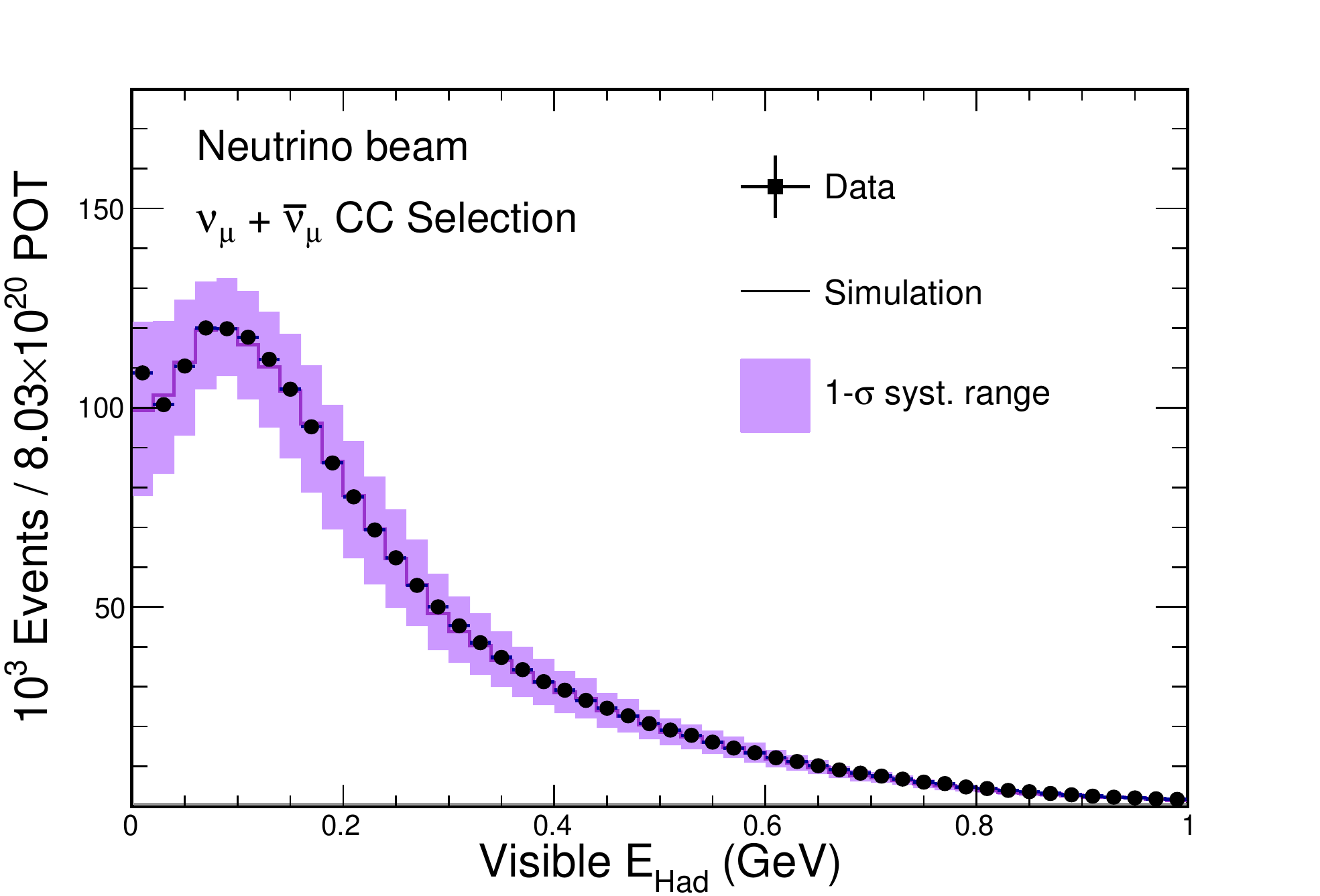}
        \label{fig:fhcq0errband} }
 \subfloat[]{
        \centering
        \includegraphics[width=.48\textwidth]{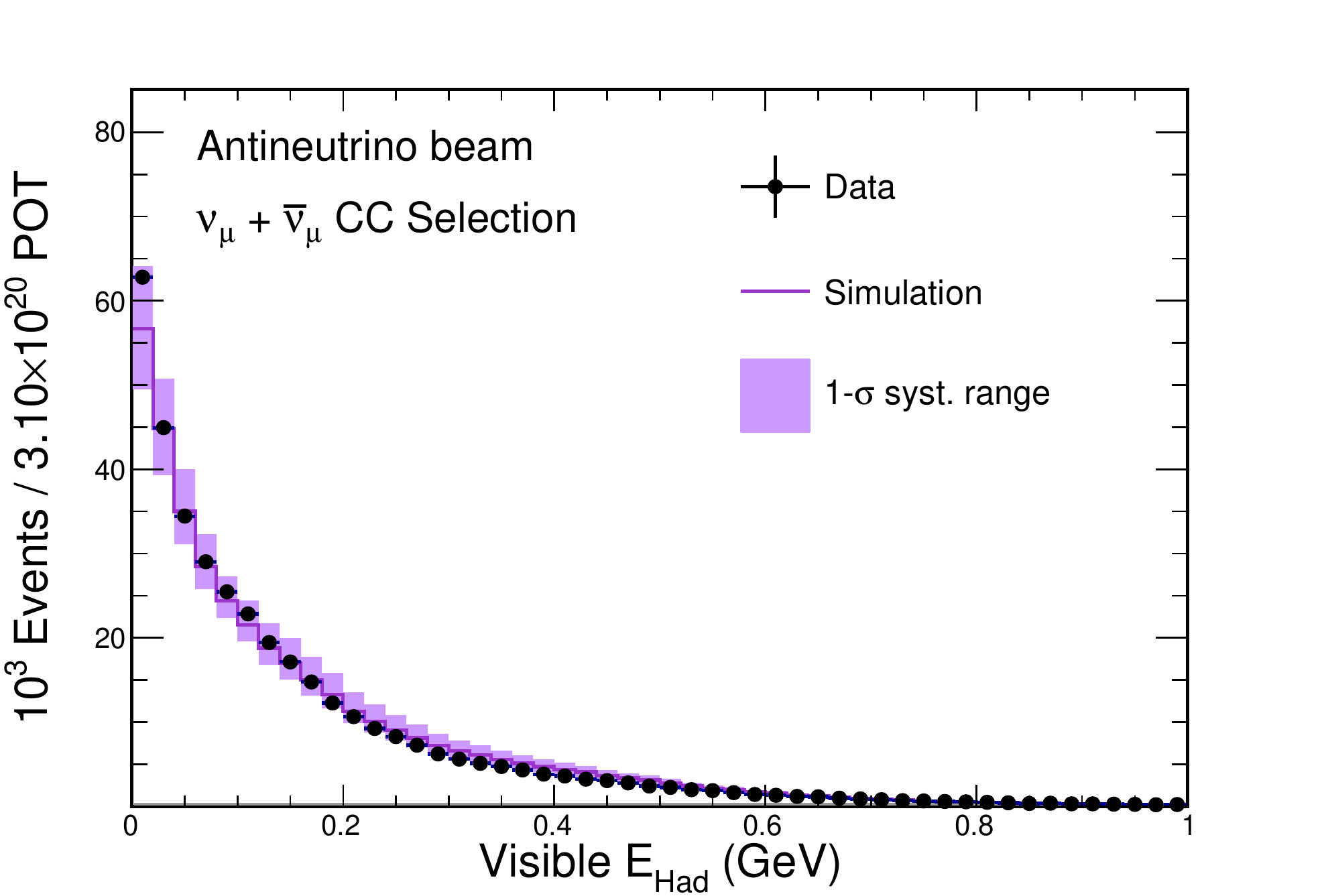}
        \label{fig:rhcq0errband} }
\quad
 \subfloat[]{
        \centering
        \includegraphics[width=.48\textwidth]{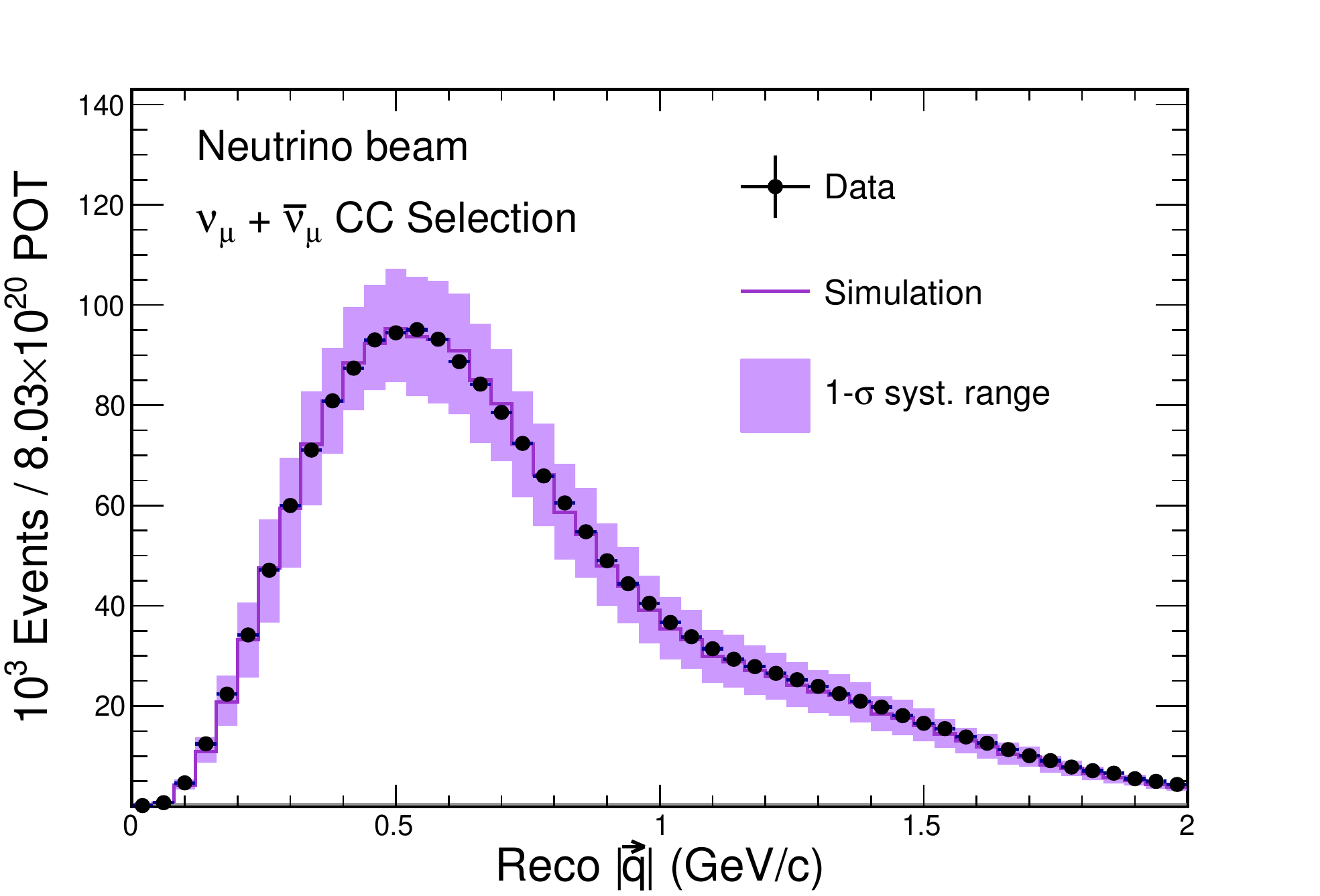}
        \label{fig:fhcq3errband} }
 \subfloat[]{
        \centering
        \includegraphics[width=.48\textwidth]{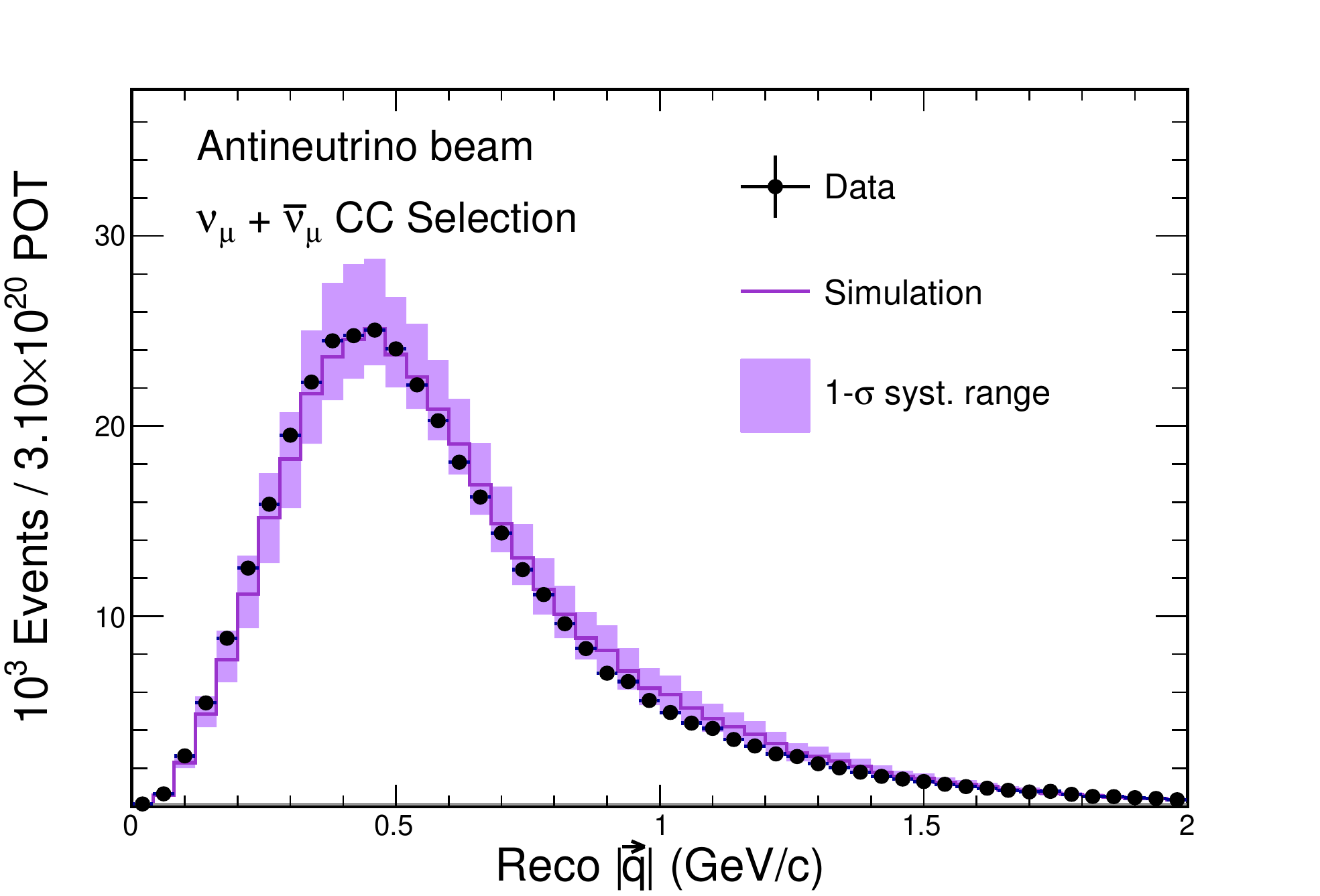}
        \label{fig:rhcq3errband} }
\quad
 \subfloat[]{
        \centering
        \includegraphics[width=.48\textwidth]{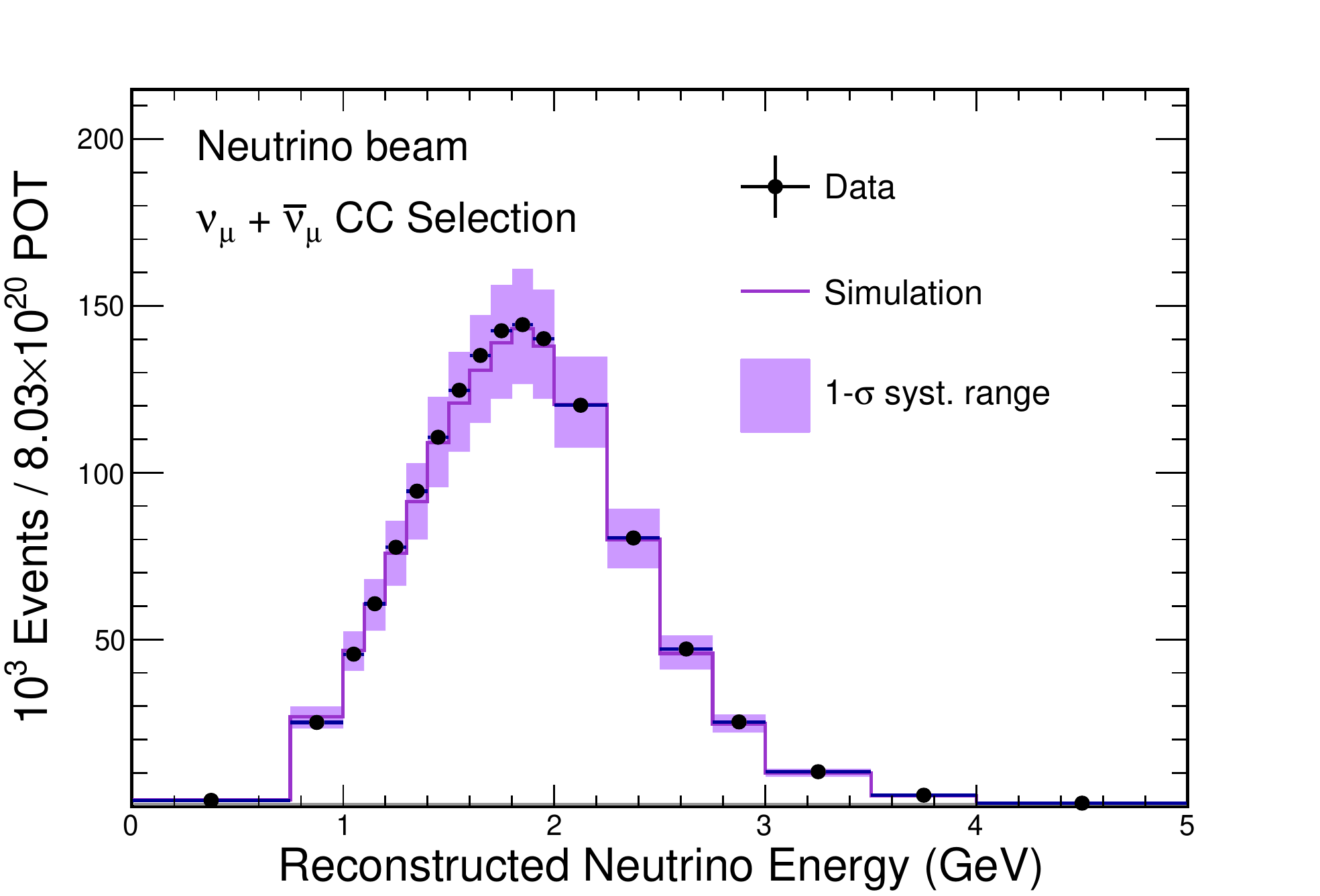}
        \label{fig:fhcnueerrband} }
 \subfloat[]{
        \centering
        \includegraphics[width=.48\textwidth]{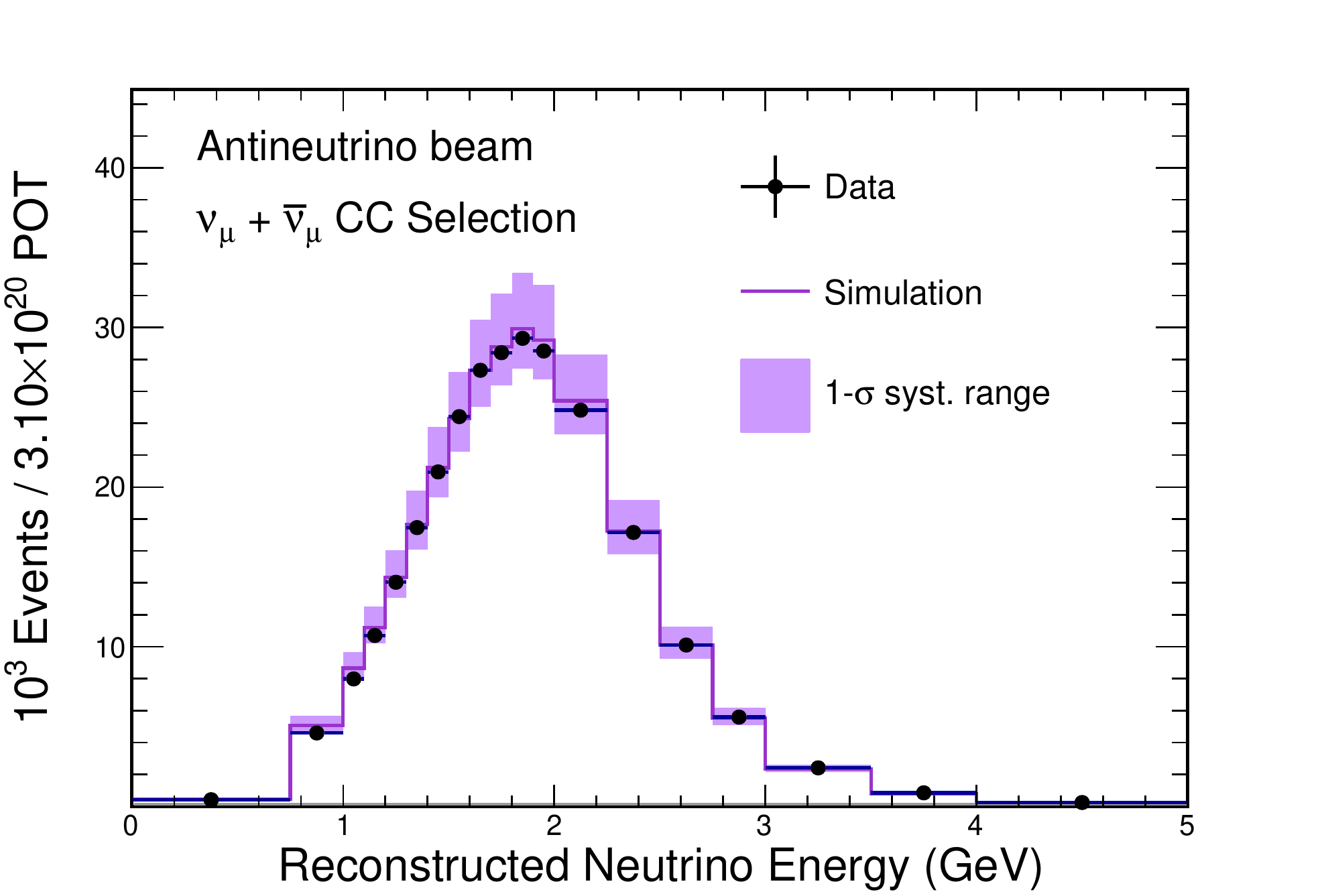}
        \label{fig:rhcnueerrband} }
    \caption{ND data compared to adjusted simulation with cross-section uncertainties represented by the shaded band.  In each bin, the $1\sigma$ deviations from nominal for each cross-section uncertainty are added in quadrature to obtain the band, which has significant bin-to-bin correlations.}
    \label{fig:errband}
\end{figure}

The combined cross-section uncertainties are shown in Fig.~\ref{fig:errband}.  The adjusted neutrino simulation agrees with data somewhat better than the antineutrino simulation, but in both cases the data lies within the uncertainty band.

\section{Comparisons to other observations}
\label{sec:other obs}

\begin{figure}[ht!]
\centering
 \subfloat[]{
\centering
\includegraphics[width=.48\textwidth]{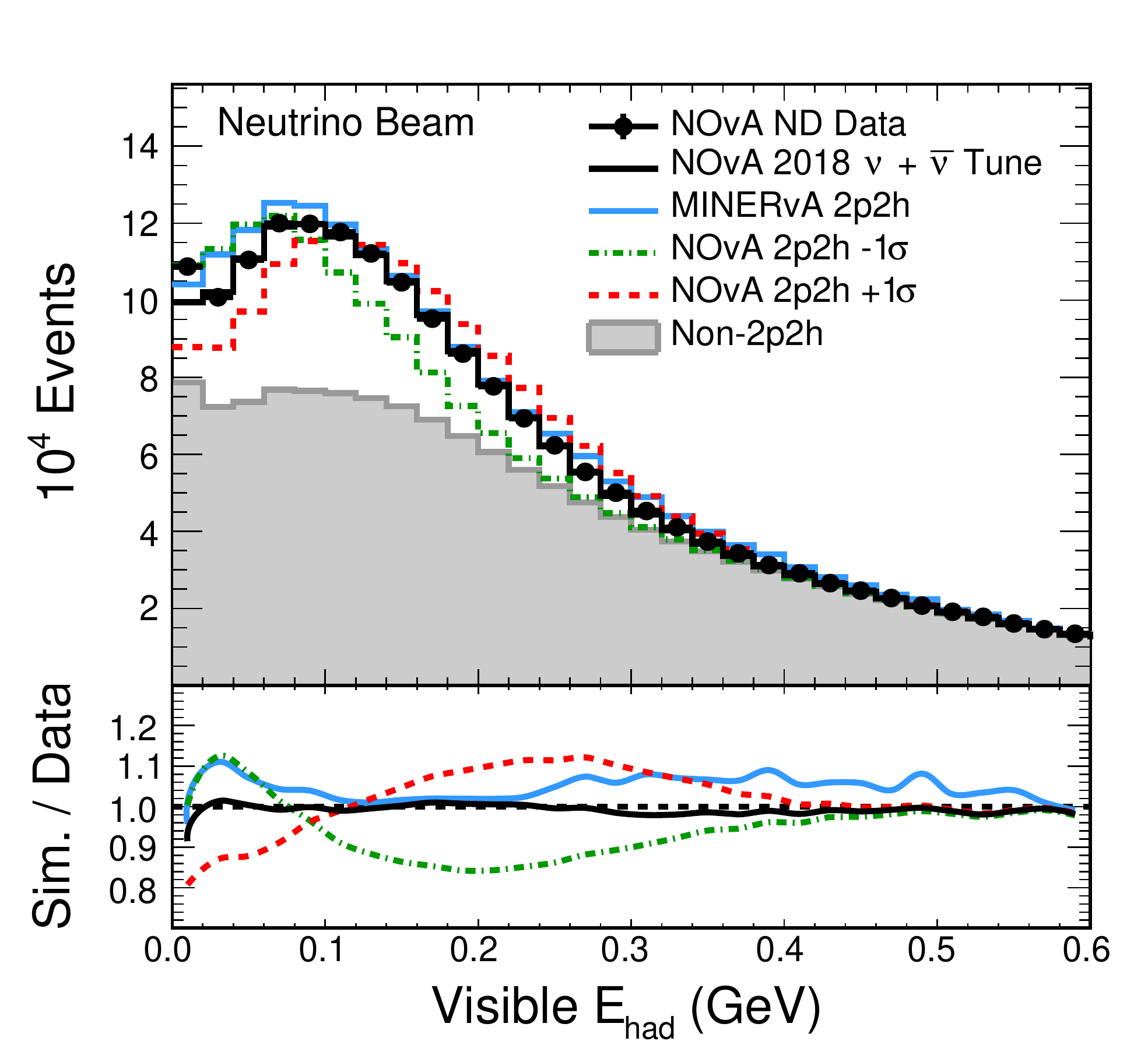}
\label{fig:minervatunefhc} }
 \subfloat[]{
\centering
\includegraphics[width=.48\textwidth]{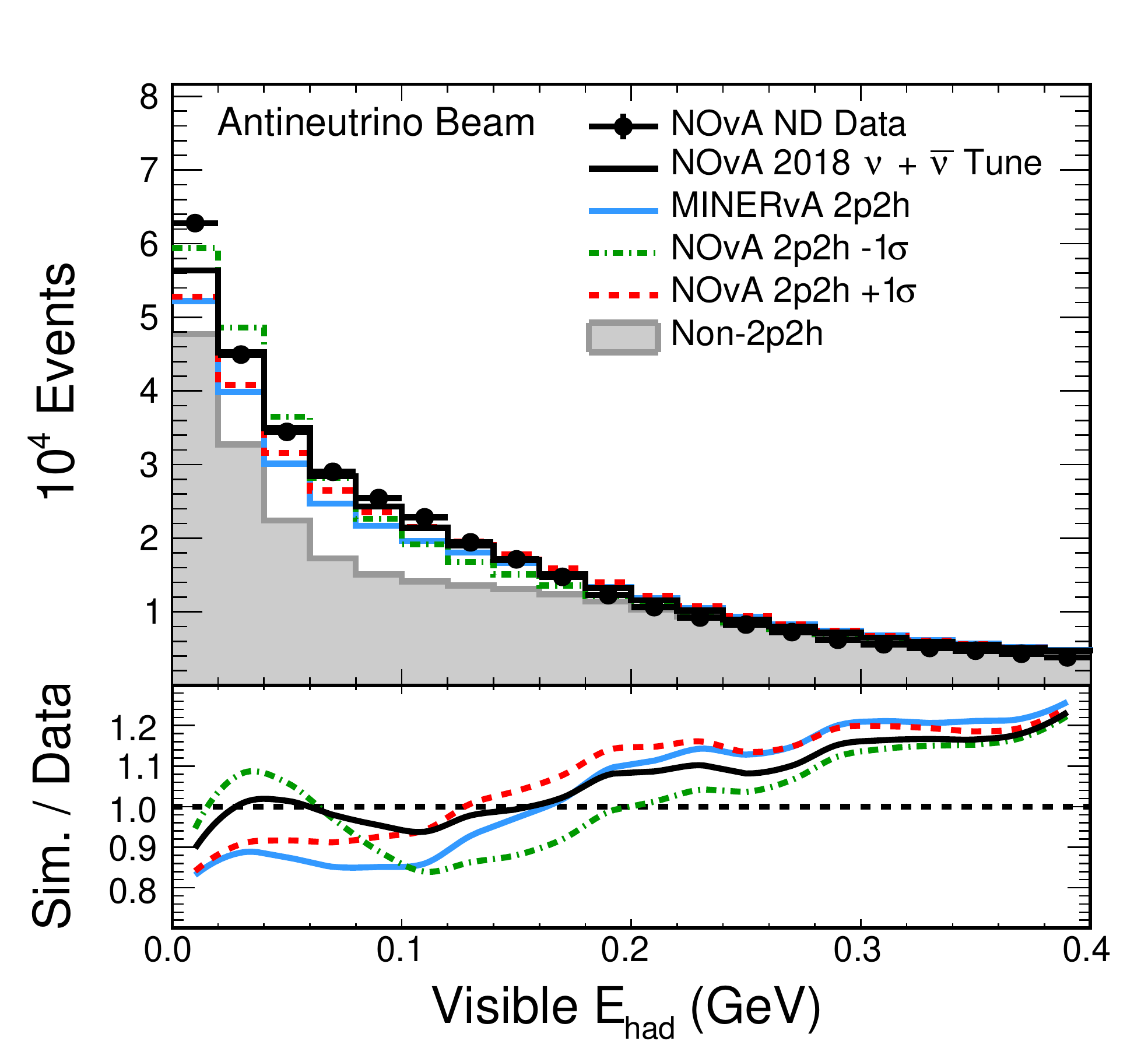}
\label{fig:minervatunerhc} }
\caption{Comparison of reconstructed visible hadronic energy distribution in ND data (black dots) to various simulations for neutrino beam (left) and antineutrino beam (right) running.  The solid black curves correspond to GENIE predictions with the full set of adjustments described in this paper, while the red and purple dotted curves are the simulation with +1 and $-1\sigma$ shifts from the 2p2h \qZqThree{} response systematic uncertainty shown in Fig.~\ref{fig:2p2hshapeuncert}, respectively.  Also shown in solid blue is the result of replacing our tuned 2p2h with MINERvA's tuned 2p2h prediction.  The shaded gray histogram represents the GENIE prediction for non-2p2h interaction channels.}
\label{fig:minervatune}
\end{figure}

As shown in Fig.~\ref{fig:finaltune} and Appendix A, the total inclusive prediction, including the 2p2h component tuned in \qZqThree{} space and fit in \recospace{}, can reproduce our observed ND distributions in numerous kinematic variables.
MINERvA, an on-axis experiment using the same neutrino beam as NOvA, has performed an analogous 2p2h tuning procedure with their inclusive neutrino-mode data set~\cite{minerva-lowrecoil}.
They use GENIE with the same QE nuclear model weights described in Sec.~\ref{subsubsec:LFG+RPA}, and apply a correction to non-resonant single pion production similar to that in the NOvA prescription, but use the \valencia/ MEC model.
In their procedure, the values of a two-dimensional Gaussian are taken as weights to the MEC prediction, and the Gaussian's parameters are fitted in order to match the observed distributions~\cite{2dgauss}.
They find good agreement with their antineutrino data using this adjusted model with no further modifications~\cite{minervanewantinu}.
The result of replacing the 2p2h component of the NOvA fully adjusted simulation with the MINERvA tuned 2p2h prediction is shown in Fig.~\ref{fig:minervatune}.
Qualitatively, the MINERvA model results in a similar overall prediction to the NOvA model, mostly falling within the 1-$\sigma$ uncertainties.

The T2K collaboration uses NEUT~\cite{neutref, neutref2} instead of GENIE to simulate neutrino interactions for their primary neutrino oscillation analysis.  In their recent work~\cite{t2k2018} they also use implementations of the \valencia/ models for the central value prediction of both QE and MEC processes.
Among the uncertainties they consider for QE is a parameterized version of the nuclear model calculations for long-range correlations that is similar to that used by NOvA and MINERvA.  
Uncertainties in the MEC model are bounded between two extreme cases: a prediction using only those MEC diagrams coupling to a $\Delta$-resonance, and a prediction removing all the $\Delta$ channels.  The T2K fit pulls this 2p2h shape uncertainty to the maximum allowed value~\cite{T2Kpulls}.
The 2p2h normalization is also pulled to be 50$\%$ larger than the default prediction.  This is consistent with the findings by NOvA and MINERvA that using an unaltered version of the \valencia/ model is insufficient to describe data.  

\section{Conclusions}

We find that modifications to the default GENIE 2.12.2 model significantly enhance the agreement between selected muon neutrino candidates in the NOvA ND data sample and simulation across a variety of kinematic variables.
Corrections to the QE and soft non-resonant single pion production predictions based on reevaluated bubble chamber measurements are included.
Improved nuclear models are also used to adjust the kinematics of QE scattering.
Furthermore, suppression at low $Q^2$ on resonant pion production is imposed as supported by observations in other experiments and our own ND data.
The Empirical MEC model in GENIE is tuned to match data in our ND.
A set of systematic uncertainties are created, addressing potential weaknesses in the models and bounding the results of our own tuning procedure with ND data.

We will continue to incorporate constraints from other measurements as well as advances in cross-section modeling into our predictions and reduce the impact of systematic uncertainty on our analyses.  Such improvements will not only benefit NOvA and other current experiments, but will be necessary for future experiments such as DUNE, which has stringent requirements on its systematic uncertainty budget~\cite{dunecdr}.

\section{Acknowledgements}
 
 This document was prepared by the NOvA collaboration using the resources of the Fermi National Accelerator Laboratory (Fermilab), a U.S. Department of Energy, Office of Science, HEP User Facility. Fermilab is managed by Fermi Research Alliance, LLC (FRA), acting under Contract No. DE-AC02-07CH11359. This work was supported by the U.S. Department of Energy; the U.S. National Science Foundation; the Department of Science and Technology, India; the European Research Council; the MSMT CR, GA UK, Czech Republic; the RAS, RFBR, RMES, RSF, and BASIS Foundation, Russia; CNPq and FAPEG, Brazil; STFC, and the Royal Society, United Kingdom; and the state and University of Minnesota. This work used resources of the National Energy Research Scientific Computing Center (NERSC), a U.S. Department of Energy Office of Science User Facility operated under Contract No. DE-AC02-05CH11231. We are grateful for the contributions of the staffs of the University of Minnesota at the Ash River Laboratory and of Fermilab.

\clearpage

\appendix

\section{Additional kinematic distributions}

\begin{figure}[ht!]
    \centering
 \subfloat[]{
        \centering
        \includegraphics[width=.48\textwidth]{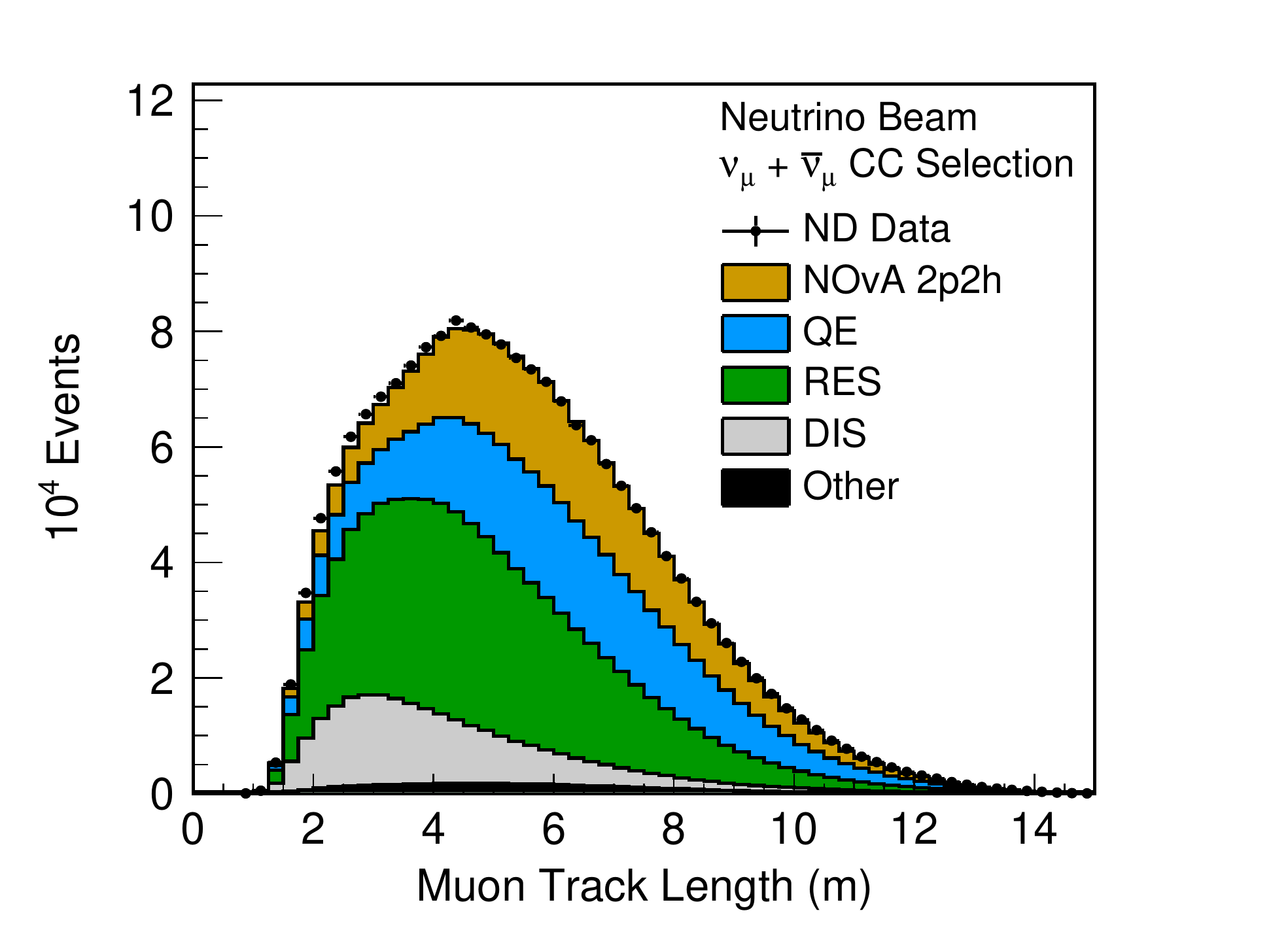}
        \label{fig:finaltunetrklenfhc} }
 \subfloat[]{
        \centering
        \includegraphics[width=.48\textwidth]{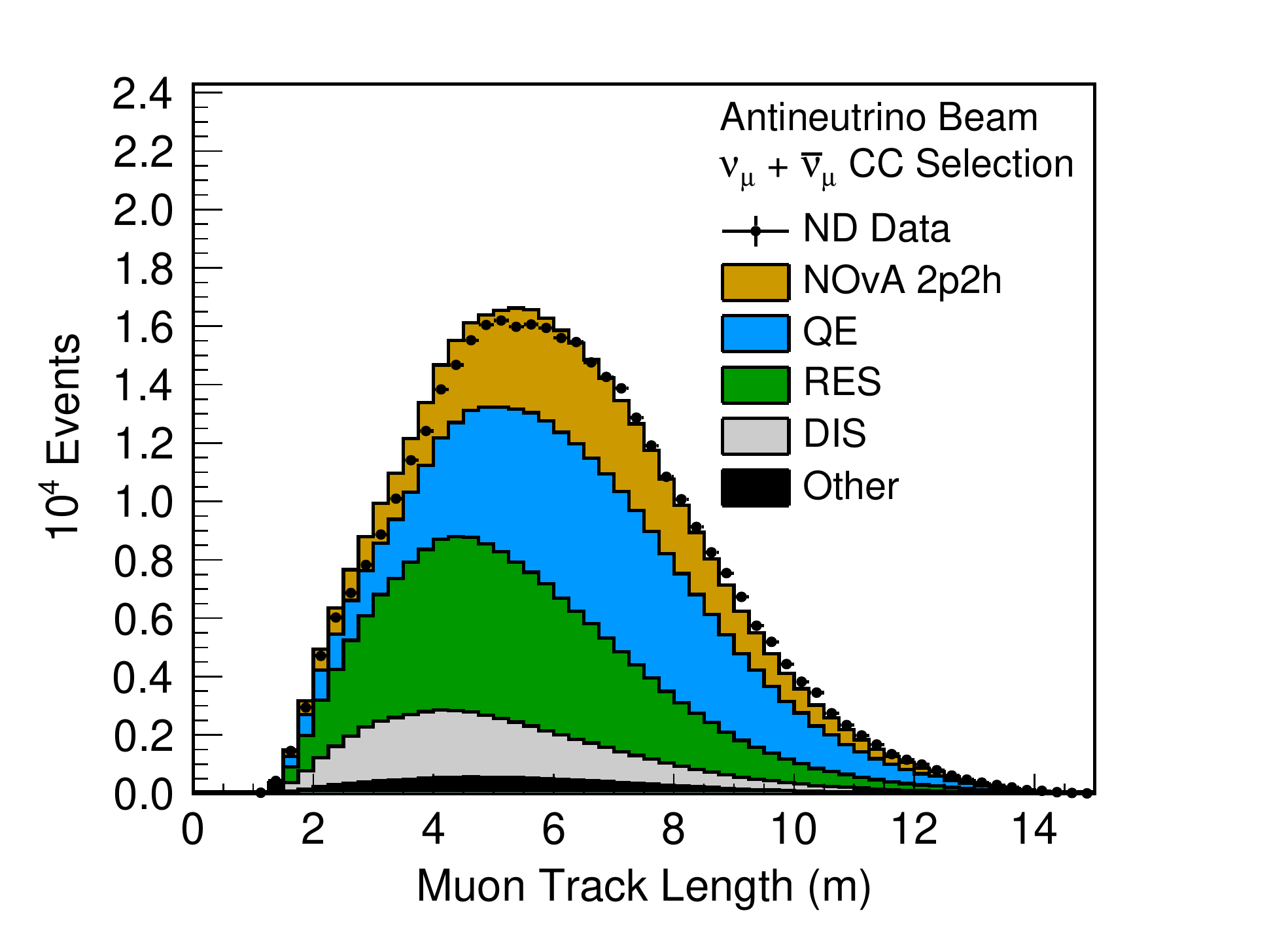}
        \label{fig:finaltunetrklenrhc} }
    \caption{Comparison of fully adjusted simulation to data in muon candidate track length, for neutrino beam (left) and antineutrino beam (right).}
    \label{fig:finaltunetrklen}
\end{figure}

\begin{figure}[ht!]
    \centering
 \subfloat[]{
        \centering
        \includegraphics[width=.48\textwidth]{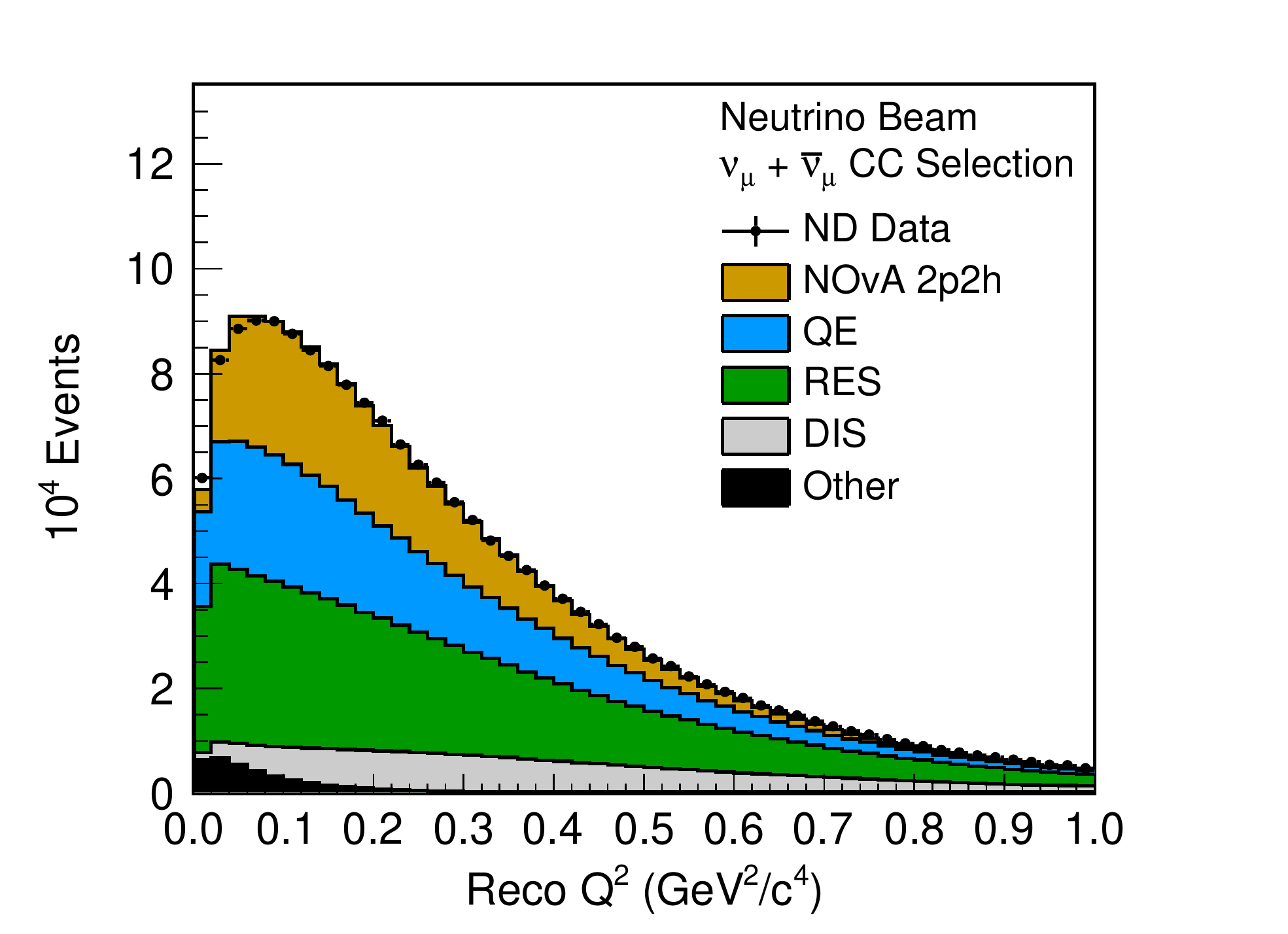}
        \label{fig:finaltuneq2fhc} }
 \subfloat[]{
        \centering
        \includegraphics[width=.48\textwidth]{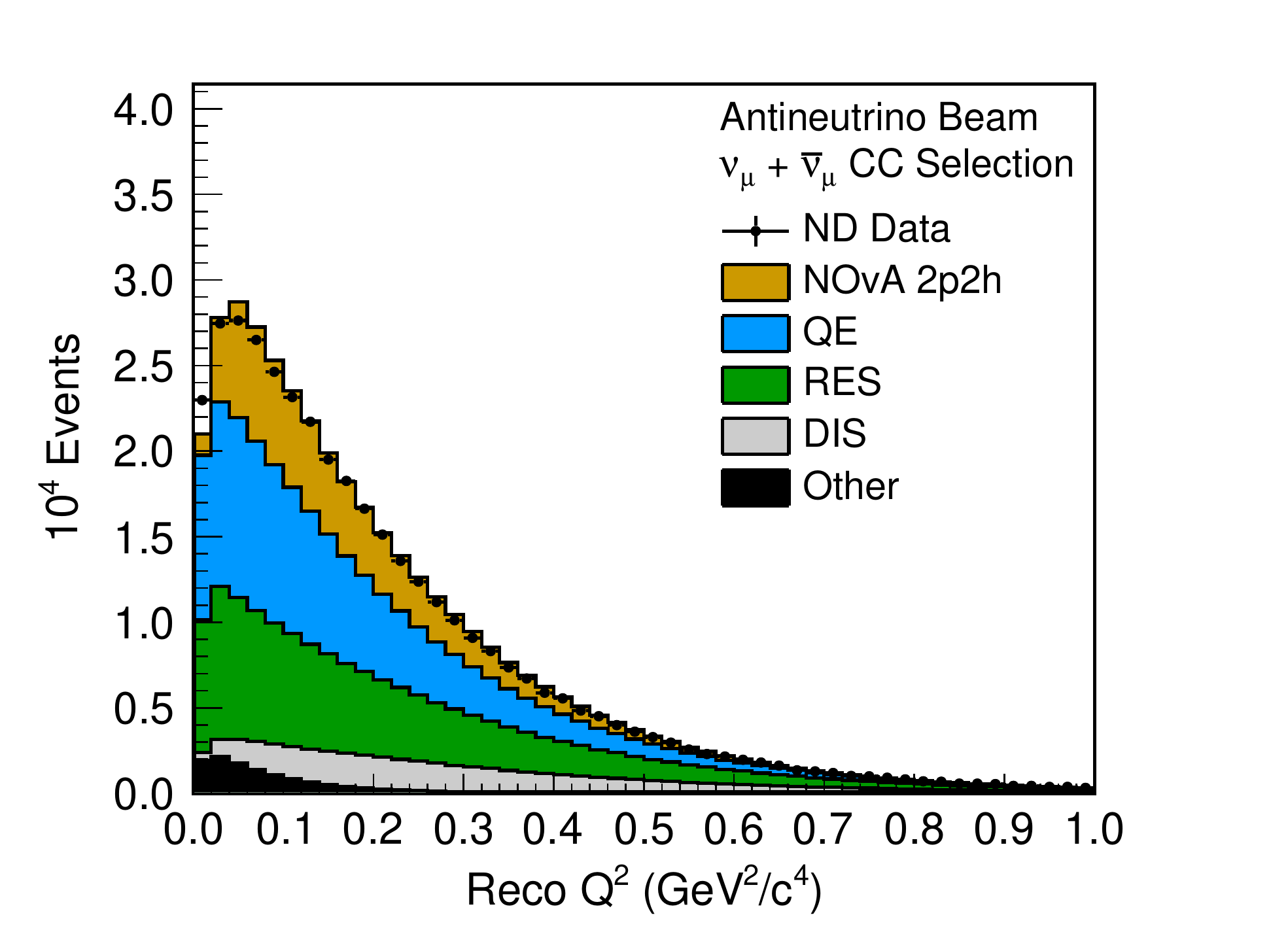}
        \label{fig:finaltuneq2rhc} }
    \caption{Comparison of fully adjusted simulation to data in reconstructed $Q^{2}$, for neutrino beam (left) and antineutrino beam (right).}
    \label{fig:finaltunetrkq2}
\end{figure}

\clearpage

\begin{figure}[ht!]
    \centering
 \subfloat[]{
        \centering
        \includegraphics[width=.44\textwidth]{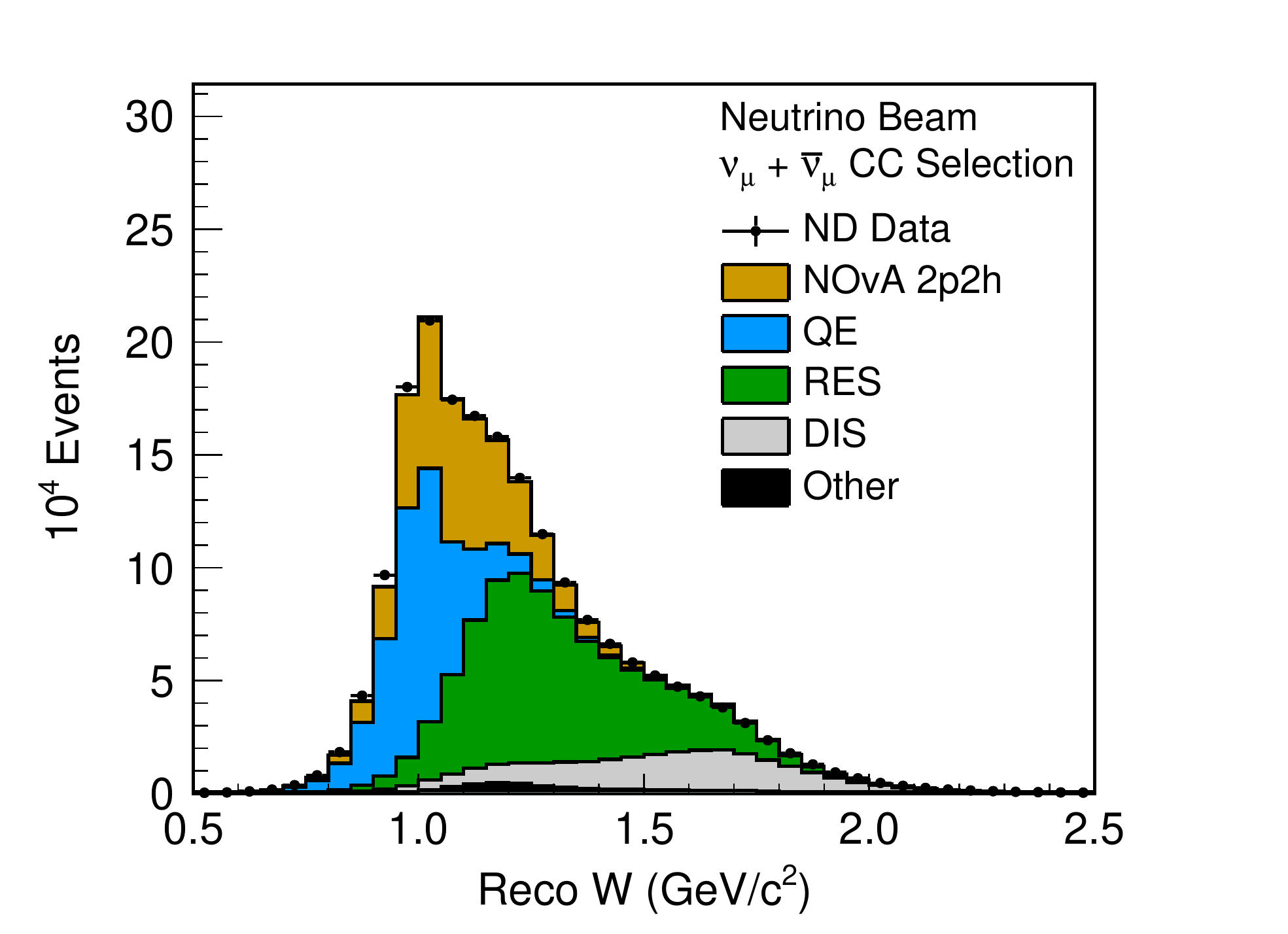}
        \label{fig:finaltunewfhc} }
 \subfloat[]{
        \centering
        \includegraphics[width=.44\textwidth]{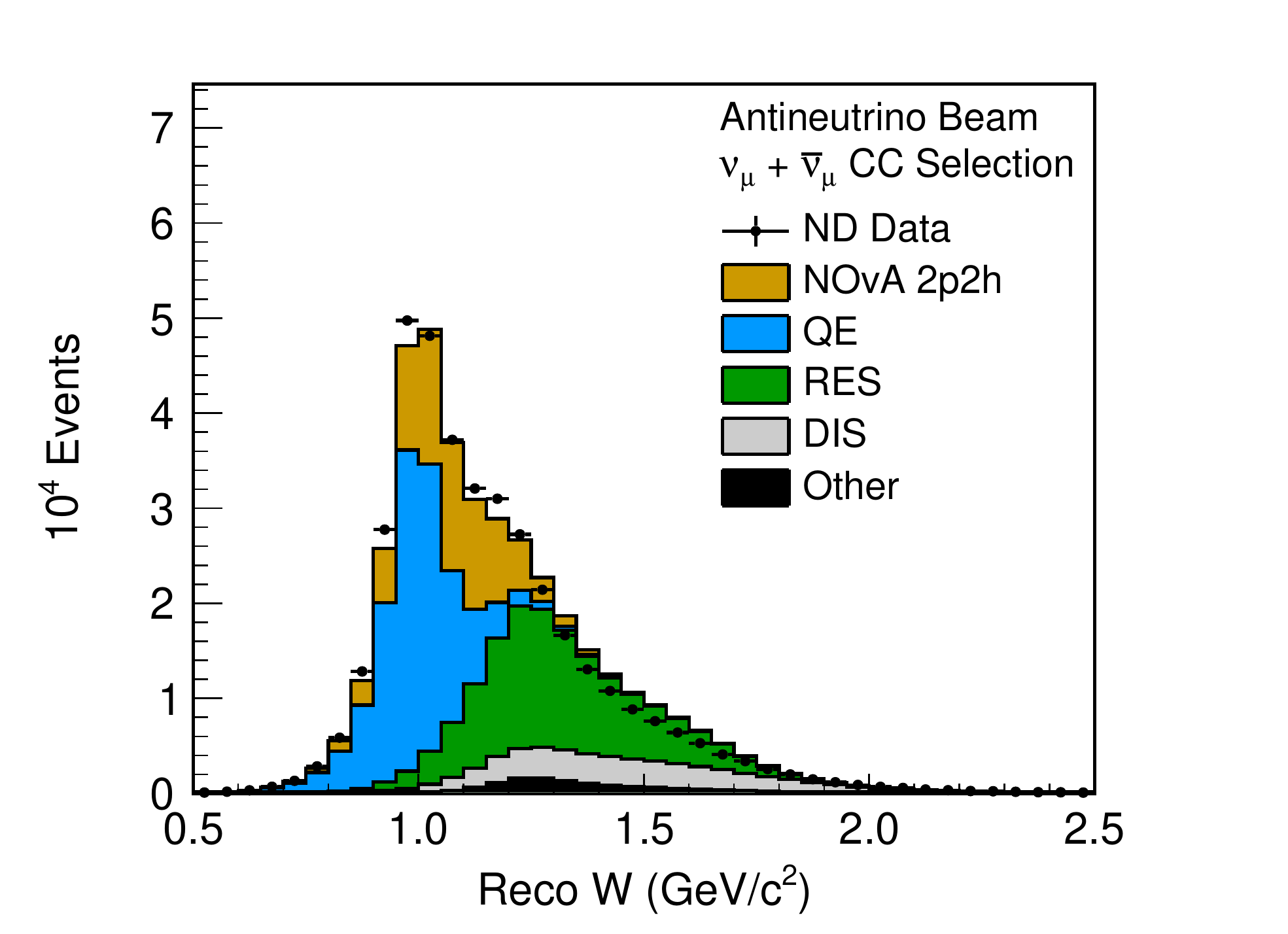}
        \label{fig:finaltunewrhc} }
    \caption{Comparison of fully adjusted simulation to data in reconstructed W, for neutrino beam (left) and antineutrino beam (right).}
    \label{fig:finaltunew}
\end{figure}

\begin{figure}[ht!]
    \centering
 \subfloat[]{
        \centering
        \includegraphics[width=.44\textwidth]{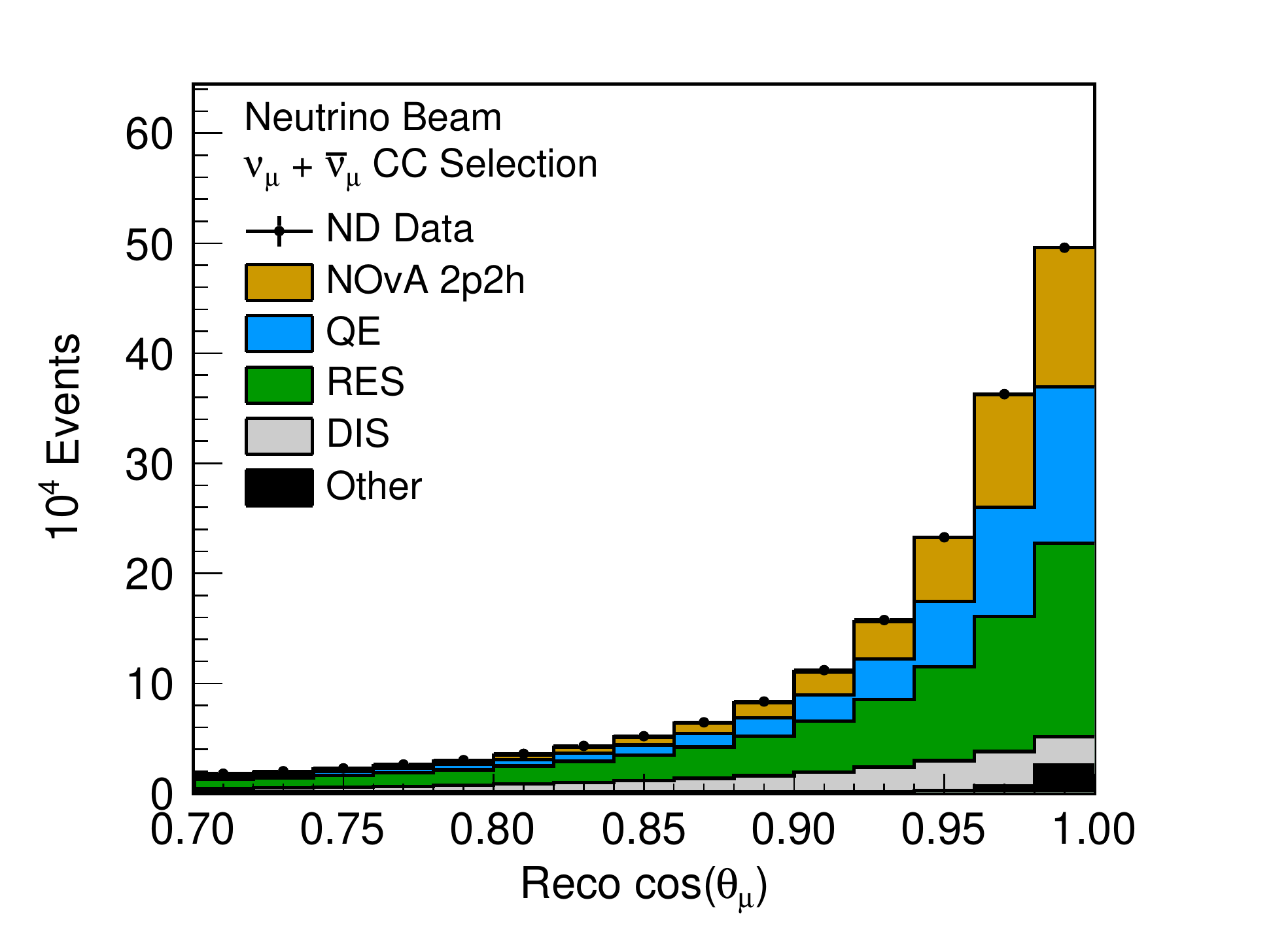}
        \label{fig:finaltunemuanglefhc} }
 \subfloat[]{
        \centering
        \includegraphics[width=.44\textwidth]{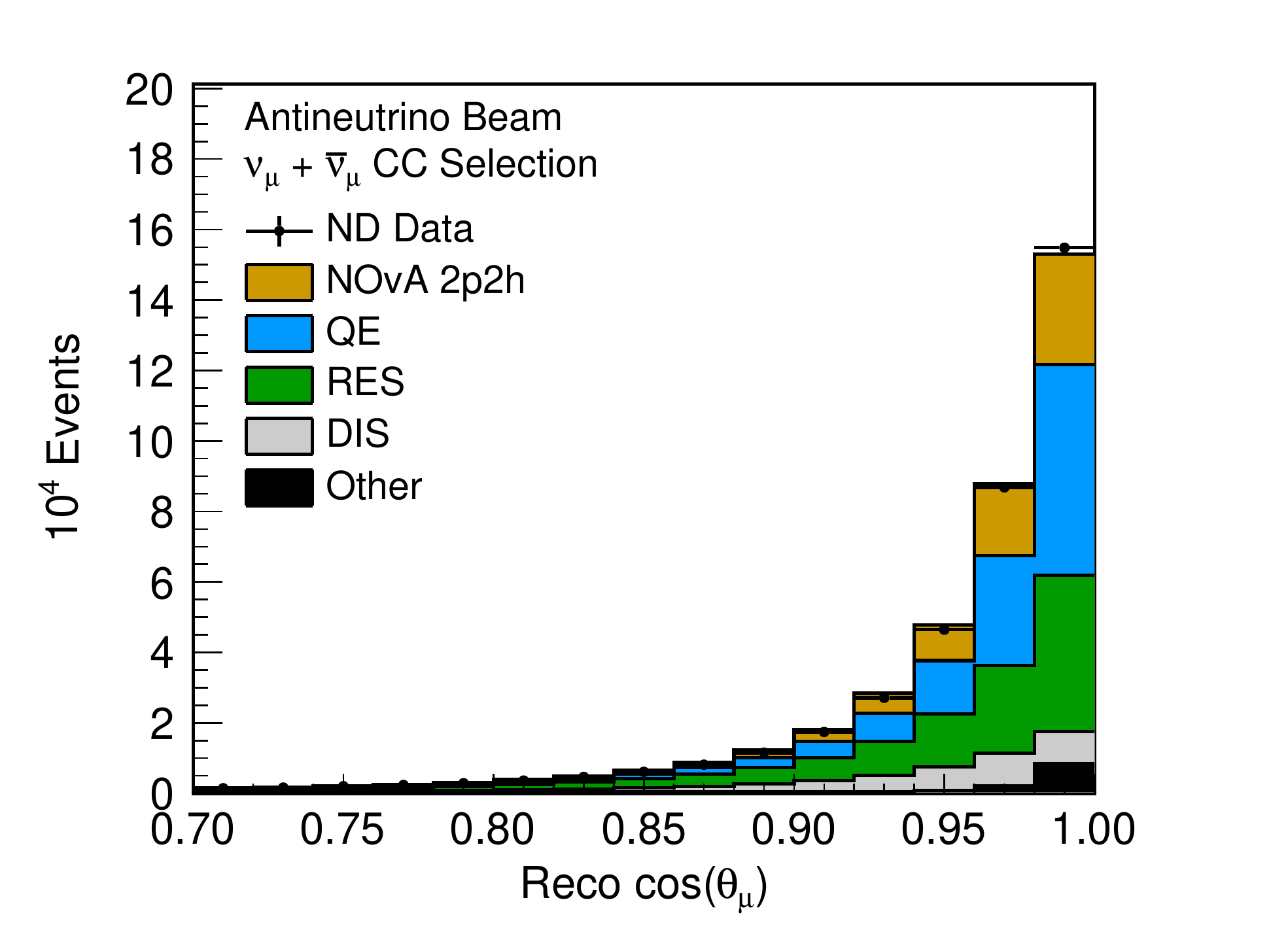}
        \label{fig:finaltunemuanglerhc} }
    \caption{Comparison of fully adjusted simulation to data in reconstructed muon candidate track opening angle, for neutrino beam (left) and antineutrino beam (right).}
    \label{fig:finaltunemuangle}
\end{figure}

\begin{figure}[ht!]
    \centering
 \subfloat[]{
        \centering
        \includegraphics[width=.44\textwidth]{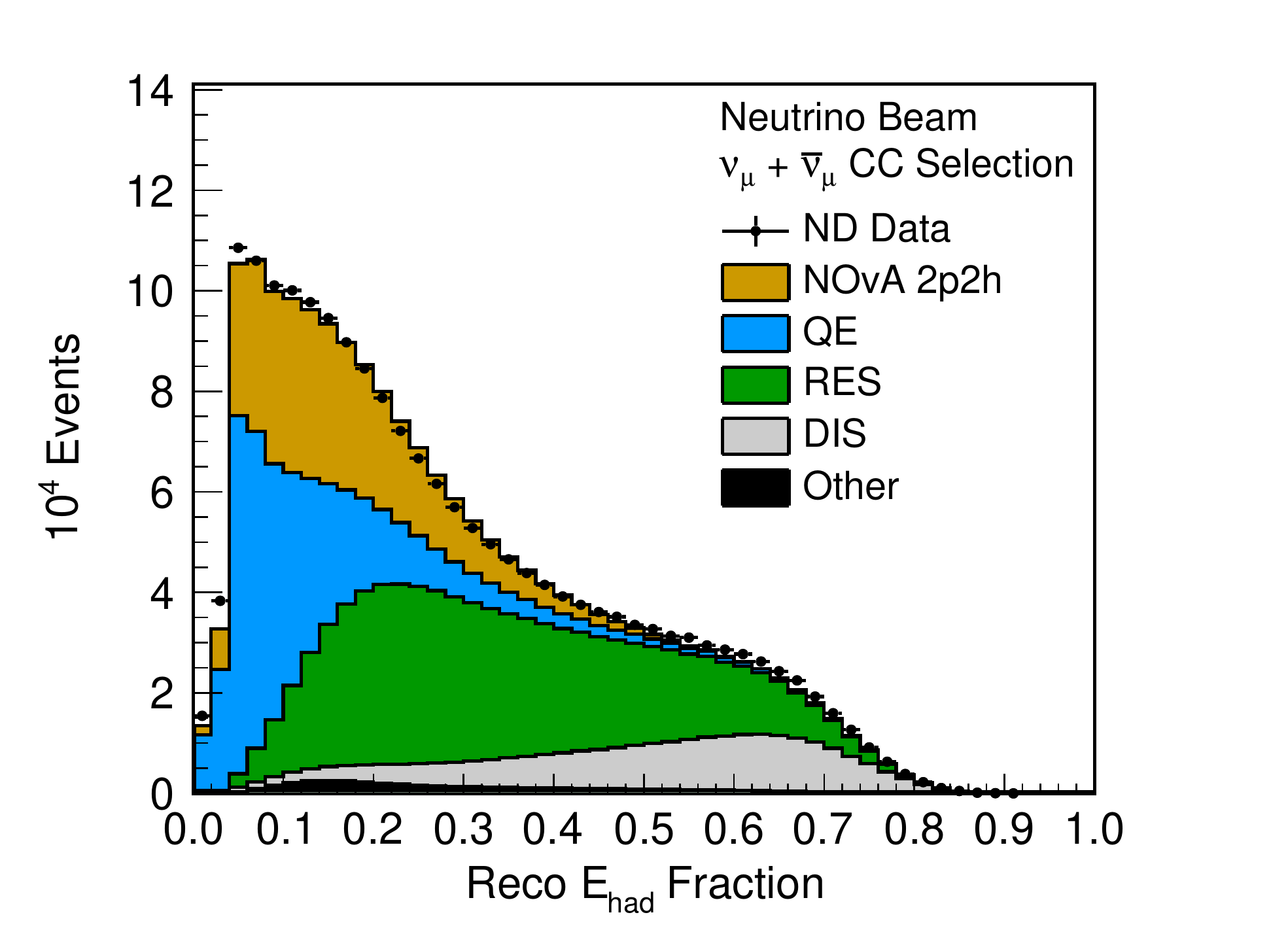}
        \label{fig:finaltunehadefracfhc} }
 \subfloat[]{
        \centering
        \includegraphics[width=.44\textwidth]{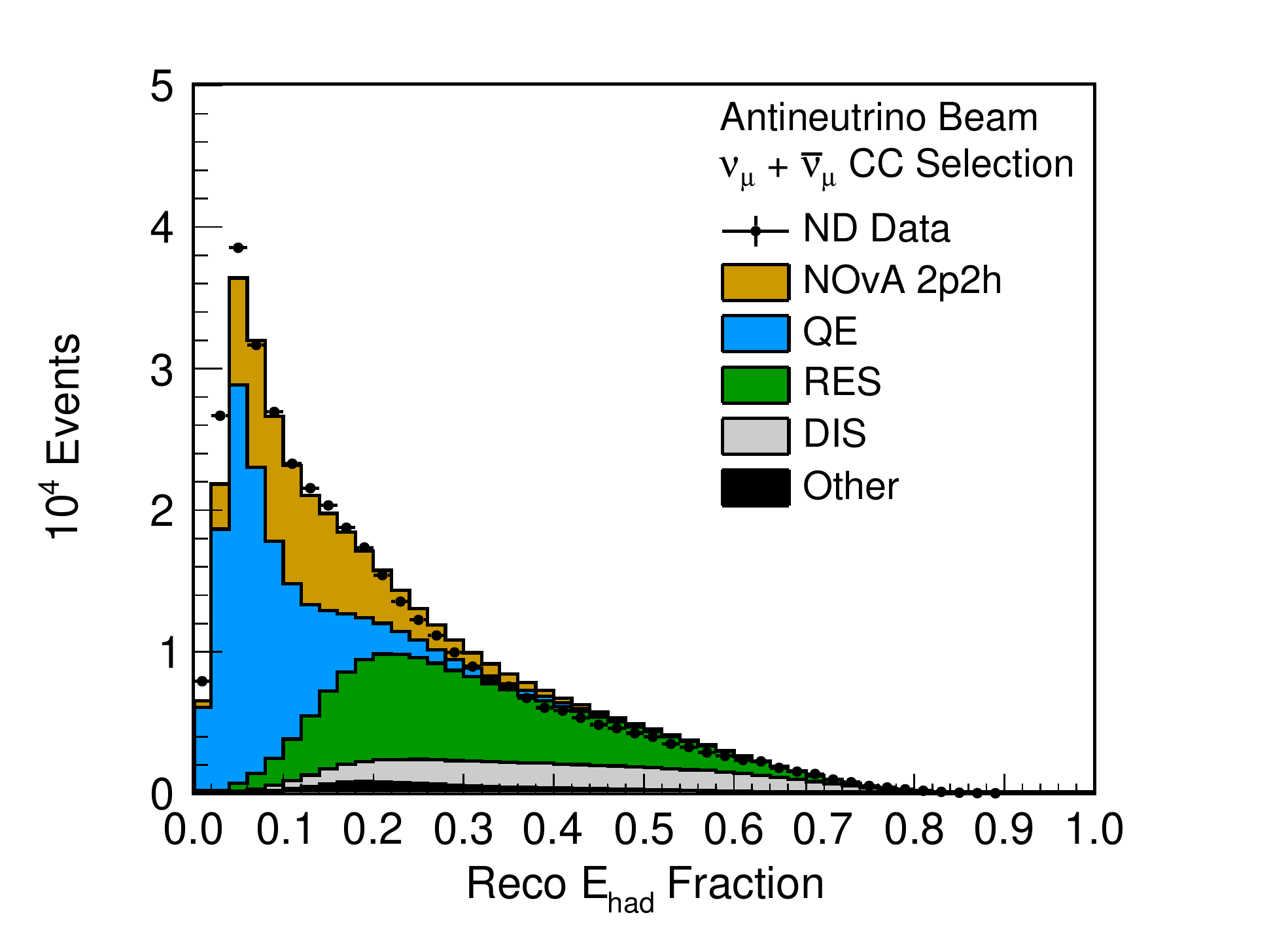}
        \label{fig:finaltunehadefracrhc} }
    \caption{Comparison of fully adjusted simulation to data in reconstructed hadronic energy fraction, for neutrino beam (left) and antineutrino beam (right).}
    \label{fig:finaltunehadefrac}
\end{figure}

\clearpage

\bibliographystyle{apsrev4-1}

\bibliography{nova_xsec_tuning}

\end{document}

%% file: authorlist.inc
\newcommand{\ANL}{Argonne National Laboratory, Argonne, Illinois 60439, 
USA}
\newcommand{\ICS}{Institute of Computer Science, The Czech 
Academy of Sciences, 
182 07 Prague, Czech Republic}
\newcommand{\IOP}{Institute of Physics, The Czech 
Academy of Sciences, 
182 21 Prague, Czech Republic}
\newcommand{\Atlantico}{Universidad del Atlantico,
Carrera 30 No. 8-49, Puerto Colombia, Atlantico, Colombia}
\newcommand{\BHU}{Department of Physics, Institute of Science, Banaras 
Hindu University, Varanasi, 221 005, India}
\newcommand{\UCLA}{Physics and Astronomy Department, UCLA, Box 951547, Los 
Angeles, California 90095-1547, USA}
\newcommand{\Caltech}{California Institute of 
Technology, Pasadena, California 91125, USA}
\newcommand{\Cochin}{Department of Physics, Cochin University
of Science and Technology, Kochi 682 022, India}
\newcommand{\Charles}
{Charles University, Faculty of Mathematics and Physics,
 Institute of Particle and Nuclear Physics, Prague, Czech Republic}
\newcommand{\Cincinnati}{Department of Physics, University of Cincinnati, 
Cincinnati, Ohio 45221, USA}
\newcommand{\CSU}{Department of Physics, Colorado 
State University, Fort Collins, CO 80523-1875, USA}
\newcommand{\CTU}{Czech Technical University in Prague,
Brehova 7, 115 19 Prague 1, Czech Republic}
\newcommand{\Dallas}{Physics Department, University of Texas at Dallas,
800 W. Campbell Rd. Richardson, Texas 75083-0688, USA}
\newcommand{\DallasU}{University of Dallas, 1845 E 
Northgate Drive, Irving, Texas 75062 USA}
\newcommand{\Delhi}{Department of Physics and Astrophysics, University of 
Delhi, Delhi 110007, India}
\newcommand{\JINR}{Joint Institute for Nuclear Research,  
Dubna, Moscow region 141980, Russia}
\newcommand{\FNAL}{Fermi National Accelerator Laboratory, Batavia, 
Illinois 60510, USA}
\newcommand{\UFG}{Instituto de F\'{i}sica, Universidade Federal de 
Goi\'{a}s, Goi\^{a}nia, Goi\'{a}s, 74690-900, Brazil}
\newcommand{\Guwahati}{Department of Physics, IIT Guwahati, Guwahati, 781 
039, India}
\newcommand{\Harvard}{Department of Physics, Harvard University, 
Cambridge, Massachusetts 02138, USA}
\newcommand{\Houston}{Department of Physics, 
University of Houston, Houston, Texas 77204, USA}
\newcommand{\IHyderabad}{Department of Physics, IIT Hyderabad, Hyderabad, 
502 205, India}
\newcommand{\Hyderabad}{School of Physics, University of Hyderabad, 
Hyderabad, 500 046, India}
\newcommand{\IIT}{Illinois Institute of Technology,
Chicago IL 60616, USA}
\newcommand{\Indiana}{Indiana University, Bloomington, Indiana 47405, 
USA}
\newcommand{\INR}{Institute for Nuclear Research of Russia, Academy of 
Sciences 7a, 60th October Anniversary prospect, Moscow 117312, Russia}
\newcommand{\Iowa}{Department of Physics and Astronomy, Iowa State 
University, Ames, Iowa 50011, USA}
\newcommand{\Irvine}{Department of Physics and Astronomy, 
University of California at Irvine, Irvine, California 92697, USA}
\newcommand{\Jammu}{Department of Physics and Electronics, University of 
Jammu, Jammu Tawi, 180 006, Jammu and Kashmir, India}
\newcommand{\Lebedev}{Nuclear Physics and Astrophysics Division, Lebedev 
Physical 
Institute, Leninsky Prospect 53, 119991 Moscow, Russia}
\newcommand{\MSU}{Department of Physics and Astronomy, Michigan State 
University, East Lansing, Michigan 48824, USA}
\newcommand{\Crookston}{Math, Science and Technology Department, University 
of Minnesota Crookston, Crookston, Minnesota 56716, USA}
\newcommand{\Duluth}{Department of Physics and Astronomy, 
University of Minnesota Duluth, Duluth, Minnesota 55812, USA}
\newcommand{\Minnesota}{School of Physics and Astronomy, University of 
Minnesota Twin Cities, Minneapolis, Minnesota 55455, USA}
\newcommand{\Mississippi}{University of Mississippi, University, Mississippi 38677, USA}
\newcommand{\Oxford}{Subdepartment of Particle Physics, 
University of Oxford, Oxford OX1 3RH, United Kingdom}
\newcommand{\Panjab}{Department of Physics, Panjab University, 
Chandigarh, 160 014, India}
\newcommand{\Pitt}{Department of Physics, 
University of Pittsburgh, Pittsburgh, Pennsylvania 15260, USA}
\newcommand{\RAL}{Rutherford Appleton Laboratory, Science and 
Technology Facilities Council, Didcot, OX11 0QX, United Kingdom}
\newcommand{\SAlabama}{Department of Physics, University of 
South Alabama, Mobile, Alabama 36688, USA} 
\newcommand{\Carolina}{Department of Physics and Astronomy, University of 
South Carolina, Columbia, South Carolina 29208, USA}
\newcommand{\SDakota}{South Dakota School of Mines and Technology, Rapid 
City, South Dakota 57701, USA}
\newcommand{\SMU}{Department of Physics, Southern Methodist University, 
Dallas, Texas 75275, USA}
\newcommand{\Stanford}{Department of Physics, Stanford University, 
Stanford, California 94305, USA}
\newcommand{\Sussex}{Department of Physics and Astronomy, University of 
Sussex, Falmer, Brighton BN1 9QH, United Kingdom}
\newcommand{\Syracuse}{Department of Physics, Syracuse University,
Syracuse NY 13210, USA}
\newcommand{\Tennessee}{Department of Physics and Astronomy, 
University of Tennessee, Knoxville, Tennessee 37996, USA}
\newcommand{\Texas}{Department of Physics, University of Texas at Austin, 
Austin, Texas 78712, USA}
\newcommand{\Tufts}{Department of Physics and Astronomy, Tufts University, Medford, 
Massachusetts 02155, USA}
\newcommand{\UCL}{Physics and Astronomy Department, University College 
London, 
Gower Street, London WC1E 6BT, United Kingdom}
\newcommand{\Virginia}{Department of Physics, University of Virginia, 
Charlottesville, Virginia 22904, USA}
\newcommand{\WSU}{Department of Mathematics, Statistics, and Physics,
 Wichita State University, 
Wichita, Kansas 67206, USA}
\newcommand{\WandM}{Department of Physics, William \& Mary, 
Williamsburg, Virginia 23187, USA}
\newcommand{\Wisconsin}{Department of Physics, University of 
Wisconsin-Madison, Madison, Wisconsin 53706, USA}
\newcommand{\deceased}{Deceased.}
\affiliation{\ANL}
\affiliation{\Atlantico}
\affiliation{\BHU}
\affiliation{\Caltech}
\affiliation{\Charles}
\affiliation{\Cincinnati}
\affiliation{\Cochin}
\affiliation{\CSU}
\affiliation{\CTU}
\affiliation{\DallasU}
\affiliation{\Delhi}
\affiliation{\FNAL}
\affiliation{\UFG}
\affiliation{\Guwahati}
\affiliation{\Harvard}
\affiliation{\Houston}
\affiliation{\Hyderabad}
\affiliation{\IHyderabad}
\affiliation{\IIT}
\affiliation{\Indiana}
\affiliation{\ICS}
\affiliation{\INR}
\affiliation{\IOP}
\affiliation{\Iowa}
\affiliation{\Irvine}
\affiliation{\JINR}
\affiliation{\Lebedev}
\affiliation{\MSU}
\affiliation{\Duluth}
\affiliation{\Minnesota}
\affiliation{\Mississippi}
\affiliation{\Panjab}
\affiliation{\Pitt}
\affiliation{\SAlabama}
\affiliation{\Carolina}
\affiliation{\SDakota}
\affiliation{\SMU}
\affiliation{\Stanford}
\affiliation{\Sussex}
\affiliation{\Syracuse}
\affiliation{\Tennessee}
\affiliation{\Texas}
\affiliation{\Tufts}
\affiliation{\UCL}
\affiliation{\Virginia}
\affiliation{\WSU}
\affiliation{\WandM}
\affiliation{\Wisconsin}

\author{M.~A.~Acero}
\affiliation{\Atlantico}

\author{P.~Adamson}
\affiliation{\FNAL}


\author{G.~Agam}
\affiliation{\IIT}

\author{L.~Aliaga}
\affiliation{\FNAL}

\author{T.~Alion}
\affiliation{\Sussex}

\author{V.~Allakhverdian}
\affiliation{\JINR}




\author{N.~Anfimov}
\affiliation{\JINR}


\author{A.~Antoshkin}
\affiliation{\JINR}



\author{L.~Asquith}
\affiliation{\Sussex}


\author{A.~Aurisano}
\affiliation{\Cincinnati}


% ...



